\documentclass[12pt]{article}
\usepackage{jheppub}
\pdfoutput=1

\usepackage{amsmath,bbm,array,amsfonts,graphicx,wrapfig,lscape,float,mathtools,multirow,longtable}
\usepackage{colortbl}
\usepackage[dvipsnames]{xcolor}

\newcommand{\be}{\begin{equation}}
\newcommand{\ee}{\end{equation}}
\newcommand{\beq}{\begin{equation}}
\newcommand{\beql}[1]{\begin{equation}\label{#1}}
\newcommand{\eeq}{\end{equation}}
\newcommand{\ba}{\begin{array}}
\newcommand{\ea}{\end{array}}
\newcommand{\bea}{\begin{eqnarray}}
\newcommand{\beal}[1]{\begin{eqnarray}\label{#1}}
\newcommand{\eea}{\end{eqnarray}}
\newcommand{\ben}{\begin{enumerate}}
\newcommand{\een}{\end{enumerate}}
\newcommand{\bean}{\begin{eqnarray*}}
\newcommand{\eean}{\end{eqnarray*}}
\newcommand{\eref}[1]{(\ref{#1})}
\newcommand{\sref}[1]{\S\ref{#1}}
\newcommand{\tref}[1]{Table~\ref{#1}}
\newcommand{\nn}{\nonumber}

\newcommand{\fref}[1]{Figure \ref{#1}}
\newcommand{\btab}[1]{\begin{tabular}{#1}}
\newcommand{\etab}{\end{tabular}}

\def \Black {\pdfsetcolor{0 0 0 1}}

\newcommand{\comment}[1]{}

\newcommand{\qed}{\nobreak \ifvmode \relax \else
      \ifdim\lastskip<1.5em \hskip-\lastskip
      \hskip1.5em plus0em minus0.5em \fi \nobreak
      \vrule height0.75em width0.5em depth0.25em\fi}

\definecolor{darkspringgreen}{rgb}{0.09, 0.45, 0.27}
\definecolor{forestgreen}{rgb}{0.13, 0.55, 0.13}
\definecolor{ablue}{RGB}{0, 156, 252}

\title{Brane Brick Models and $2d$ $(0,2)$ Triality}

\author[a,b]{Sebasti\'an Franco,}
\author[c,d,e,f]{Sangmin Lee,}
\author[g]{Rak-Kyeong Seong}

\affiliation[a]{
Physics Department, The City College of the CUNY \\
160 Convent Avenue, New York, NY 10031, USA}

\affiliation[b]{The Graduate School and University Center, The City University of New York  \\
365 Fifth Avenue, New York NY 10016, USA }

\affiliation[c]{
Center for Theoretical Physics, Seoul National University, Seoul 08826, Korea
}

\affiliation[d]{
Department of Physics and Astronomy, Seoul National University, Seoul 08826, Korea
}

\affiliation[e]{
College of Liberal Studies, Seoul National University, Seoul 08826, Korea
}

\affiliation[f]{
School of Natural Sciences, Institute for Advanced Study, Princeton, NJ 08540, USA
}

\affiliation[g]{
School of Physics, Korea Institute for Advanced Study, Seoul 02455, Korea
}

\emailAdd{sfranco@ccny.cuny.edu}
\emailAdd{sangmin@snu.ac.kr}
\emailAdd{rakkyeongseong@gmail.com}

\preprint{
\begin{flushright}
CCNY-HEP-16-01 \\
SNUTP-15-012 \\
KIAS-P15062
\end{flushright}
}

\abstract{
We provide a brane realization of $2d$ $(0,2)$ Gadde-Gukov-Putrov triality in terms of brane brick models. These are Type IIA brane configurations that are T-dual to D1-branes over singular toric Calabi-Yau 4-folds. Triality translates into a local transformation of brane brick models, whose simplest representative is a cube move. We present explicit examples and construct their triality networks. We also argue that the classical mesonic moduli space of brane brick model theories, which corresponds to the probed Calabi-Yau 4-fold, is invariant under triality. Finally, we discuss triality in terms of phase boundaries, which play a central role in connecting Calabi-Yau 4-folds to brane brick models. 
}

\begin{document}

\maketitle

\section{Introduction}

While $2d$ $\mathcal{N}=(0,2)$ theories have only two supercharges, their dynamics is under considerable control due to chirality, holomorphy and anomalies. In recent years, we have witnessed remarkable developments in our understanding of these theories. They include: $c$-extremization \cite{Benini:2012cz,Benini:2013cda}, exact CFT description of the low energy limit of SQCD-like theories \cite{Gadde:2014ppa}, detailed studies of renormalization group (RG) flows \cite{Gadde:2015kda}, connections to $4d$ $\mathcal{N}=1$ theories via dimensional reduction \cite{Kutasov:2013ffl,Kutasov:2014hha} and to $6d$ $(2,0)$ theories by compactification on 4-manifolds \cite{Gadde:2013sca}. In addition, it has been discovered that $2d$ $(0,2)$ theories exhibit IR dualities \cite{Gadde:2013lxa} similar to Seiberg duality in $4d$ $\mathcal{N}=1$ gauge theories \cite{Seiberg:1994pq}. This low energy equivalence is called {\it triality}, since it is only after three of these transformations acting on the same gauge group that we return to the original theory.

Embedding quantum field theories in string or M-theory provides a powerful grip on their dynamics. This approach has been particularly helpful for understanding, and in some cases uncovering, quantum field theory dualities in various dimensions and with different amounts of supersymmetry. This program has been recently extended to $2d$ $(0,2)$ gauge theories, following the pioneering work in \cite{GarciaCompean:1998kh}. In \cite{Franco:2015tna},  a systematic procedure for deriving the $2d$ $(0,2)$ gauge theories on the worldvolume of D1-branes probing  generic singular toric Calabi-Yau (CY) 4-folds was developed.\footnote{Other interesting recent approaches for constructing $2d$ $(0,2)$ theories include: stacks of D5-branes wrapped over 4-cycles of resolved/deformed conifolds fibered over a 2-torus \cite{Tatar:2015sga}, compactifications on Riemann surfaces of $4d$ $\mathcal{N}=1$ quiver gauge theories on D3-branes over CY$_3$ singularities \cite{Benini:2015bwz} and F-theory compactifications on singular, elliptically fibered CY 5-folds \cite{Schafer-Nameki:2016cfr}.} The general structure of this infinite class of theories was then studied in detail. The brane engineering of these gauge theories was taken a further step forward in \cite{Franco:2015tya}, which introduced a new type of Type IIA brane configurations, denoted {\it brane brick models}, which are related to the D1-branes over toric CY$_4$ singularities by T-duality. Brane brick models fully encode the gauge theories on the worldvolume of the D1-branes and extremely simplify the connection to the probed geometries.

The previous discussion leads to a natural question: is there a brane realization of  $2d$ $(0,2)$ triality? This is the central topic we set to explore in this article, in which we will explain how triality is nicely realized in the context of brane brick models.

In this article we will see that, in general, brane brick models associate a class of $2d$ $(0,2)$ quiver gauge theories to every toric CY$_4$ singularity. The multiple gauge theories within a class can be constructed using several methods developed in \cite{Franco:2015tna,Franco:2015tya}. They turn out to be related by triality, which is realized as a local transformation of the brane brick models.\footnote{In a sense, the situation is reminiscent of the early investigations of {\it toric duality} \cite{Feng:2000mi,Feng:2001xr}. Several $4d$ $\mathcal{N}=1$ gauge theories were constructed for a given toric CY$_3$ singularity. It was later realized that the transformation relating different theories associated to the same CY$_3$ was precisely Seiberg duality \cite{Beasley:2001zp,Feng:2001bn}.} The simplest example of this transformation is a cube move.  The probed CY$_4$ corresponds to the mesonic moduli space of the gauge theories and it is thus common to all brane brick models related by triality. As shown in \cite{Gadde:2014ppa,Gadde:2015kda}, $2d$ $(0,2)$ SQCD theories with different ranks of the three flavor groups exhibit triality as an exact IR equivalence even when the classical mesonic moduli spaces are different. Our brane brick models are, however, similar to the case of SQCD with equal ranks of the flavor groups, for which the three phases share the same moduli space.

This paper is organized as follows. 
Section \sref{sec:brane-brick} reviews the basic concepts of the brane brick models introduced in \cite{Franco:2015tya}.
Section \sref{sec:triality-review} reviews the original proposal of \cite{Gadde:2013lxa} for triality for $2d$ $(0,2)$ supersymmetric QCD and certain quiver generalizations.
Section \sref{sec:brick-triality} explains how triality is implemented in terms of brane brick models. 
Section \sref{sec:examples} presents explicit examples, for which we construct part of their triality networks \cite{Gadde:2013lxa}. 
In section \sref{sec:mesonic-moduli}, we show in examples that triality preserves the classical mesonic moduli space and sketch a general proof of its invariance for general brane brick models.
In section \sref{sec:phase-boundary}, we examine triality from the perspective of phase boundaries, which bridge CY$_4$ singularities and brane brick models.
Section \sref{sec:conclusion} offers conclusions and directions for future work.  
In the two appendices, we collect detailed data for the explicit examples used in the main text.

\section{Brane Brick Models \label{sec:brane-brick}}

This section contains a lightning review of brane brick models. Its primary goal is to introduce the basic concepts and nomenclature. It is not intended to be complete, and we encourage the reader to look at \cite{Franco:2015tna,Franco:2015tya} for in depth presentations.

Brane brick models were introduced in \cite{Franco:2015tya} as a powerful tool for studying the $2d$ $(0,2)$ quiver gauge theories that arise on the worldvolume of D1-branes probing toric CY$_4$ singularities. A brane brick model is a Type IIA brane configuration of D4-branes suspended from an NS5-brane. The NS5-brane extends along the $(01)$ directions and wraps a holomorphic surface (i.e. four real dimensions) $\Sigma$ embedded into the $(234567)$ directions. The directions $(246)$ are periodically identified to form a $T^3$. The coordinates $(23)$, $(45)$ and $(67)$ are pairwise combined to form three complex variables $x$, $y$ and $z$ of which $(246)$ are the arguments. The surface $\Sigma$ is given by the zero locus of the Newton polynomial associated to the toric diagram of the CY$_4$: $P(x,y,z)=0$. Stacks of D4-branes extend along $(01)$ and are suspended inside the voids cut out by $\Sigma$ on the $T^3$ given by the $(246)$ directions. The $2d$ $(0,2)$ gauge theory lives in the $(01)$ directions, which are common to all the branes. \tref{tbconfig} summarizes the basic ingredients of a brane brick model.

\begin{table}[ht!!]
\centering
\begin{tabular}{c|cccccccccc}
\; & 0 & 1 & 2 & 3 & 4 & 5 & 6 & 7 & 8 & 9\\
\hline
\text{D4} & $\times$ & $\times$ & $\times$ & $\cdot$ & $\times$ & $\cdot$ & $\times$ & $\cdot$ & $\cdot$ & $\cdot$ \\
\text{NS5} & $\times$ & $\times$ & \multicolumn{6}{c}{----------- \ $\Sigma$ \ ------------} & $\cdot$ & $\cdot$\\
\end{tabular}
\caption{Brane brick models are Type IIA configurations with D4-branes suspended from an NS5-brane that wraps a holomorphic surface $\Sigma$.}
\label{tbconfig}
\end{table}

Most of the non-trivial information concerning a brane brick model is captured by its skeleton on the $T^3$. 
For brevity, we will refer to both the full brane configuration and this simpler object as the brane brick model. Every brane brick model fully encodes a $2d$ $(0,2)$ gauge theory according to the dictionary in \tref{tbrick}. Bricks correspond to $U(N)$ gauge groups.\footnote{All ranks are equal in the T-dual of a stack of regular D1-branes at the CY$_4$. More generally, allowing for fractional D1-branes can lead to brane brick models in which bricks have different numbers of D4-branes.} There are two types of faces. First, there are oriented faces, which correspond to chiral fields. In addition, there are unoriented faces, each of which represents a Fermi field $\Lambda$ and its conjugate $\bar{\Lambda}$. Fermi faces are always 4-sided. Finally, edges of the brane brick model are associated to monomials in $J$- or $E$-terms. For detailed discussions of $2d$ $(0,2)$ theories, including their supermultiplet structure and the construction of their Lagrangians in $(0,2)$ superspace, we refer to \cite{Witten:1993yc,GarciaCompean:1998kh,Gadde:2013lxa,Kutasov:2013ffl}.

\begin{table}[h]
\centering
\resizebox{0.9\hsize}{!}{
\begin{tabular}{|l|l|}
\hline
\ \ \ \ \ \ {\bf Brane Brick Model} \ \ \ \ \ & \ \ \ \ \ \ \ \ \ \ \ \ \ \ \ \ \ \ \ \ {\bf Gauge Theory} \ \ \ \ \ \ \ \ \ \ \ \ 
\\
\hline\hline
Brick  & Gauge group \\
\hline
Oriented face between bricks & Chiral field in the bifundamental representation \\
$i$ and $j$ & of nodes $i$ and $j$ (adjoint for $i=j$) \\
\hline
Unoriented square face between & Fermi field in the bifundamental representation \\
bricks $i$ and $j$ & of nodes $i$ and $j$ (adjoint for $i=j$) \\
\hline
Edge  & Plaquette encoding a monomial in a \\ 
& $J$- or $E$-term \\
\hline
\end{tabular}
}
\caption{
Dictionary between brane brick models and $2d$ gauge theories.
\label{tbrick}
}
\end{table}

\begin{figure}[h]
\begin{center}
\resizebox{0.7\hsize}{!}{
\includegraphics[width=8cm]{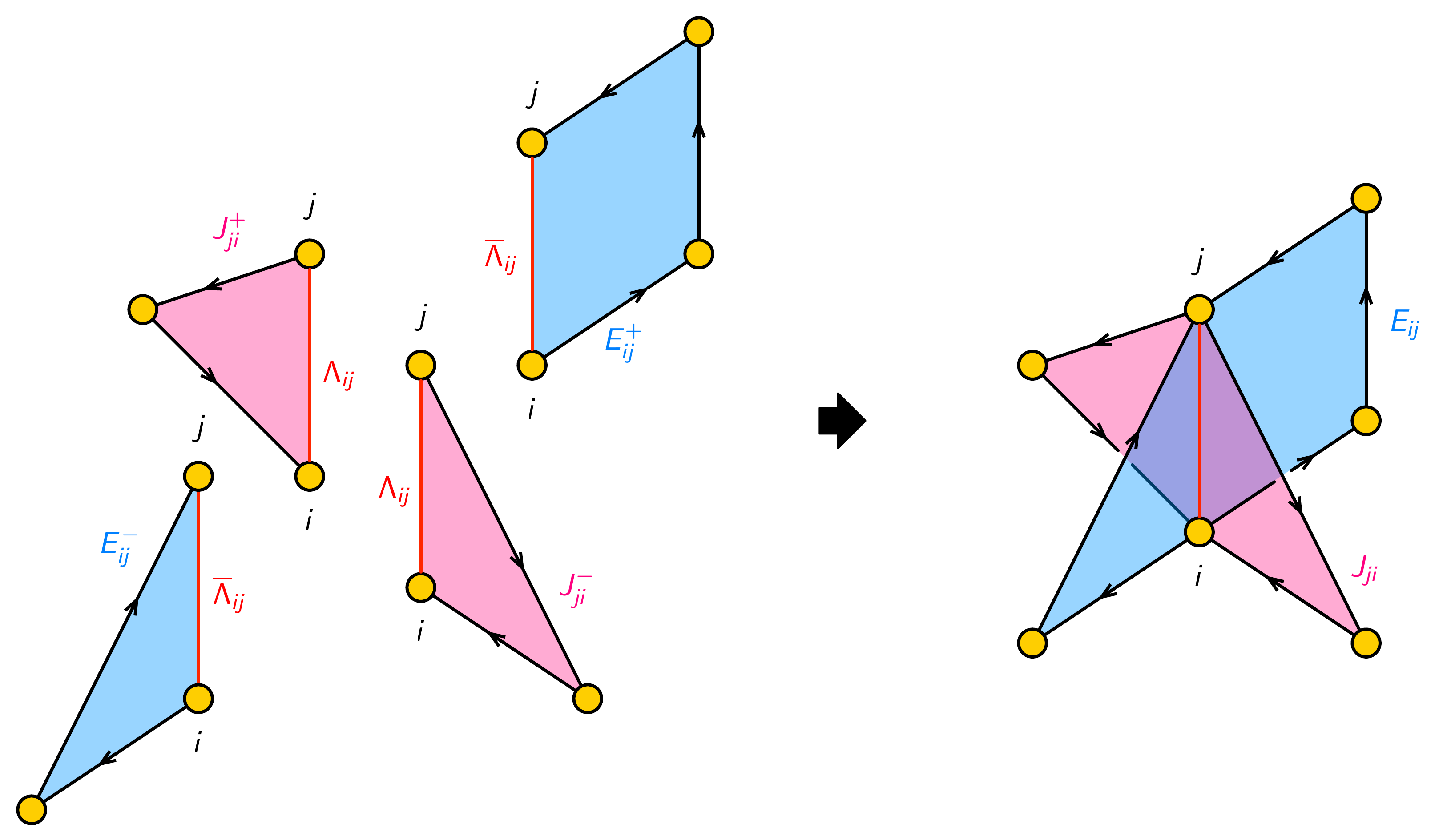}
}
\caption{
The four plaquettes $(\Lambda_{ij}, J^{\pm}_{ji})$ and $(\overline\Lambda_{ij}, E^\pm_{ij})$ associated to a Fermi field $\Lambda_{ij}$. The $J$- and $E$-terms are $J_{ji}=J^{+}_{ji}- J^{-}_{ji}=0$ and $E_{ij}=E_{ij}^{+}- E_{ij}^{-}=0$, respectively, where $J_{ji}^\pm$ and $E_{ij}^{\pm}$ are holomorphic monomials in chiral fields.
\label{fplaquettes}}
 \end{center}
 \end{figure}

Brane brick models are in one-to-one correspondence with {\it periodic quivers} on $T^3$. Periodic quivers also fully capture the gauge symmetry, matter content, and $J$- and $E$-terms of a $2d$ $(0,2)$ theory \cite{Franco:2015tna}. The latter correspond to \textit{minimal plaquettes}. A plaquette is a closed loop in the quiver consisting of an oriented path of chiral fields and a single Fermi field. Every Fermi field in this class of theories is associated to two pairs of minimal plaquettes as shown in \fref{fplaquettes}, which translates into monomial relations from vanishing $J$- and $E$-terms. This is known as the \textit{toric condition} and is a general property of the gauge theories on D1-branes over toric CY$_4$ singularities \cite{Franco:2015tna}.

The brane brick model can be regarded as a tropical limit of the {\it coamoeba} projection of $\Sigma$ onto $(\arg(x),\arg(y),\arg(z))$.\footnote{For this reason, in the restricted context of orbifolds of $\mathbb{C}^4$, this object has been referred to as a {\it tropical coamoeba} \cite{2010arXiv1001.4858F}.} 
The alternative projection of $\Sigma$ onto $(|x|,|y|,|z|)$ is called the {\it amoeba projection}. The amoeba approaches infinity along ``legs" that are in one-to-one correspondence with edges of the toric diagram. Along each of these asymptotic legs, the coamoeba simplifies and reduces to a $2d$-plane in $T^3$ that is orthogonal to the corresponding edge of the toric diagram. We refer to these planes as {\it phase boundaries}. We point the reader to \cite{Franco:2015tya} for the details of this construction.\footnote{More generally, phase boundaries are $2d$ surfaces, not necessarily planes, whose homology on $T^3$ is determined by the corresponding edge in the toric diagram \cite{Franco:2015tya}.}

\begin{figure}[h]
\begin{center}
\resizebox{0.65\hsize}{!}{
\includegraphics[width=8cm]{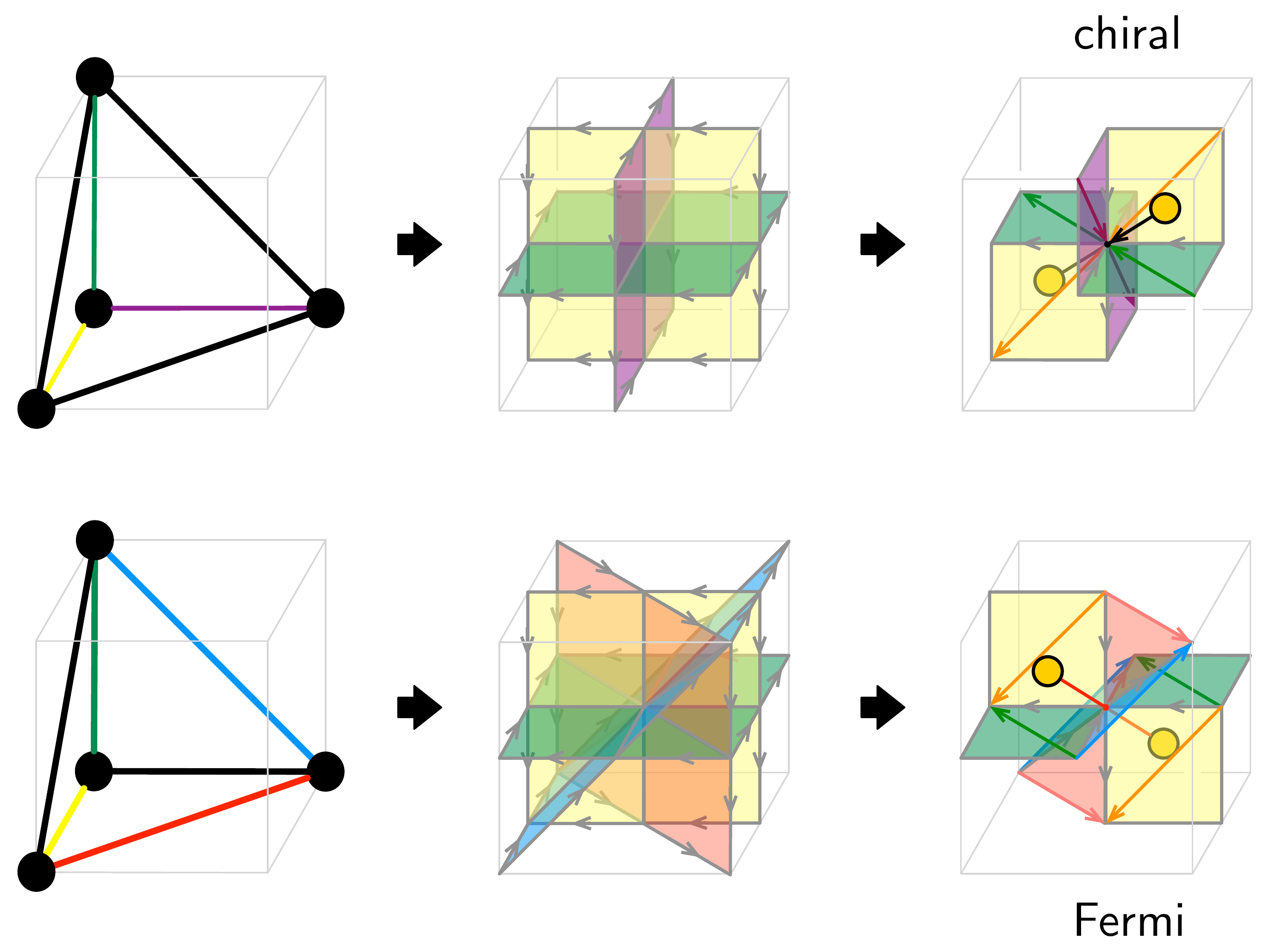}
}
\caption{
Phase boundaries are in one-to-one correspondence with edges of the toric diagram of the Calabi-Yau 4-fold. Certain point intersections of phase boundaries give rise to chiral or Fermi fields, depending on whether they are oriented or alternating.
\label{ffieldplanes}}
 \end{center}
 \end{figure}

Brane brick models can be reconstructed from phase boundaries. This procedure is called the {\it fast inverse algorithm} and makes it possible to go from toric diagrams to brane brick models \cite{Franco:2015tya}. Chiral and Fermi fields arise at point intersections between them. Phase boundaries divide the neighborhood of every intersection point into a collection of cones. In \cite{Franco:2015tya}, we introduced a prescription for assigning an orientation to every phase boundary. The distinction between chiral and Fermi fields depends on the orientation properties of the cones at the corresponding intersection. Chiral field are associated to {\it oriented intersections}, which are defined as those containing two opposite oriented cones. An oriented cone is one for which all phase boundaries on its boundary, which might be a subset of all the ones participating in the intersection, are oriented towards the intersection or away from it. On the other hand, Fermi fields correspond to {\it alternating intersections}, which are those that contain a pair of {\it alternating cones}. Alternating cones are cones in which the orientations of the line intersections between consecutive pairs of phase boundaries alternate between going into and away from the intersection. The chiral and Fermi fields in the periodic quiver associated to the two types of intersections are aligned with the corresponding oriented and alternating cones, respectively. \fref{ffieldplanes} presents two examples of this construction. It is possible for a point intersection of phase boundaries to be neither oriented nor alternating. In this case, it does not correspond to any field in the gauge theory.

Remarkably, brane brick models not only encapsulate the entire $2d$ $(0,2)$ gauge theory data, but also substantially simplify the connection to the probed CY$_4$ geometry. We have already seen glimpses of this beautiful relation in our discussion of the interplay between toric diagrams, phase boundaries and brane brick models. The connection to geometry becomes even more tantalizing in terms of a new type of combinatorial objects denoted {\it brick matchings}. A brick matching is defined as a collection of chiral, Fermi and conjugate Fermi fields that contribute exactly once to every plaquette in the theory, while satisfying some additional simple rules (see section \sref{section_review_fast_forward_algorithm} for the complete definition).
Brick matchings are in one-to-one correspondence with the GLSM fields describing the classical mesonic moduli space of the gauge theory, namely the CY$_4$ singularity. They are the key ingredients of the {\it fast forward algorithm}, a powerful method for obtaining the probed geometry \cite{Franco:2015tya}, which we review in section \sref{section_review_fast_forward_algorithm}.

\section{$2d$ $(0,2)$ SQCD and Triality \label{sec:triality-review}}

In this section, we review triality for $2d$ $(0,2)$ theories. This is a low energy equivalence between gauge theories originally introduced in \cite{Gadde:2013lxa}.

\begin{figure}[h]
\begin{center}
\resizebox{0.9\hsize}{!}{
\includegraphics[width=8cm]{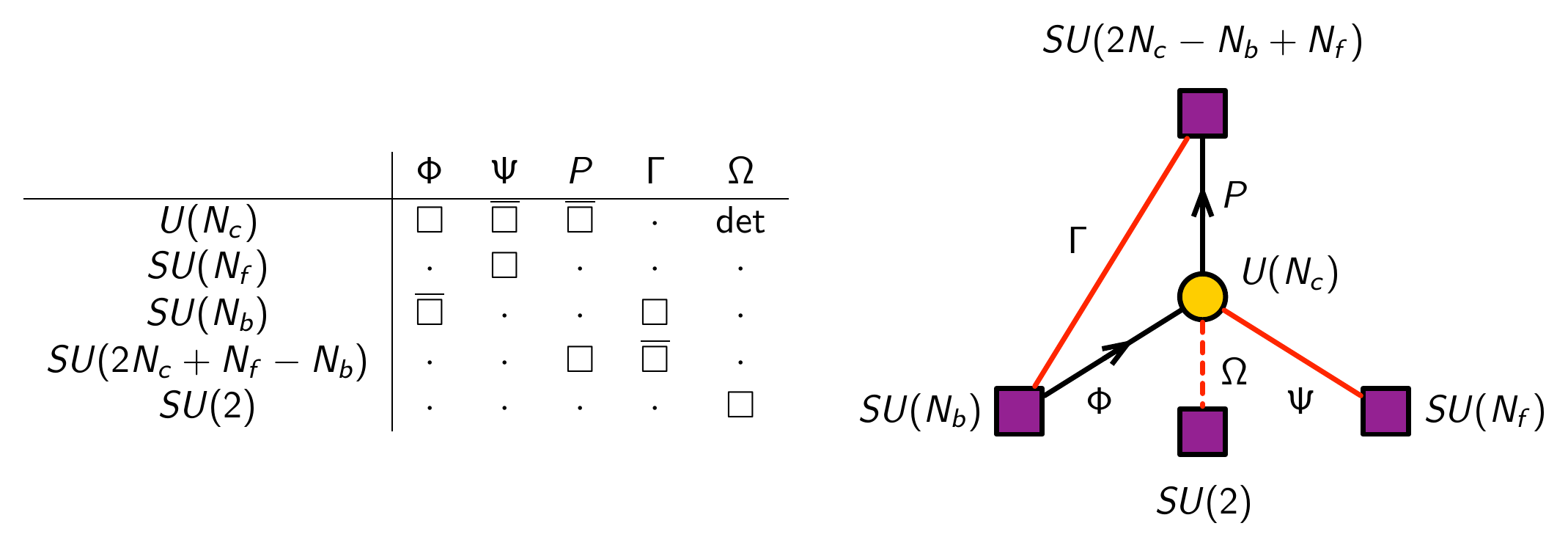}
}
\caption{
The quiver diagram for $2d$ $(0,2)$ SQCD (original theory $D$). Square nodes indicate flavor symmetry groups.
\label{ftheoryA}}
 \end{center}
 \end{figure}

Let us consider $2d$ $(0,2)$ SQCD with $U(N_c)$ gauge group. This theory has $N_b$ chiral fields $\Phi$ in the fundamental representation of $U(N_c)$ and $N_f$ Fermi fields $\Psi$ in the antifundamental representation. The $\Phi$ contribute $\frac{1}{2}N_b$, the $\Psi$ contribute $-\frac{1}{2}N_f$ and the vector multiplet contributes $-N_c$ to the $SU(N_c)^2$ anomaly. The resulting $-\frac{1}{2}(2N_c + N_b - N_f)$ anomaly can be cancelled by introducing $2N_c + N_b - N_f$ chiral multiplets $P$ in the antifundamental representation of $U(N_c)$. These three types of fields give rise to an $SU(N_b)\times SU(N_f) \times SU(2N_c + N_b - N_f)$ global symmetry. In addition, we introduce a Fermi field $\Gamma$, which is a singlet of the gauge symmetry and transforms in the bifundamental representation of the global $SU(N_b) \times SU(2N_c + N_b - N_f)$. Finally, in order to cancel the anomaly in the $U(1)$ part of $U(N_c)$, we introduce two Fermi multiplets $\Omega$ in the determinant representation of $U(N_c)$. The theory can be represented by the quiver diagram shown in \fref{ftheoryA}.\footnote{In this article we adopt the convention that the head and tail of the arrow associated to a chiral field correspond to fundamental and antifundamental representations, respectively.} We include a $J$-term for the Fermi field $\Gamma$: $J_\Gamma=\Phi P$. This term corresponds to the triangular plaquette in the quiver and is also sometimes referred to as a $\Phi P \Gamma$ superpotential.

\begin{figure}[H]
\begin{center}
\resizebox{1\hsize}{!}{
\includegraphics[width=8cm]{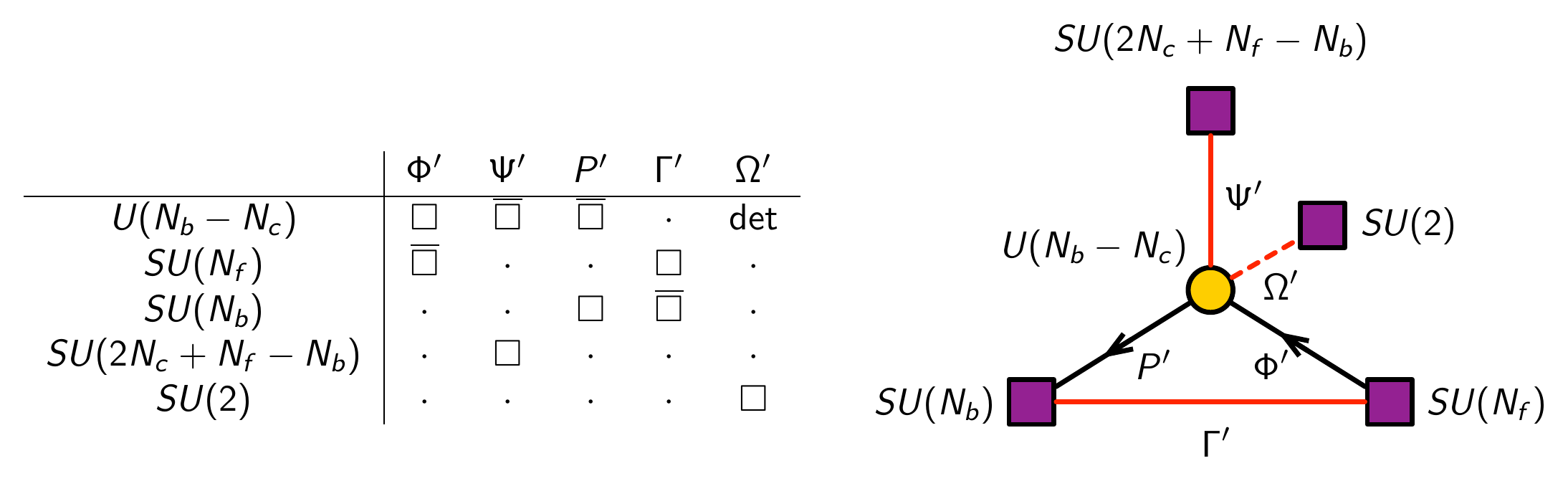}
}
\caption{
The quiver diagram for the dual of $2d$ $(0,2)$ SQCD (dual theory $D^\prime$). 
\label{ftheoryB}}
 \end{center}
 \end{figure}

Triality turns the original theory into the one shown in \fref{ftheoryB}. The new gauge group is $U(N_c')$, with $N_c'=N_b-N_c$. The structure of the new theory is identical to the original one up to a counterclockwise 120$^\circ$ rotation. This means that the fundamental chirals $\Phi$, antifundamental chirals $P$ and fundamental Fermis $\Psi$ are replaced by antifundamental chirals $P'$, fundamental Fermis $\Psi'$ and fundamental chirals $\Phi'$, respectively. Borrowing the terminology of $4d$ Seiberg duality, these fields are the analogues of magnetic flavors. This theory also requires a pair of Fermi fields $\Omega'$ in the determinant representation of $U(N_c')$ to cancel the $U(1)^2$ anomaly. The new Fermi singlet $\Gamma'$ is a mesonic field that, in terms of the electric flavors, is given by $\Gamma'=\Phi \Psi$. The dual flavors and the Fermi meson are coupled by a superpotential $\Phi' P' \Gamma'$. The disappearance of the original Fermi singlet $\Gamma$ can be understood as follows. The original chiral fields combine into a chiral field meson $M=\Phi P$ in the dual theory, which transforms in the bifundamental representation of $SU(2N_c + N_b - N_f) \times SU(N_b)$. The original superpotential becomes $M\Gamma$, giving a mass for $M$ and $\Gamma$, which can thus be integrated out and disappear at low energies.

Let us call the original theory in \fref{ftheoryA} theory $D$ and the dual theory in \fref{ftheoryB} theory $D^\prime$. Applying the triality transformation once more, we obtain a third theory $D^{\prime\prime}$. Its gauge group is $U(N_c^{\prime\prime})$, with $N_c^{\prime\prime}=N_b^\prime - N_c^\prime = N_f - N_b + N_c$. As before, its quiver diagram is obtained from the one for $D^\prime$ by a counterclockwise 120$^\circ$ rotation.  It is only after a third dualization that we obtain $N_c^{\prime\prime\prime} = N_b^{\prime\prime} - N_{c}^{\prime\prime} = N_c$ and we return to the original theory. The fact that this is an IR equivalence between three theories motivates calling it a triality. 

\begin{figure}[H]
\begin{center}
\resizebox{0.8\hsize}{!}{
\includegraphics[width=8cm]{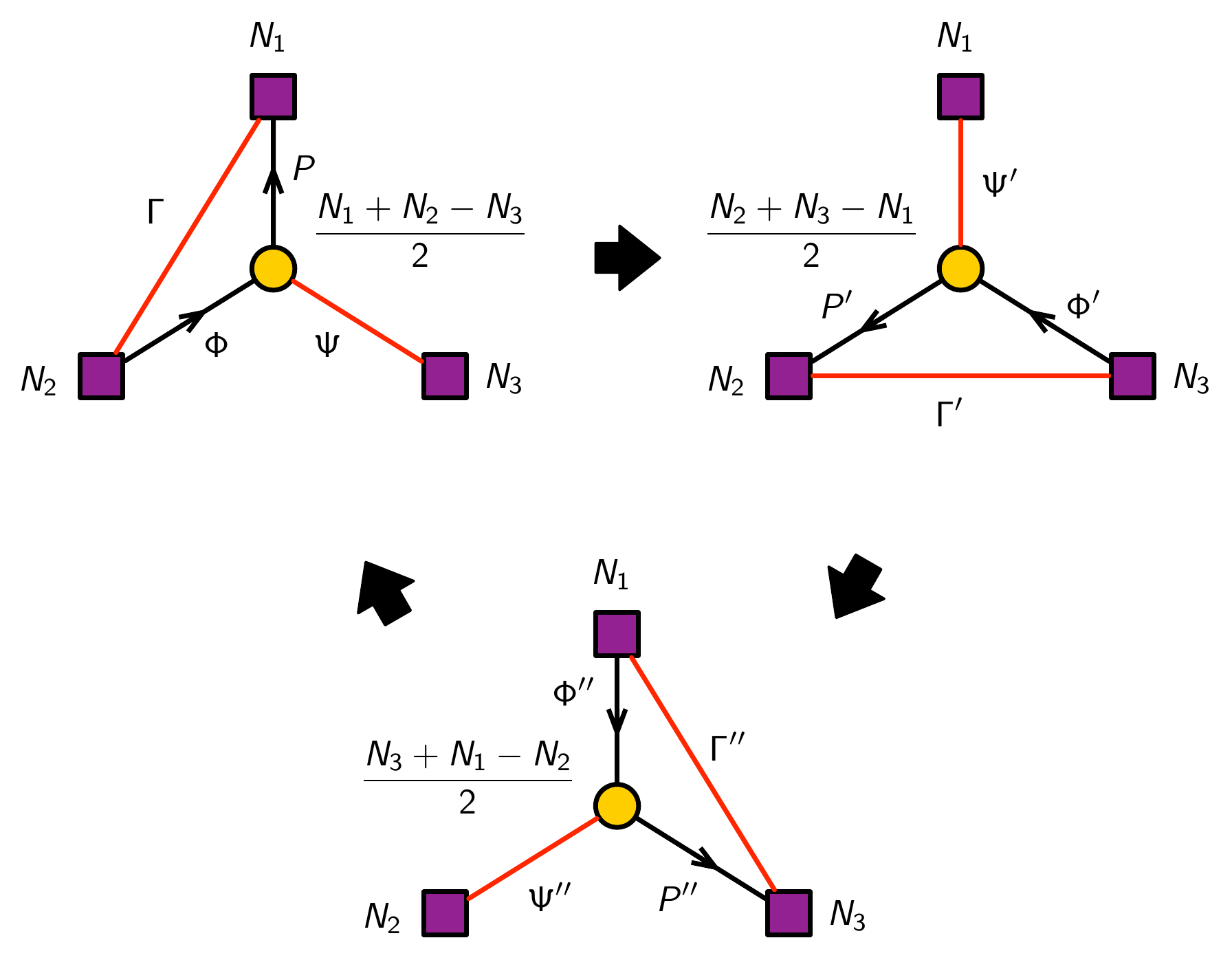}
}
\caption{
The triality loop for $2d$ $(0,2)$ SQCD.
\label{ftrialitytree}}
 \end{center}
 \end{figure}

Relabeling the ranks of the flavor symmetry groups as $N_1\equiv 2N_c-N_b+N_f$, $N_2\equiv N_b$ and $N_3\equiv N_f$, the rank of the gauge group becomes $\frac{N_i + N_j - N_k}{2}$ and the triality chain can be thought of as a cyclic permutation of $N_1$, $N_2$ and $N_3$. This is shown in \fref{ftrialitytree}. From now on we omit the Fermi fields in the determinant representation of the gauge group. They are absent if the gauge group is $SU(N_c)$ instead of $U(N_c)$, since they are not required for anomaly cancellation. While the theories on D1-branes we will study have $U(N_c)$ gauge groups, such fields are not required because abelian anomalies are cancelled by a generalized Green-Schwarz mechanism via interactions with bulk RR fields \cite{Mohri:1997ef}. 

Substantial evidence for triality in $2d$ $(0,2)$ SQCD was presented in \cite{Gadde:2013lxa} by matching the flavor symmetry anomalies, central charges, and the equivariant indices. Furthermore, an exact CFT description of the low energy physics of SQCD was given in \cite{Gadde:2014ppa} and an exact beta function for the K\"ahler modulus   was determined in \cite{Gadde:2015kda}.

\paragraph{Triality for quiver gauge theories.} 

In \cite{Gadde:2013lxa}, triality was generalized to a special class of quiver gauge theories with multiple gauge and flavor nodes. In this class of theories, all chiral and Fermi fields are in bifundamental representations, aside from possible $\Omega$ Fermi multiplets in determinant representations for cancellation of $U(1)^2$ anomalies. Furthermore, 
only $J$-terms are non-trivial. They are all quadratic, namely they come from cubic plaquettes, and every chiral field participates in at least one of them.

The transformation of such a theory under triality can be summarized as follows. Consider acting on gauge node $k$. All other nodes in the quiver remain unaltered, while the rank of node $k$ becomes
\begin{align}
N'_k = \sum_{j\neq k} n_{jk}^\chi N_j - N_k \,, 
\label{rank-rule}
\end{align}
where $n^\chi_{jk}$ is the number of chiral fields from node $j$ to node $k$.

The field content around node $k$ is modified according to the following rules:
\begin{align}
\begin{split}
&\mbox{(1) Replace each of $(\rightarrow k)$, $(\leftarrow k)$, $(\textcolor{red}{\text{ --- }}  k)$ by  $(\leftarrow k)$, $( \textcolor{red}{\text{ --- }}  k)$, $(\rightarrow k)$, respectively.}
\\
&\mbox{(2) For each subquiver $i\rightarrow k \rightarrow j$, add a new chiral field $i\rightarrow j$.}
\\
&\mbox{(3) For each subquiver $i\rightarrow k \textcolor{red}{\text{ --- }} j$, add a new Fermi field $i \textcolor{red}{\text{ --- }}  j$.}
\\
&\mbox{(4) Remove all chiral-Fermi pairs generated in the previous steps.}
\end{split}
\label{triality-rule}
\end{align}
The fields generated at steps 2 and 3 can be regarded as composites of those in the original theory. We will thus often refer to them as {\it mesonic} fields.

Later in this paper we will extend triality to the class of theories associated to brane brick models. In particular, such theories can have a richer structure of $J$- and $E$-terms. 

\paragraph{Triality networks and triality loops.} Acting with triality on all the gauge groups of a quiver generates a family of IR equivalent gauge theories that can be neatly organized into a {\it triality network} \cite{Gadde:2013lxa}. Triality networks can contain {\it triality loops}, namely closed sequences of triality transformations that return to the original theory. The simplest example of a triality loop, which is present in every theory, is the triangular loop associated to three consecutive triality transformations on the same gauge group. \fref{ftrialitytree} shows the triality loop for $2d$ $(0,2)$ SQCD.

\paragraph{Chiral conjugation.} We define chiral conjugation as a global operation on a quiver that reverses the directions of all chiral fields and exchanges $J$- and $E$-terms. Let us denote the local triality at node $k$ by $\tau_k$ and chiral conjugation by $\gamma$. Clearly, 
$\tau_k^3=1$ and $\gamma^2=1$. Upon inspection of the transformation rule \eref{triality-rule}, we note that $\tau_k$ and $\gamma$ satisfy 
\begin{align}
\gamma \tau_k \gamma = \tau_k^{-1} = \tau_k^2\,.
\end{align}
As a result, the triality loop associated to three consecutive triality transformations on the same gauge group of the chiral conjugate configuration is isomorphic to the original one, except that the orientation of the loop is reversed.

\section{Triality for Brane Brick Models \label{sec:brick-triality}} 

The goal of this section is to introduce a triality proposal for brane brick models. We will not consider triality transformations of arbitrary gauge groups. Instead, we will focus on those such that the resulting gauge theories are also described by brane brick models. We refer to such theories as {\it toric phases}. In order to shape our proposal, we will combine a natural generalization of the basic triality transformation with various desired properties for the resulting theories.

\paragraph{Ranks.} 

Let us first consider the ranks of gauge groups. Toric phases associated to $N$ regular D1-branes on a toric CY 4-fold must have all ranks equal to $N$.\footnote{Notice that dual phases with unequal ranks might exist, as we briefly mention in section \sref{sec:triality-review}. Such theories, however, are not described by brane brick models.} What types of nodes can in principle be dualized such that, the resulting rank remains equal to $N$? It is reasonable to assume that the transformation rule for ranks of the basic triality continues to apply for these more general theories. Specializing \eref{rank-rule} to a node $k$ in a toric phase, we get
\beq
N_k'=n_{k,in}^\chi N-N\, .
\eeq
In order to obtain $N_k'=N$ we need $n_{k,in}^\chi=2$, i.e. nodes with only two ingoing chiral field arrows.

The cancellation of non-abelian anomalies further constraints the dualized node. As discussed in \cite{Franco:2015tna,Franco:2015tya}, the cancellation of $SU(N_k)^2$ anomalies requires that 
\begin{align}
\sum_{j\neq k} (n^\chi_{jk} N_j + n^\chi_{kj} N_j - n^F_{kj} N_j) + 2 (a_k^\chi - a_k^F) N_k = 2 N_k  \,.
\label{sun-anomaly}
\end{align}
When all gauge nodes have equal ranks, the relation simplifies to 
\begin{align}
n_k^\chi - n_k^F = 2  \,.
\label{anomaly-simple}
\end{align}
Combining \eqref{sun-anomaly} with $n_{k,in}^\chi=2$, we conclude that the dualized node must have
\beq
n_{k,out}^\chi=n_k^F \, .
\eeq 
Summarizing what we have learnt so far: in order to remain within toric phases, we have to dualize nodes with $n_{k,in}^\chi=2$ and $n_{k,out}^\chi=n_k^F \geq 2$. The lower bound follows from the fact that, in order to avoid SUSY breaking in this class of theories, both $n_{k,in}^\chi$ and $n_{k,out}^\chi$ must be greater or equal than 2.

\begin{figure}[ht!!]
\begin{center}
\includegraphics[height=4.5cm]{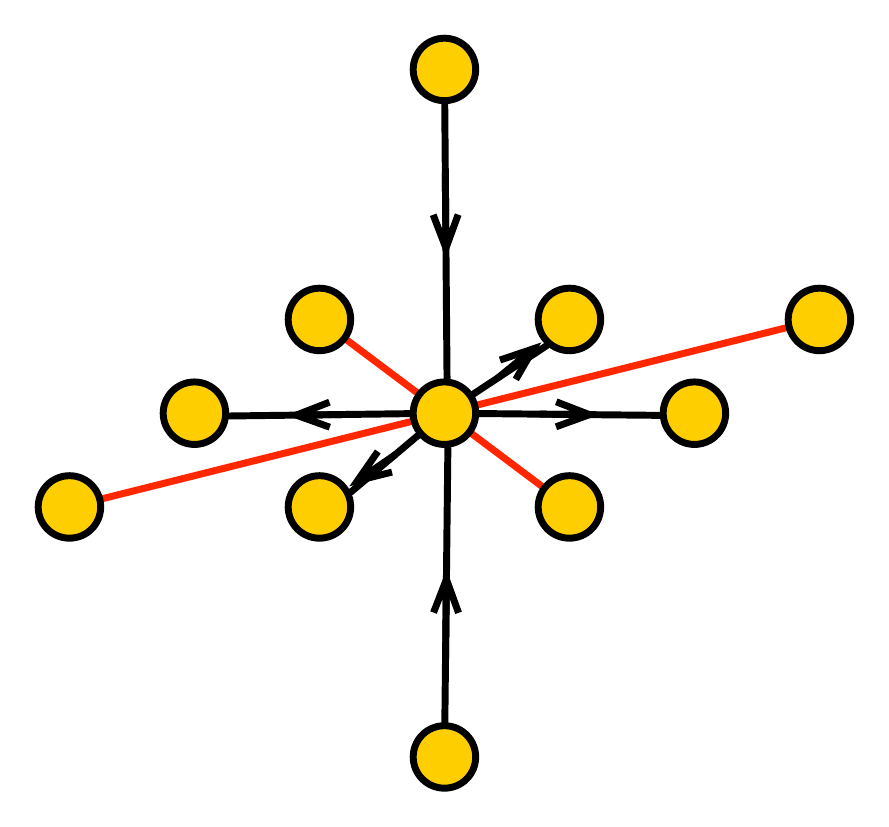}
\caption{General configuration for a node in the periodic quiver whose dualization leads to another toric phase. The example we show corresponds to $n_{k,out}^\chi = n_k^F =4$.}
\label{basic_node_triality}
 \end{center}
 \end{figure}

\paragraph{The Quiver.} 
Let us now consider the transformation of the quiver. We also assume that the fields charged under the dualized node and the new mesons obey the rules in \eref{triality-rule}. In order for the dual theory to correspond to a brane brick model, we should not generate mesons that correspond to lines crossing over the dualized node or interlaced loops of fields in the periodic quiver. A natural solution to this problem is given by a local configuration of the general form shown in \fref{basic_node_triality} for the case of $n_{k,out}^\chi = n_k^F =4$

\begin{figure}[ht!!]
\begin{center}
\includegraphics[height=4.5cm]{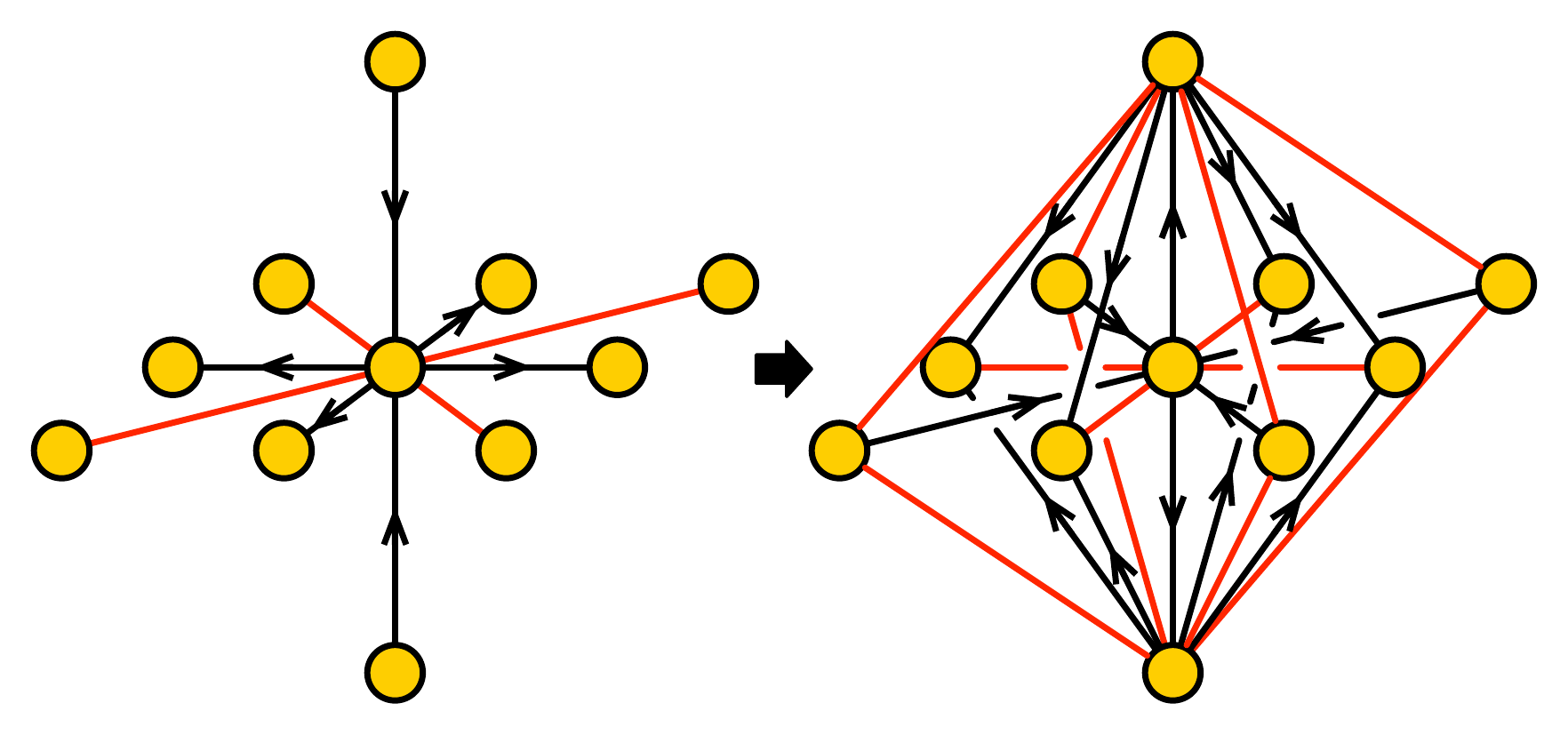}
\caption{Local transformation of the quiver under triality for a node with $n_{k,out}^\chi = n_k^F =4$. The initial configuration is such that the dual is also a toric phase, i.e. it continues to be well described by a periodic quiver.}
\label{local_triality_quiver}
 \end{center}
 \end{figure}

The orientations of chiral fields in this configuration are such that, as illustrated in \fref{local_triality_quiver}, the dual theory does not have mesons going over the dualized node or interlaced loops. Furthermore, we chose the outgoing chiral fields and the Fermi fields to alternate along an equatorial plane. This alternation facilitates the construction of plaquettes, and hence it is a natural structure to arise in toric theories.\footnote{It would be interesting to determine whether more general configurations are possible.}
 
\paragraph{Plaquettes.} 
So far, we have only explained how the quiver transforms under a triality transformation. A full definition of triality for toric phases also requires a rule for obtaining the $J$- and $E$-terms in the dual theory. We propose the following prescription for doing so: after performing the local transformation of the dualized node shown in \fref{local_triality_quiver}, the $J$-  and $E$-terms are those that follow from the resulting periodic quiver. This prescription will be illustrated in numerous examples in section \sref{sec:examples}, for which the $J$- and $E$-terms are explicitly presented in the two appendices.

As reviewed in section \sref{sec:triality-review}, a triality prescription for a wide class of quiver theories was introduced in \cite{Gadde:2013lxa}. Our proposal applies to a different class of quiver gauge theories, hence considerably extending the range of applicability of triality. Our theories have more general $J$-terms, non-trivial $E$-terms and are engineered in terms of branes.

\paragraph{Loops of toric phases and cubic nodes.} 

Following our previous discussion, starting from a toric phase and dualizing any node of the general form shown in \fref{basic_node_triality} with $n_{k,in}^\chi=2$ and $n_{k,out}^\chi=n_k^F \geq 2$, we obtain a new toric phase. According to the rules in \eref{triality-rule}, $n_{k,in}^{\chi \prime}$ in the dual theory is equal to $n_k^F$ of the original one. Hence, a second dualization on the same node will result in a toric phase only if $n_k^F=2$. This implies that nodes with $n_{k,in}^\chi= n_{k,out}^\chi = n_k^F =2$ are not just the simplest configurations that can be dualized to obtain a toric phase. They are also special because it is only for them that the triality loops of three consecutive dualizations on the same node exclusively involve toric phases. The corresponding bricks in the brane brick models have six faces, as shown on the left of \fref{triality-simple-brick}. We thus refer to these nodes as {\it cubic nodes}.

For the aforementioned reasons, most of our discussion in coming sections will focus on triality of cubic nodes. An explicit example of a toric phase obtained by a triality transformation of a node with $n_{k,in}^\chi>2$ will be discussed in section \sref{sq111ex}.

\subsection{Triality and Brane Brick Models} 

We now rephrase our previous discussion from the viewpoint of brane brick models, focusing on triality transformations of cubic nodes.

\begin{figure}[H]
\begin{center}
\includegraphics[height=4.5cm]{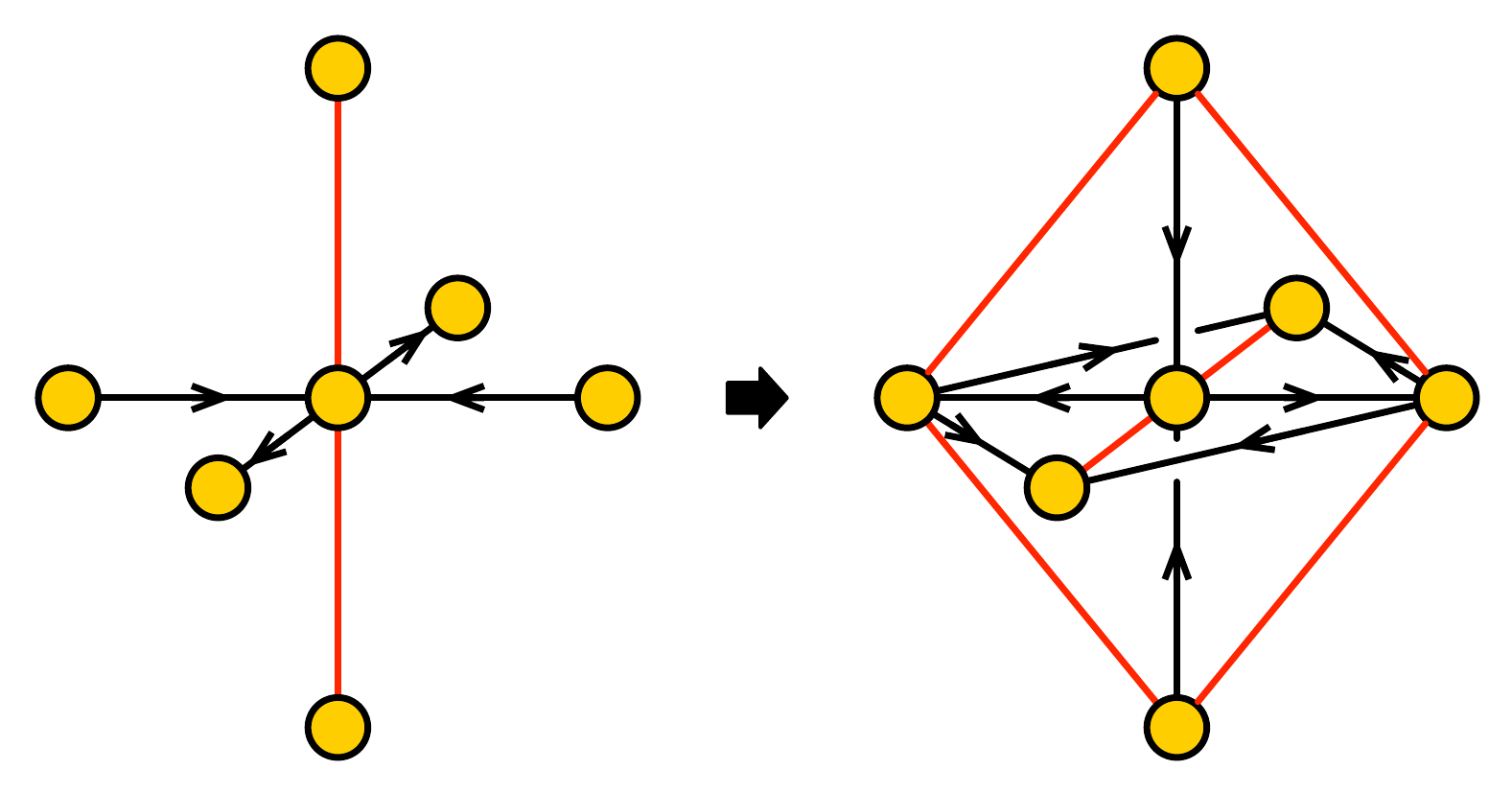}
\caption{
Local triality action on a cubic node in the periodic quiver.
\label{triality-simple-quiver}}
 \end{center}
 \end{figure}

A cubic node has six nearest neighbors. For simplicity, let us begin by assuming that, before the triality move, there are no fields connecting the nearest neighbors among themselves. The triality move on the cubic node is shown in 
\fref{triality-simple-quiver}. 

\begin{figure}[ht]
\begin{center}
\includegraphics[width=12cm]{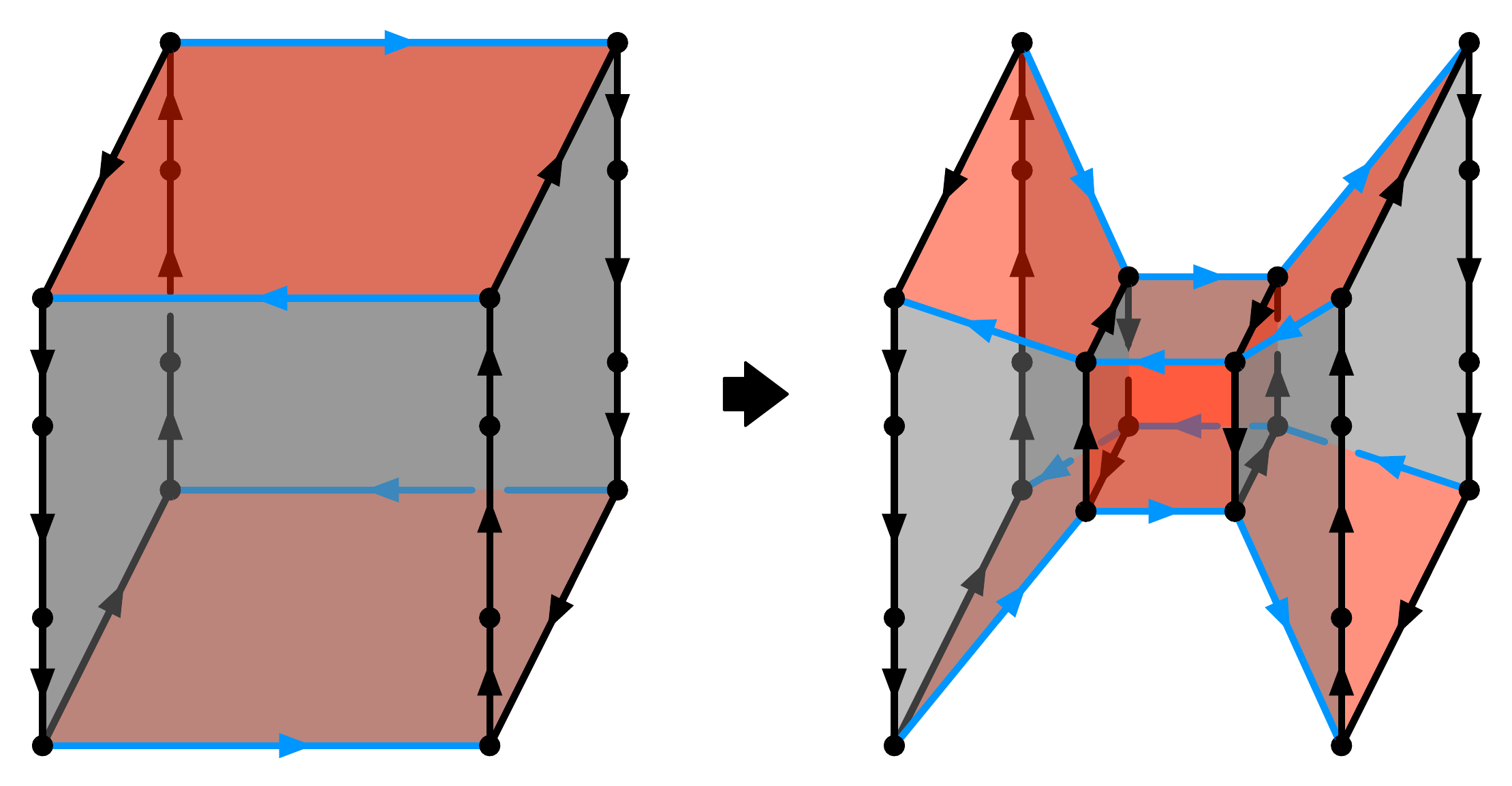}
\caption{
Local triality action on a cubic brick.
\label{triality-simple-brick}}
 \end{center}
 \end{figure}

Translating \fref{triality-simple-quiver} into a brane brick model by 
the usual graph dual method, we arrive at \fref{triality-simple-brick}. 
As explained in \cite{Franco:2015tya}, it is convenient to keep track of the orientation of brick faces by assigning orientations to their edges. 

Recall that Fermi faces are always quadrilaterals. Chiral faces are less restricted. We will {\em not} assume that the chiral faces of the cube before triality are necessarily simple squares, by which we mean four sided. Instead, we will allow the common boundary between two chiral faces to carry a sequence of oriented edges. \fref{triality-simple-brick} shows examples of composite boundaries containing three vertical edges.

Under these conditions, the local brick configuration after triality is determined uniquely as shown on the right of \fref{triality-simple-brick}. The original cube is replaced by a new, smaller, cube consisting of chirals and Fermis determined by the triality rule. In addition, eight new ``diagonal" faces are produced, which connect a subset of the edges of the original cube to edges in the new cube. The edges are oriented such that four of the diagonal faces are Fermis and the other four are chirals. 
Note that, in order for the diagonal mesonic chiral faces to have even number of edges, 
the number of edges constituting the composite boundary in the initial cube should be odd. 
 
Faces associated to fields that are neutral under the dualized gauge group may undergo modifications. Consider the blue edges of the cubic brick shown in \fref{triality-simple-brick}. Unlike the other eight edges in the cube, they are not connected to new mesonic faces in the dual theory. Faces outside of the cube that are initially attached to the blue edges remain connected to them. In the process, each of them gains two new edges as shown in \fref{triality-simple-brick}. If a face glued to such an edge was a $(2k)$-gon before triality, it would become a $(2k+2)$-gon afterwards. This is perfectly fine for a chiral face, but seems problematic for a Fermi face. Interestingly, the apparent problem occurs if and only if the periodic quiver contains {\it nested plaquettes}. Nested plaquettes refer to the case in which the chiral fields in a plaquette are a subset of those in a larger plaquette (see \cite{Franco:2015tya} for a discussion). In the brane brick model perspective, nested plaquettes arise when more than one Fermi faces share a common edge. Establishing how to modify \fref{triality-simple-brick} in the presence of nested plaquettes would require a more refined analysis. In all examples considered in this article, this phenomenon arises only when we dualize a node that leads to a toric phase but that it is not cubic. Since we will mainly focus on triality moves on cubes, we will not delve into the subtleties associated to nested plaquettes. 

\begin{figure}[H]
\begin{center}
\includegraphics[width=8.5cm]{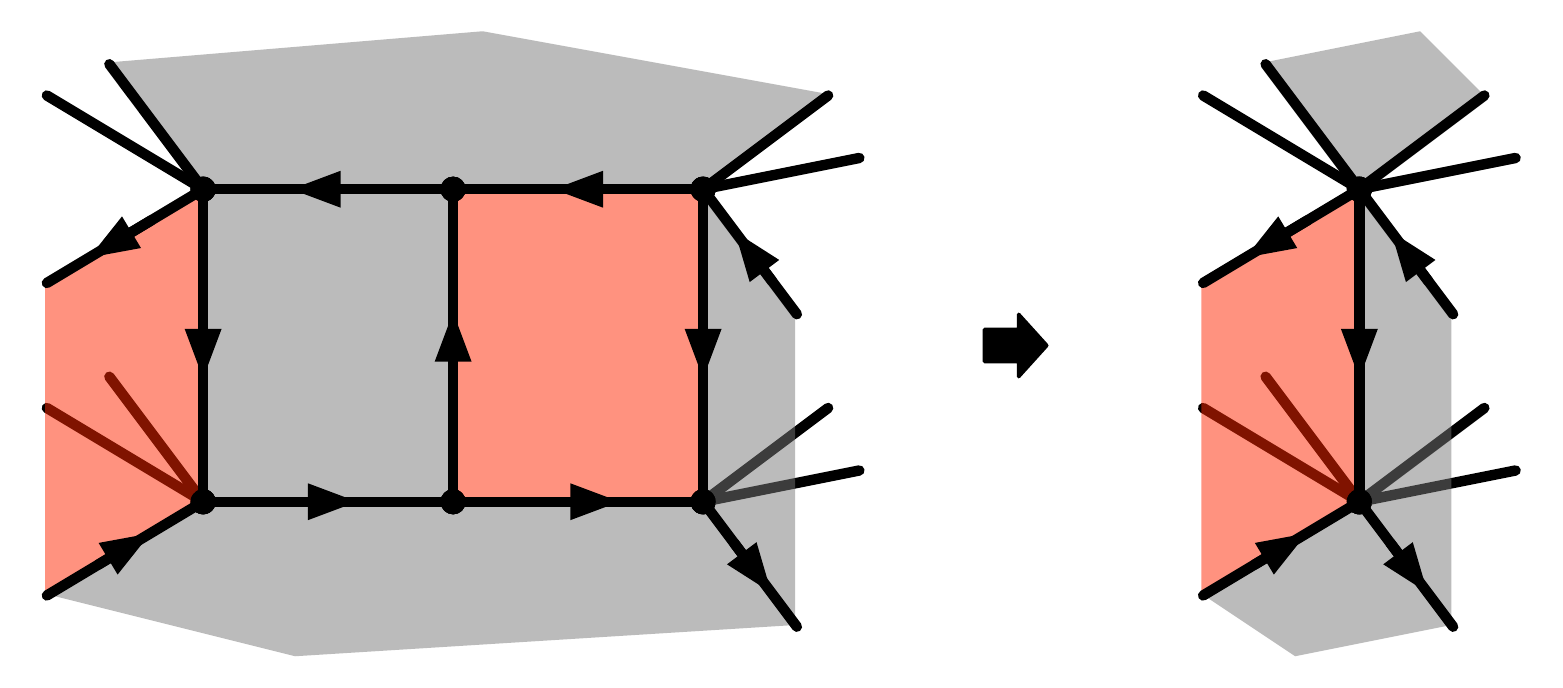}
\caption{
Integrating out a massive chiral-Fermi pair in a brane brick model.
\label{bf-cancellation}}
 \end{center}
 \end{figure}

Triality can also give rise to chiral-Fermi massive pairs. As explained in \cite{Franco:2015tya}, such a pair corresponds to a Fermi and a chiral faces connected by an edge to which no other face is attached. Massive pairs can be integrated out at low energies, simplifying the brane brick model. This process is illustrated in \fref{bf-cancellation}. 

All the discussion in this section also applies if we relax our initial assumption on the absence of fields running between the six nearest neighbors. Moreover, in all the explicit examples considered in the next section, these pre-exiting fields form massive pairs with mesonic fields and disappear at low energies.

\subsection{Transformation of $J$- and $E$-Terms}

$2d$ $(0,2)$ theories must satisfy the {\it vanishing trace condition}
\beq
\sum_{a} \mathrm{tr} \left( J^a E_a \right) = 0 \,.
\eeq
In \cite{Franco:2015tna}, partial resolution was used to show that all brane brick models satisfy this condition. Here we would like to take a complementary approach to this problem. We will show that, for brane brick models, triality preserves the vanishing trace condition.

We find it convenient to introduce a formal object for bookkeeping $J$- and $E$-terms, 
\begin{align} 
\Omega = \sum_{a} \mathrm{tr} \left( \Lambda_a J^a + \bar{\Lambda}^a E_a \right)\,. 
\end{align}
Using a formal inner product, $(\Lambda_a)^i{}_j \cdot (\bar{\Lambda}^b)^k{}_l = \delta_{a}^b \delta^i{}_l \delta^k_j$, where $i,j,k,l$ are (anti)fundamental indices at appropriate gauge nodes, we can rewrite the vanishing trace condition as
\begin{align} 
\Omega \cdot \Omega = \sum_{a} \mathrm{tr} \left( J^a E_a \right) = 0 \,. 
\end{align}

\begin{figure}[H]
\begin{center}
\includegraphics[width=11cm]{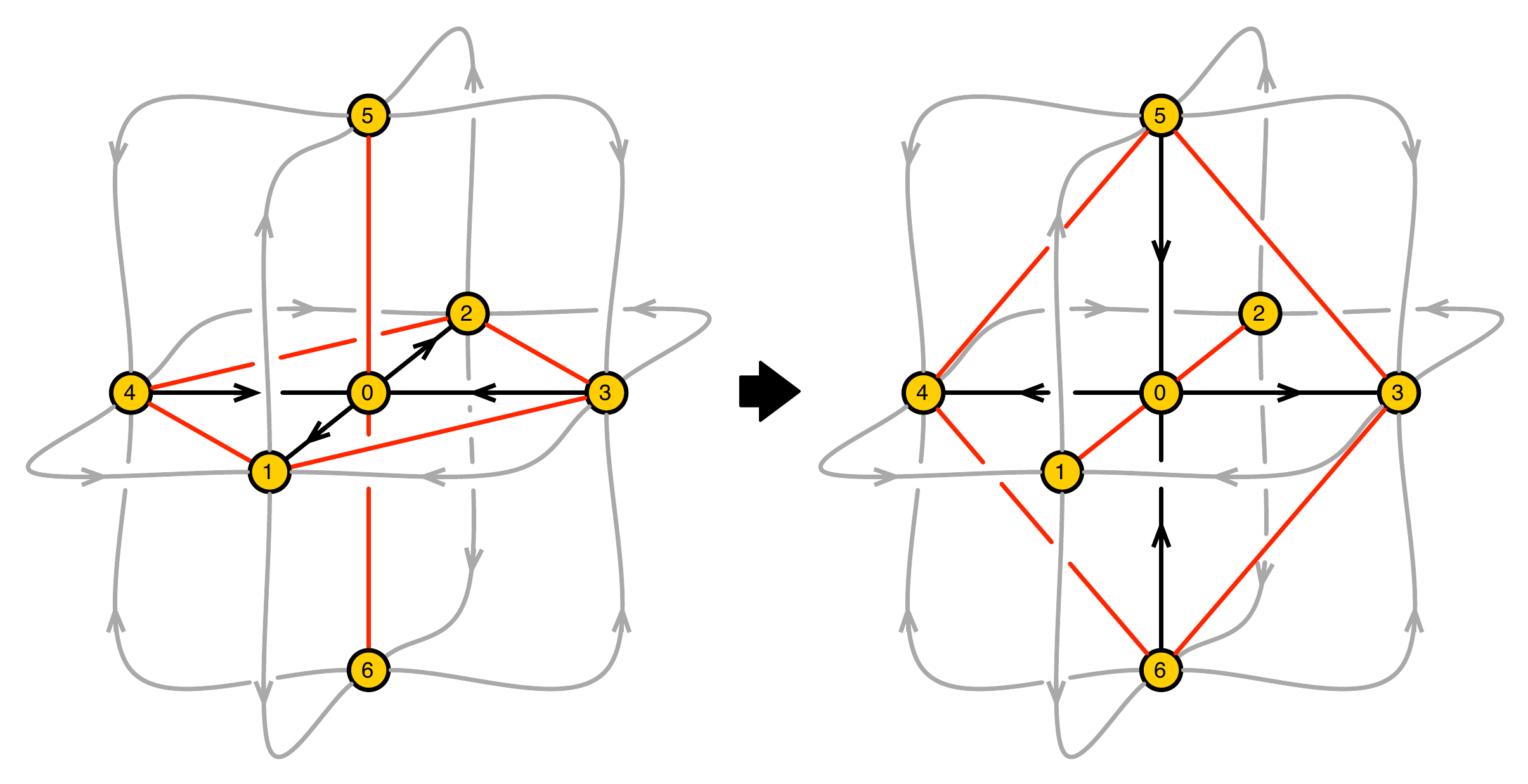}
\caption{
Transformation of $J$- and $E$-terms. The curved grey arrows represent oriented chains of chiral fields.
\label{EJ-a}}
 \end{center}
 \end{figure}

For simplicity, let us begin with the local triality move shown in \fref{EJ-a}. In the figure, we included open chains of chiral fields connecting pairs of adjacent neighbors to the cubic node. We denote them by $O_{ij}$. 

Let us assume that the plaquettes near the cubic nodes are simply those we can recognize in \fref{EJ-a}. The formal sum $\Omega$ before the triality is 
\begin{align}
\begin{split}
\Omega = \Omega_0 &
+ \Lambda_{13} ( X_{30}X_{01} - O_{31} ) + \Lambda_{31} (O_{15}O_{53} - O_{16} O_{63})
\\
& 
+ \Lambda_{14} ( X_{40}X_{01} - O_{41} ) - \Lambda_{41} (O_{15}O_{54} - O_{16} O_{64})
\\
& 
+ \Lambda_{23} ( X_{30}X_{02} - O_{32} ) - \Lambda_{32} (O_{25}O_{53} - O_{26} O_{63})
\\
& 
+ \Lambda_{24} ( X_{40}X_{02} - O_{42} ) + \Lambda_{42} (O_{25}O_{54} - O_{26} O_{64})
\\
& 
+ \Lambda_{05} ( O_{53} X_{30} - O_{54} X_{40} ) - \Lambda_{50} (X_{01} O_{15} - X_{02} O_{25})
\\
& 
+ \Lambda_{06} ( O_{63}X_{30} - O_{64} X_{40} ) + \Lambda_{60} (X_{01}O_{16} - X_{02} O_{26}) \,.
\end{split} 
\end{align}
Here, $\Omega_0$ collects the contributions from plaquettes whose Fermi fields are not shown in the figure. We suppressed the ``tr" symbol, but the cyclic ordering in a monomial should be understood. We also suppressed the bar from $\bar{\Lambda}$, with the identification $\Lambda_{ji} = \bar{\Lambda}_{ij}$. 

A simplifying feature of the local quiver in \fref{EJ-a} is that 
the tr$(E_aJ^a)$ terms involving the four chiral and six Fermi fields shown explicitly cancel among themselves, leaving
\begin{align}
\begin{split}
\Omega \cdot \Omega &= \Omega_0 \cdot \Omega_0 - (O_{31}O_{15}O_{53}) + (O_{25}O_{53}O_{32}+O_{41}O_{15}O_{54}+O_{63}O_{31}O_{16})
\\
&\qquad\qquad \quad - (O_{16} O_{64}O_{41} + O_{32}O_{26} O_{63} + O_{54}O_{42}O_{25}) + O_{42}O_{26} O_{64} \,.
\end{split} 
\label{Omega-triality}
\end{align}
For the theory to be consistent, this sum should vanish. 

In the local quiver of \fref{EJ-a}, the triality acts as a cyclic permutation 
on node labels: $(123456) \rightarrow (345612)$. But, the $O_{ij} O_{jk}O_{kl}$ terms in \eqref{Omega-triality} are grouped in cyclically invariant combinations. So, if $\Omega\cdot \Omega$ vanishes before triality, it should still vanish after the triality. 

\begin{figure}[H]
\begin{center}
\includegraphics[width=12cm]{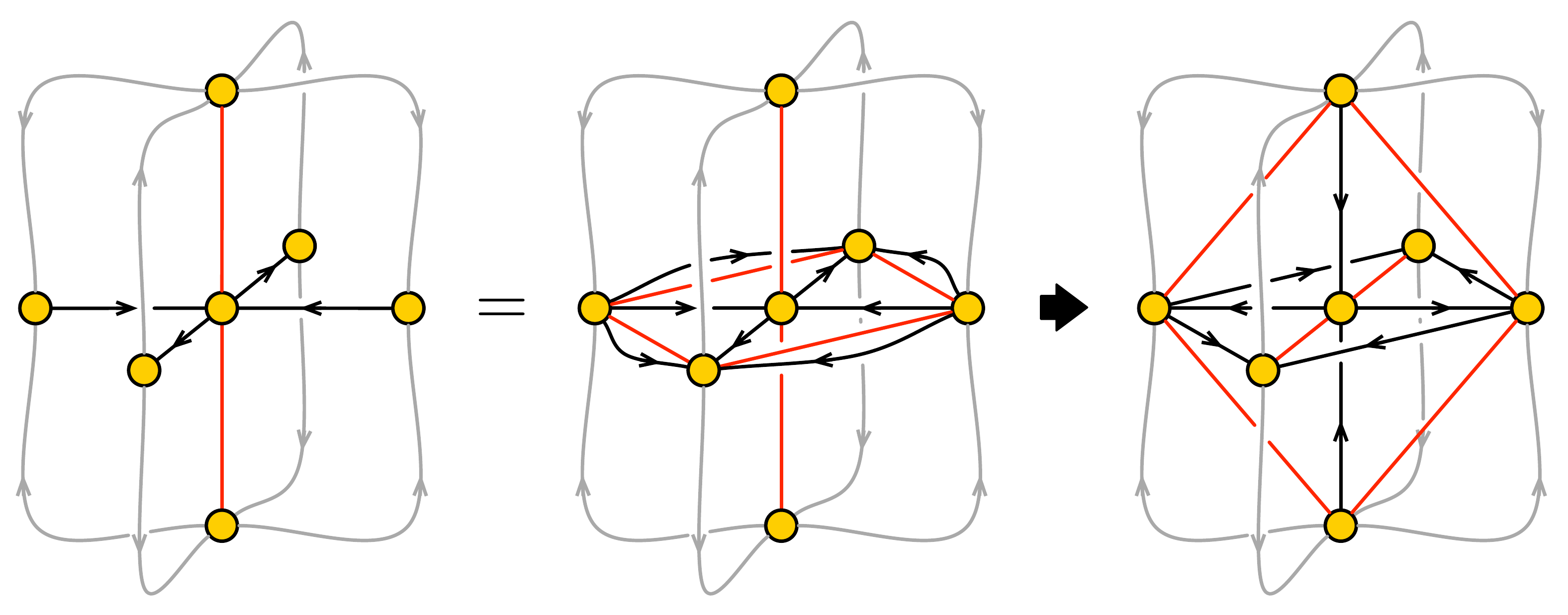}
\caption{
Massive chiral-Fermi pairs added to the initial configuration in \fref{triality-simple-quiver} so that the transformation rule of \fref{EJ-a} can be applied directly.
\label{EJ-b}}
 \end{center}
 \end{figure}

There are local quiver configurations that look more complicated than \fref{EJ-a}. For example, at first sight, the left quiver in \fref{triality-simple-quiver} does not seem to allow for an augmentation by external chains of chiral fields $O_{ij}$ with obvious assignments of plaquettes. 
Fortunately, by integrating in/out chiral-Fermi pairs, we can bring an arbitrary local quiver into the simple form of \fref{EJ-a}. As an illustration, applying this idea to \fref{triality-simple-quiver}, we arrive at \fref{EJ-b}.

\section{Examples \label{sec:examples}} 

\begin{figure}[H]
\begin{center}
\resizebox{0.85\hsize}{!}{
\includegraphics[trim=0cm 0cm 0cm 0cm,totalheight=10 cm]{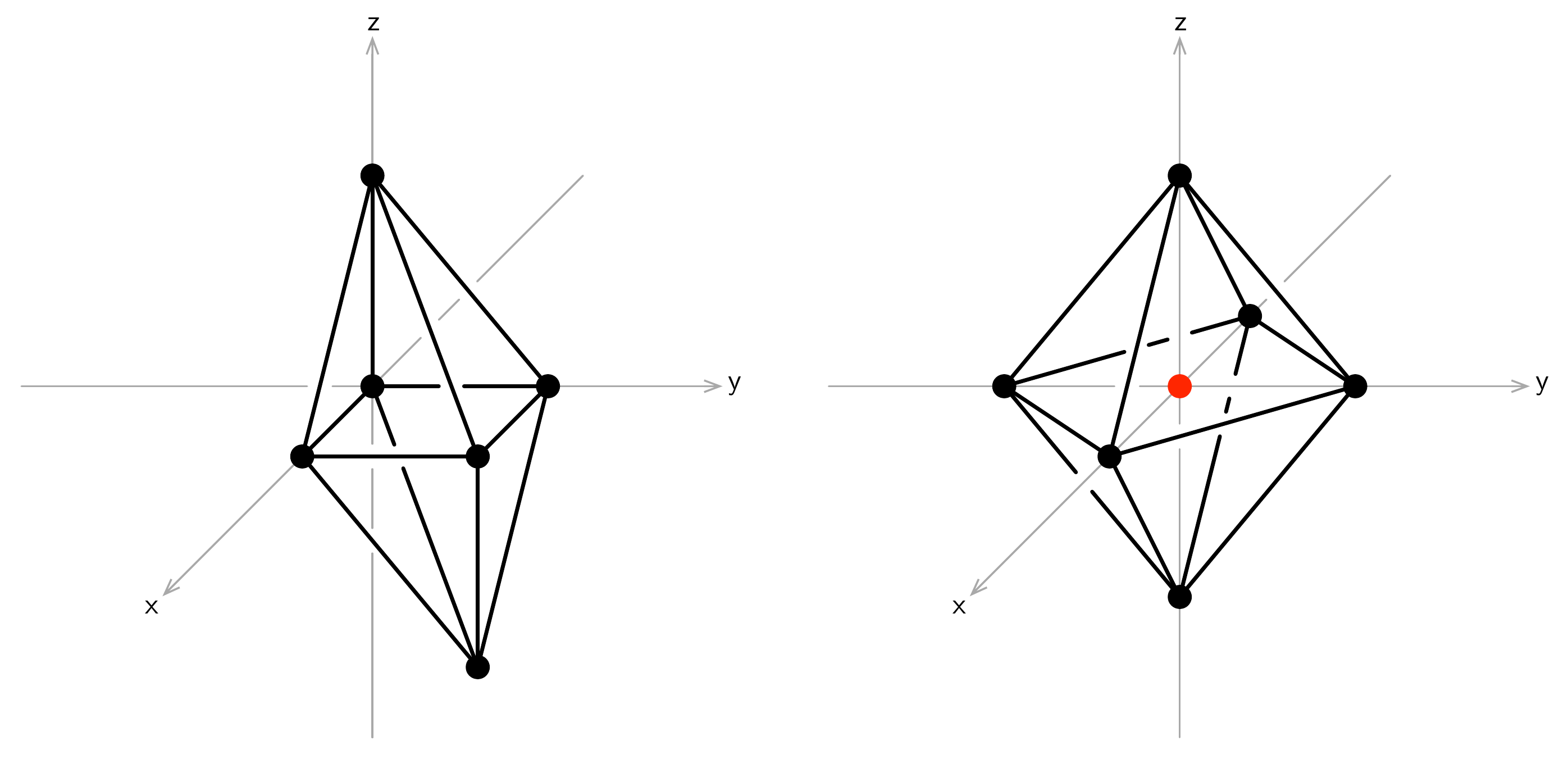}
}  
\caption{
The toric diagrams for $Q^{1,1,1}$ (left) and $Q^{1,1,1}/\mathbb{Z}_2$ (right).
 \label{fq111-toric}}
\end{center}
\end{figure}

In this section we investigate some explicit examples. As mentioned earlier, we primarily focus on triality acting on cubic nodes. In order to do so, it is necessary to find toric CY 4-folds that give rise to brane brick models containing cubic bricks. Fortunately, a theory of this type was already identified in earlier work \cite{Franco:2015tna,Franco:2015tya}. It corresponds to D1-branes probing the cone over $Q^{1,1,1}$. In order to generate an additional example with a richer triality structure, we will simply consider the $Q^{1,1,1}/\mathbb{Z}_2$ orbifold. Following the general construction for orbifolds of toric geometries introduced in \cite{Franco:2015tna,Franco:2015tya}, a brane brick model for this orbifold is obtained by taking one for $Q^{1,1,1}$ and appropriately doubling the size of the unit cell. \fref{fq111-toric} shows the toric diagrams for both geometries.

To keep our discussion succinct, we will mostly phrase it in terms of periodic quivers. The brane brick models for all the theories studied in this section are presented in the appendices, which collect additional detailed information about these theories, such as explicit expressions for their $J$- and $E$-terms.

\subsection{Triality Network for $Q^{1,1,1}$ \label{sq111ex}}

\begin{figure}[h]
\begin{center}
\resizebox{0.53\hsize}{!}{
\includegraphics[trim=0cm 0cm 0cm 0cm,totalheight=10 cm]{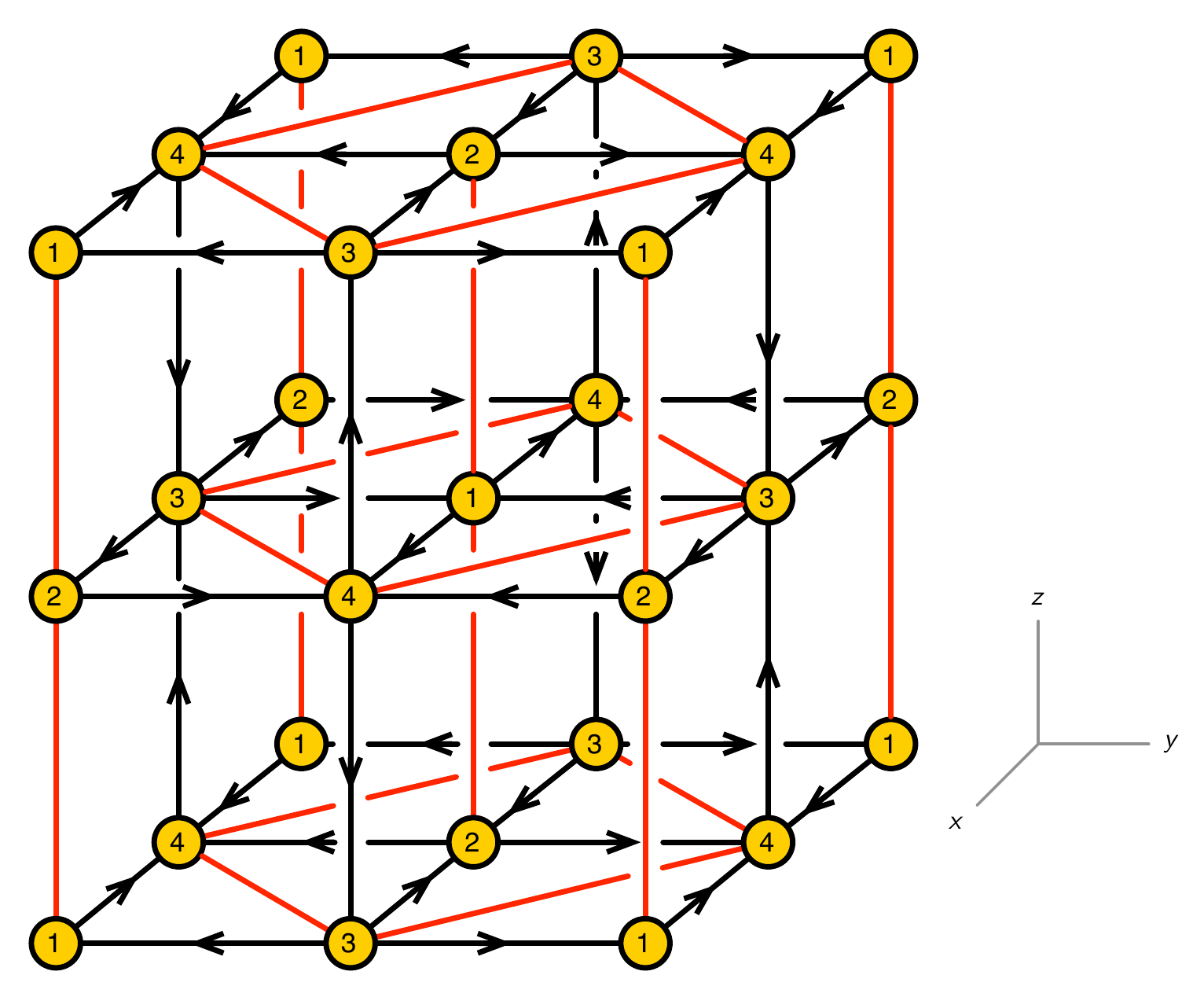}
}  
\caption{
Periodic quiver for phase A of $Q^{1,1,1}$. Notice that the region represented has twice the volume of the unit cell.
\label{q111a-quiver-temp}}
 \end{center}
 \end{figure}

Let us start from the periodic quiver shown in \fref{q111a-quiver-temp}. This theory was introduced in \cite{Franco:2015tna,Franco:2015tya}, where it was shown that it corresponds to D1-branes probing the cone over $Q^{1,1,1}$. We will refer to it as the {\it asymmetric phase} (A), since it is not manifestly invariant under the octahedral symmetry permuting the three axes. This is indeed a symmetry of the underlying geometry, as it follows from the toric diagram.

The brane brick model for this theory is presented in appendix \sref{Q111-detail}. The periodic quiver contains two types of nodes. Nodes 1 and 2 correspond to cubic bricks, while nodes 3 and 4, to octagonal cylinders. 

\begin{figure}[H]
\begin{center}
\includegraphics[height=11cm]{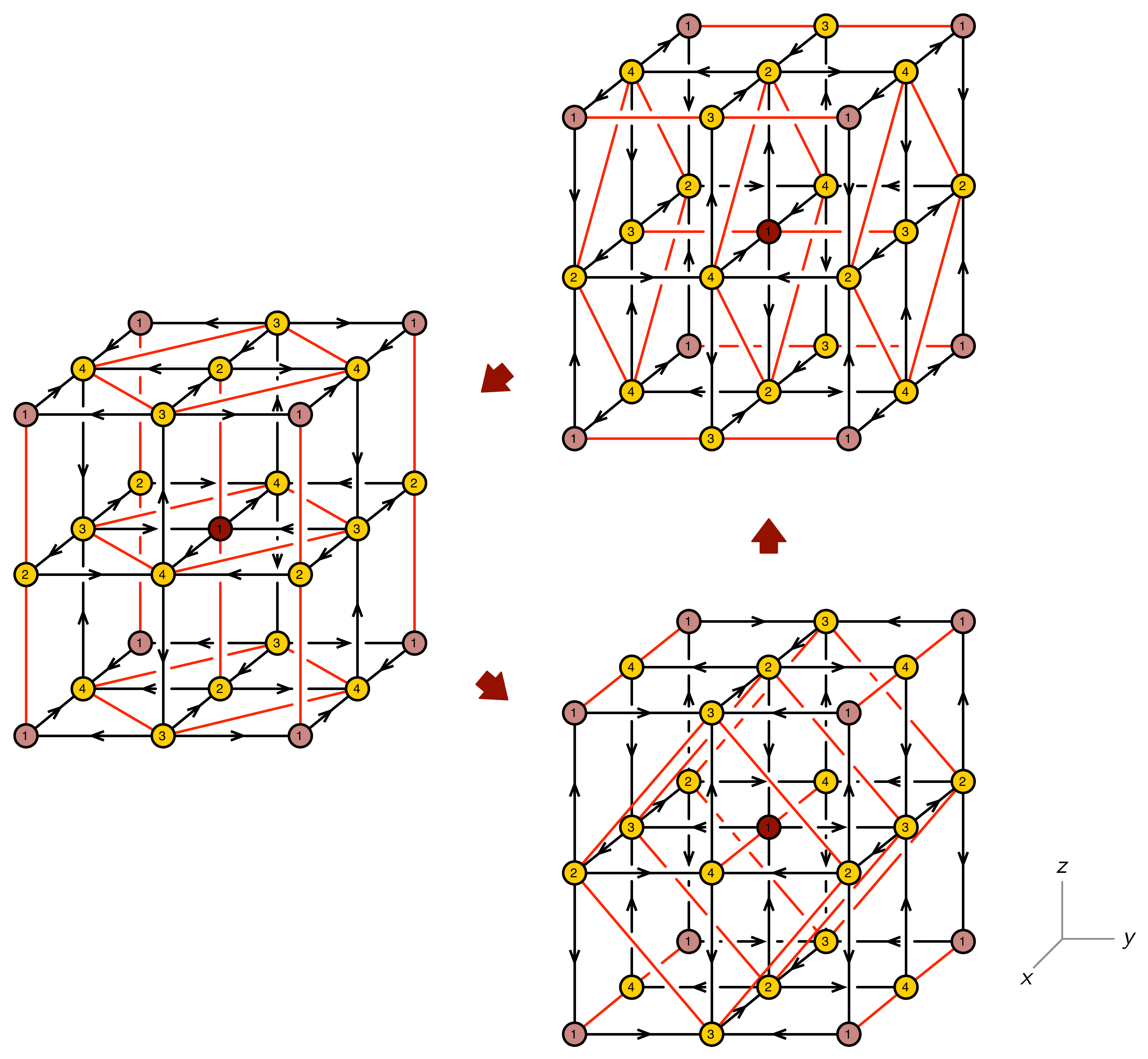}
\caption{Action of three consecutive triality transformations on the cubic node 1 of phase A of $Q^{1,1,1}$. It simply amounts to a cyclic permutation of the three $T^3$ directions.
\label{Q111-cube}
} 
 \end{center}
 \end{figure}

Let us consider the action of triality on a cubic node, say 1. The resulting theory is identical to the original one up to a cyclic permutation of the three axes. \fref{Q111-cube} shows the triality loop arising from three consecutive dualizations of node 1. 

\begin{figure}[ht!!]
\begin{center}
\includegraphics[height=12cm]{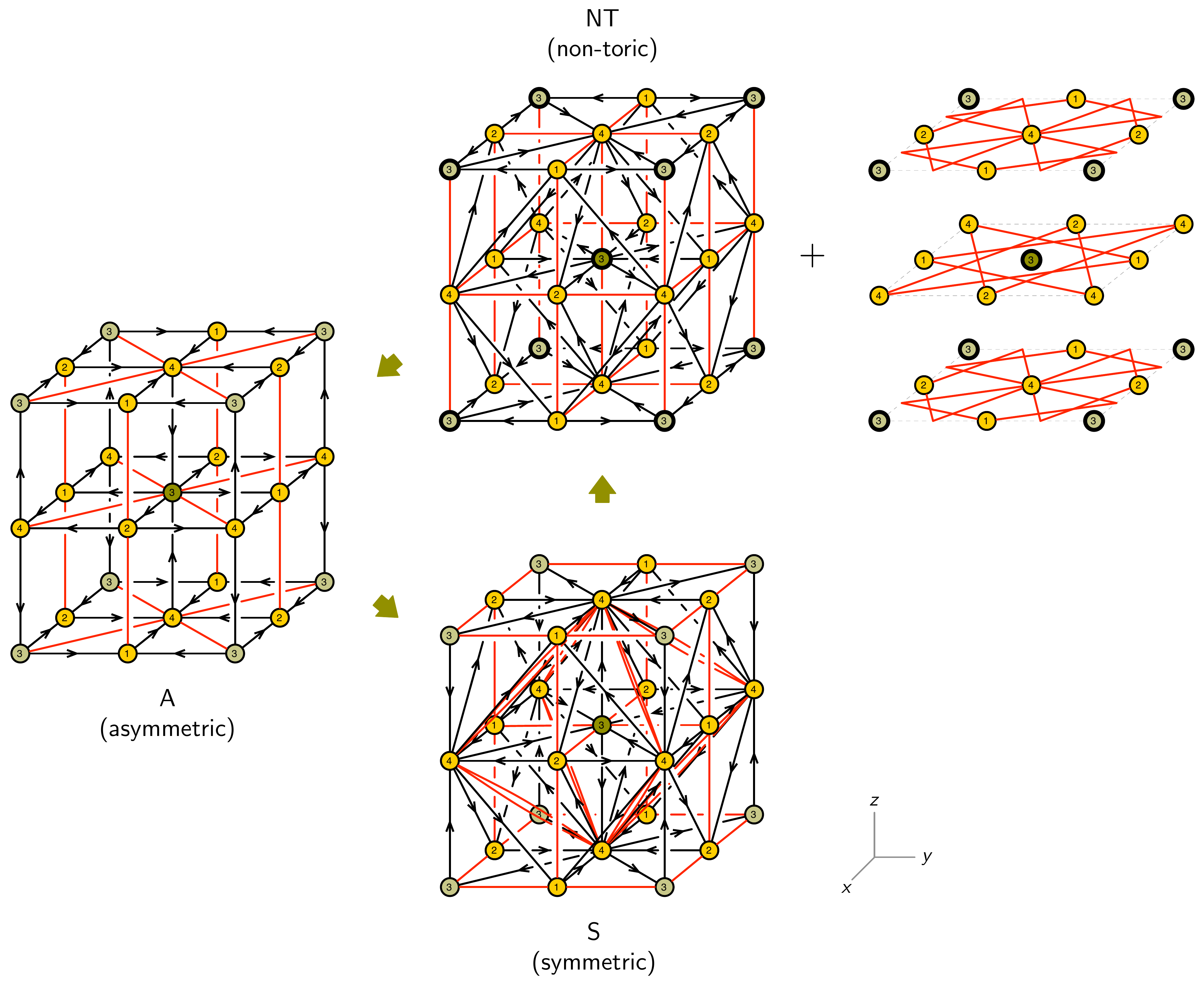}
\caption{Action of three consecutive triality transformations on the node 3 of phase A of $Q^{1,1,1}$. The symmetric phase S is a new toric phase. Notice that in this periodic quiver the Fermi lines attached to node 4 are double. The $U(3N)$ gauge node in phase NT is distinguished by a thick boundary.
\label{Q111-oct}
}
 \end{center}
 \end{figure}

Let us now consider what happens when dualizing node 3. Interestingly, while this node is not cubic, it is of the general form discussed in section \sref{sec:brick-triality} and shown in \fref{local_triality_quiver}. We thus know that triality on it takes phase A to another toric phase. The new theory is shown at the bottom right of \fref{Q111-oct}. It is indeed a toric phase and, furthermore, it is invariant under the permutation of the coordinate axes. We hence refer to it as the {\it symmetric phase} (S).\footnote{It is interesting to remark that phase S can also be directly determined from the $Q^{1,1,1}$ geometry using the general methods introduced in \cite{Franco:2015tna,Franco:2015tya}, without reference to triality.} It is important to note that in this periodic quiver, the multiplicity of the Fermi lines that are adjoints of node 4 is two. This special feature follows from how the periodic quiver captures $J$- and $E$-terms for this theory, is related to details of its brane brick model and is discussed in detail in appendix \sref{Q111-detail}. 

Continuing with a second dualization on node 3 generates a non-toric phase (NT), which we show at the top right of \fref{Q111-oct}. The rank of node 3 in this theory is $3N$. Since the theory is non-toric, it is not fully captured by a periodic quiver, i.e. there is no known prescription for reading the $J$- and $E$-terms directly from it. The quiver diagram in \fref{Q111-oct} should be understood merely as a representation of the gauge symmetry and matter content of the theory. A third dualization on node 3 returns to the toric phase A. \fref{Q111-oct} shows the triangular loop obtained by three consecutive dualizations on node 3. 
In order to place the dualized node at the center, we have shifted the periodic quivers with respect to \fref{q111a-quiver-temp} by half a period in the $x$ direction.
It is straightforward to verify that the three theories are free of non-abelian anomalies, which is in fact guaranteed by the triality transformation rules.

\begin{figure}[ht!!]
\begin{center}
\includegraphics[height=4.4cm]{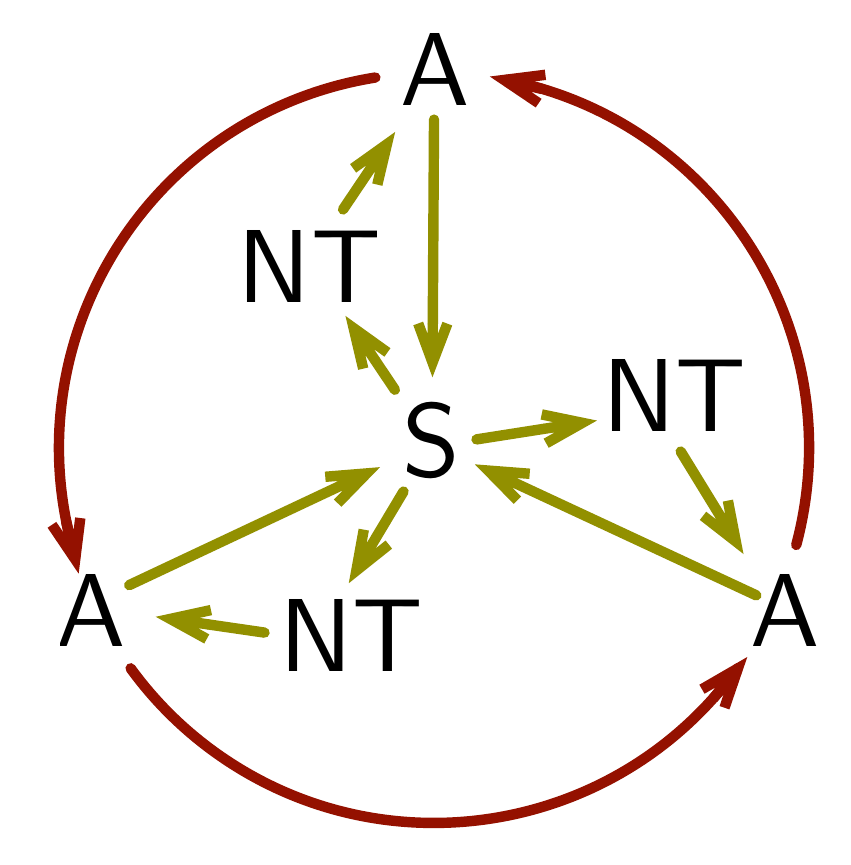}
\caption{A piece of the triality network for $Q^{1,1,1}$, involving only dualizations of nodes 1 and 3.
}
\label{Q111-network}
 \end{center}
 \end{figure}

We can combine our previous analyses to construct a piece of the triality network for $Q^{1,1,1}$ involving only dualizations on nodes 1 and 3. The result is shown in \fref{Q111-network} and consists of the triality loop in \fref{Q111-cube} and three permutations of the triality loop in \fref{Q111-oct}.  Since phase S is invariant under the permutation of coordinate axes, it sits at the intersection of the triality loops of all the three versions of phase A. 

The brane brick model for phase A contains another cube (node 2) and another octagonal cylinder (node 4). Their behavior under triality is identical to the cases we considered, up to chiral conjugation. Of course, we may also perform triality moves on other nodes of phases S and NT, generating even more non-toric phases. We will not pursue them in this paper.

\subsection{Triality Network for $Q^{1,1,1}/\mathbb{Z}_2$}
 
\begin{figure}[ht]
\begin{center}
\resizebox{0.5\hsize}{!}{
\includegraphics[trim=0cm 0cm 0cm 0cm,totalheight=10 cm]{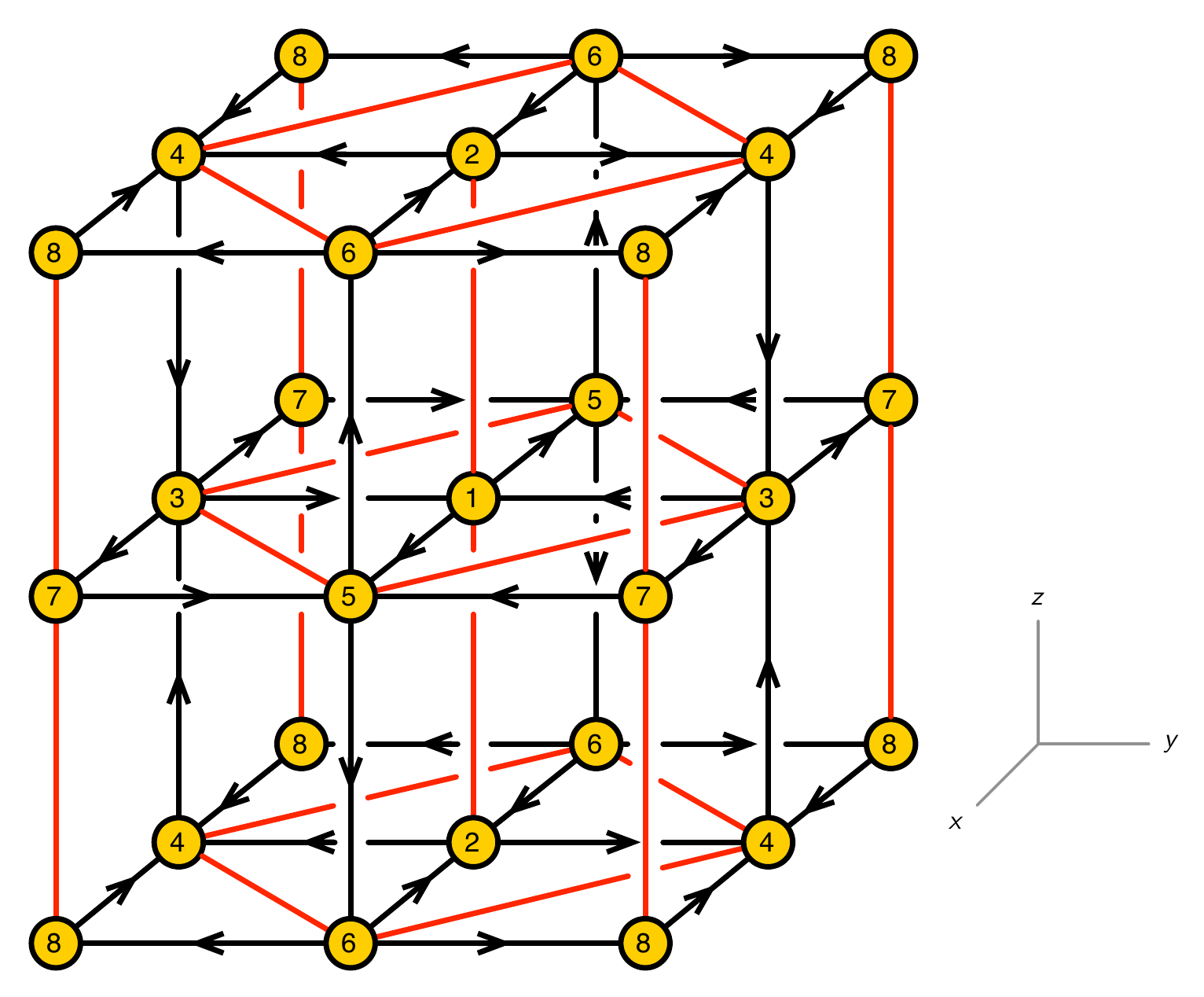}
}  
\caption{
Periodic quiver for phase A of $Q^{1,1,1}/\mathbb{Z}_2$.
\label{fq111a1}}
\end{center}
\end{figure}

We now perform a similar study for $Q^{1,1,1}/\mathbb{Z}_2$. Our starting point is the theory defined by the periodic quiver in \fref{fq111a1}. Since it is simply related to \fref{q111a-quiver-temp} by doubling the size of the unit cell, we also refer to it as phase A. 

For simplicity, we will only consider triality acting on cubic nodes. Even with this restriction, the triality network has a much richer structure of toric phases than the one for $Q^{1,1,1}$ due to the larger number of gauge groups.\footnote{As for $Q^{1,1,1}$, additional toric phases can be generated by dualizing certain non-cubic nodes. For brevity, we will not discuss such theories.}
  
\begin{figure}[ht!!]
\begin{center}
\includegraphics[height=3.5cm]{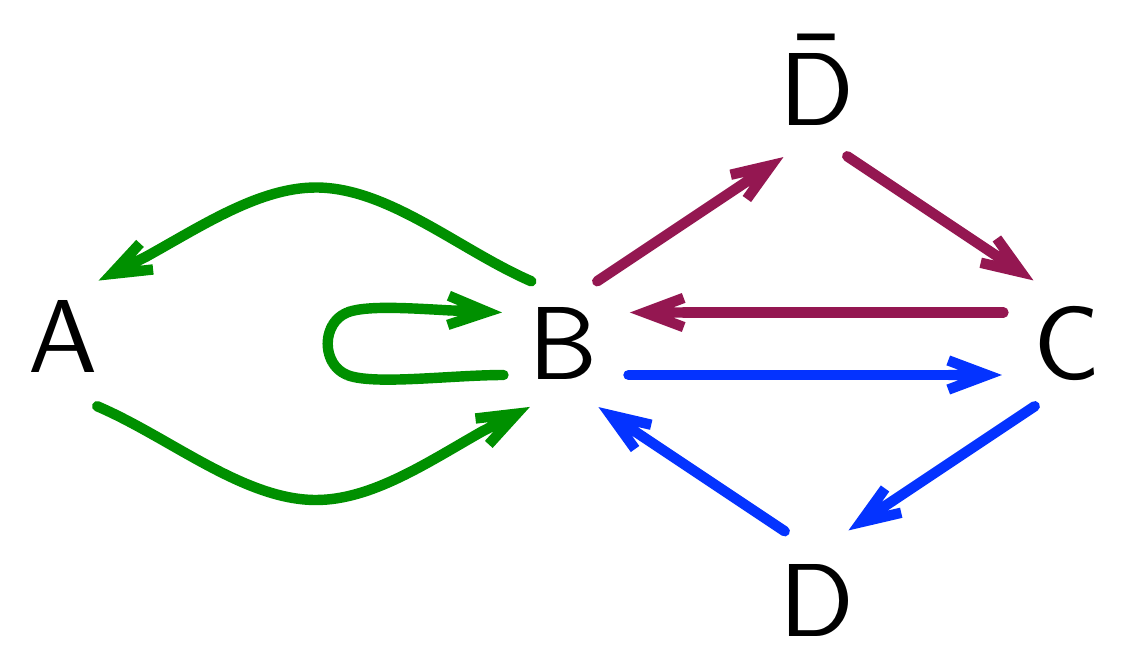}
\caption{A piece of the triality network for $Q^{1,1,1}/\mathbb{Z}_2$ containing phase A and restricting to dualizing cubic nodes. Theories differing by node relabeling are identified. 
\label{Q11-Z2-network}}
 \end{center}
 \end{figure}

\begin{figure}[ht!!]
\begin{center}
\includegraphics[height=16cm]{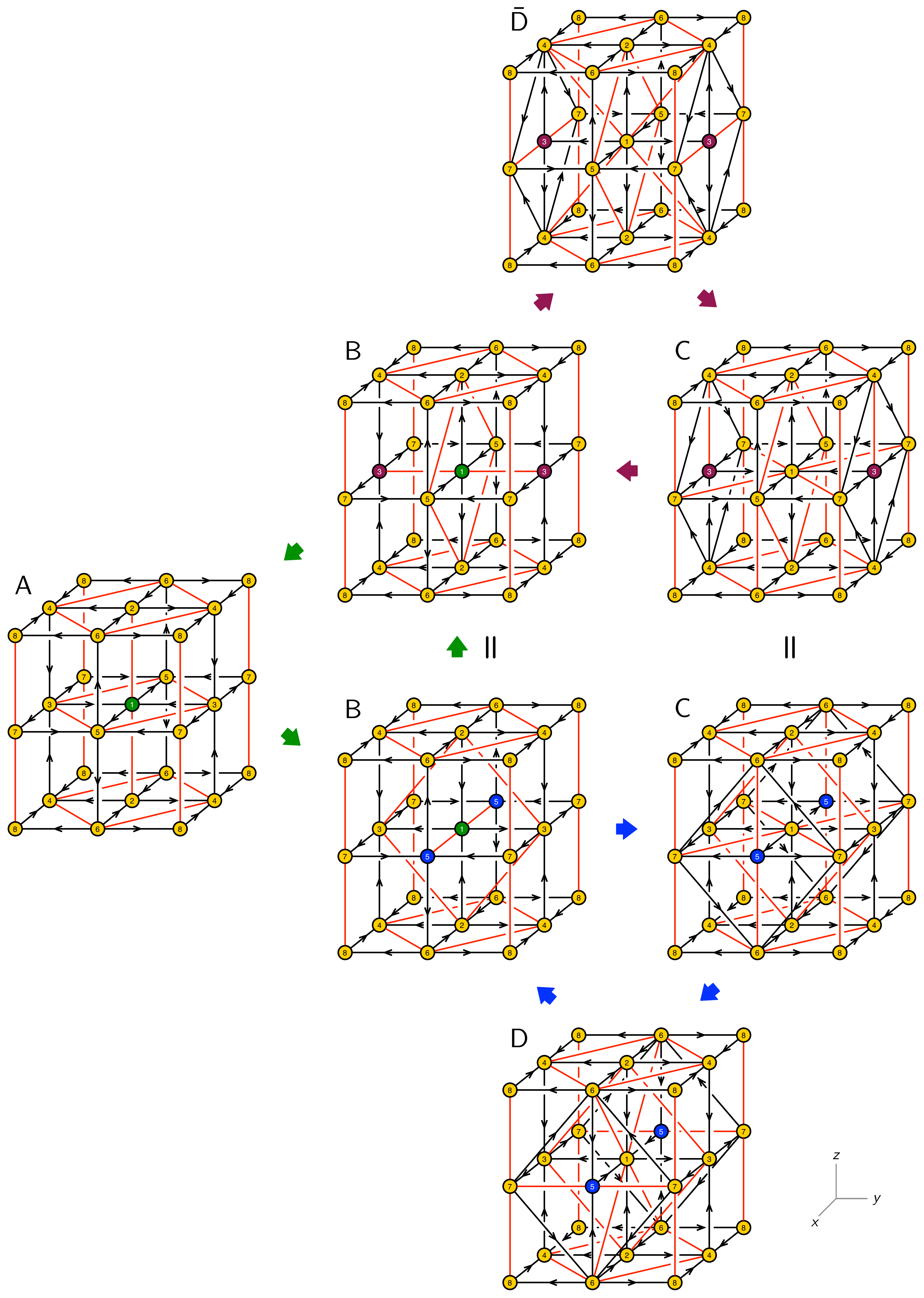}
\caption{A detailed version of the triality network for $Q^{1,1,1}/\mathbb{Z}_2$ shown in \fref{Q11-Z2-network}. On each triality loop, the node on which the triality operates is given a different color.
\label{Q11-Z2-network-full}}
 \end{center}
 \end{figure}

In order to simplify the presentation of results, we will identify theories that differ by a relabeling of the nodes in their periodic quivers. Restricting to theories connected to phase A by a sequence of dualizations on cubic nodes, the triality network for $Q^{1,1,1}/\mathbb{Z}_2$ contains five distinct phases: A, B, C, D, $\bar{\textrm{D}}$. As the notation suggests, phases A, B, C are self-conjugate under chiral conjugation, whereas D and  $\bar{\textrm{D}}$ are conjugate to each other. These theories form three minimal triality loops: (A-B-B), (B-C-D), (B-$\bar{\textrm{D}}$-C). \fref{Q11-Z2-network} shows a brief summary of the network and \fref{Q11-Z2-network-full} provides further details. Detailed information regarding all these phases, including their brane brick models, is presented in appendix \sref{Q111-Z2-detail}.

\section{The Mesonic Moduli Space \label{sec:mesonic-moduli}}

In this section we will see that triality on brane brick models preserves their classical mesonic moduli space. This moduli space is the toric CY$_4$ transverse to the probe D1-branes. A more interesting perspective on this fact is as follows. Brane brick models assign $2d$ $(0,2)$ gauge theories to toric CY$_4$ singularities. In general, a toric CY$_4$ does not give rise to a single gauge theory, but to a class of them. Remarkably, at least for the large class of explicit examples we have considered, all theories within this class are in fact related by triality.\footnote{When the $2d$ theories have $(2,2)$ SUSY, they are actually related by the $(2,2)$ duality of \cite{Benini:2014mia}, instead of triality.} It would be interesting to find a general proof of this statement, if it is indeed true as evidence suggests.

The mesonic moduli space can be computed using the {\it forward algorithm} developed in \cite{Franco:2015tna}. This calculation is in principle a straightforward exercise. However, it becomes extremely demanding already for theories such as the $Q^{1,1,1}/\mathbb{Z}_2$ phases discussed earlier. These complications are overcome by the {\it fast forward algorithm}, which was introduced in \cite{Franco:2015tya} and we summarize below.

\subsection{The Fast Forward Algorithm}

\label{section_review_fast_forward_algorithm}

We now briefly review the fast forward algorithm. Its key ingredients are the so-called {\it brick matchings}. Brick matchings are in one-to-one correspondence with GLSM fields in the toric description of the CY$_4$ singularity. A crucial feature that makes them extremely powerful is that they are defined combinatorially. To do so, it is useful to complete $J_a$- and $E_a$-terms into pairs of plaquettes by multiplying them by the corresponding $\Lambda_a$ or $\bar{\Lambda}_a$. A brick matching is then defined as a collection of chiral, Fermi and conjugate Fermi fields that contribute to every plaquette exactly once as follows:

\begin{itemize}
\item[1.] For every Fermi field pair $(\Lambda_a,\bar{\Lambda}_a)$, the chiral fields in the brick matching cover {\it either} each of the two $J_a$-term plaquettes {\it or} each of the two $E_a$-term plaquettes exactly once. 

\item[2.] 
If the chiral fields in the brick matching cover the plaquettes associated to the $J_a$-term, then $\bar{\Lambda}_a$ is included in the brick matching.

\item[3.] 
If the chiral fields in the brick matching cover the plaquettes associated to the $E_a$-term, then $\Lambda_a$ is included in the brick matching. 
\end{itemize}

The position of every brick matching in the toric diagram is determined by the intersections between the chiral field faces in the brick matching and the edges of the unit cell of the brane brick model, $\gamma_a$ ($a=x,y,z$). The integer-valued coordinates, $(n_x,n_y,n_z)\in \mathbb{Z}^3$, of a brick matching $p$ are 
given by 
\beal{es200a1}
n_a(p) = \sum_{X_{ij} \in p} \langle X_{ij}, \gamma_a \rangle ~,~ 
\eea
$a=x,y,z$, where the angle brackets indicate the standard intersection number between an orientated surface and an oriented line, contributing $\pm 1$ or $0$ to the overall coordinate $n_a$. In general, more than one brick matching can be mapped to the same point of the toric diagram. 

The chiral and Fermi field content of brick matchings can be efficiently encoded in terms of the {\it brick matching matrix} $P_{\Lambda\bar{\Lambda}}$ (see \cite{Franco:2015tya} for details). Rows and columns in this matrix correspond to quiver fields and brick matchings, respectively. An entry is equal to 1 if the corresponding field is contained in the brick matching and 0 otherwise.  We refer to the restriction of $P_{\Lambda\bar{\Lambda}}$ to chiral fields as the $P$-matrix. Since only chiral fields contain scalar components, the $P$-matrix is sufficient for studying the mesonic moduli space. Chiral fields should be expressed in terms of brick matchings as follows
\beq
X_i = \prod_\mu p_\mu^{P_{i\mu}} \, .
\eeq

\subsection{A Detailed Example}

We have computed the classical mesonic moduli space for all the examples presented in section \sref{sec:examples} using the fast forward algorithm. Our results are collected in the appendices and confirm that the theories connected by triality have the same mesonic moduli space. 

\begin{figure}[H]
\begin{center}
\resizebox{0.4\hsize}{!}{
\includegraphics[trim=0cm 0cm 0cm 0cm,totalheight=10 cm]{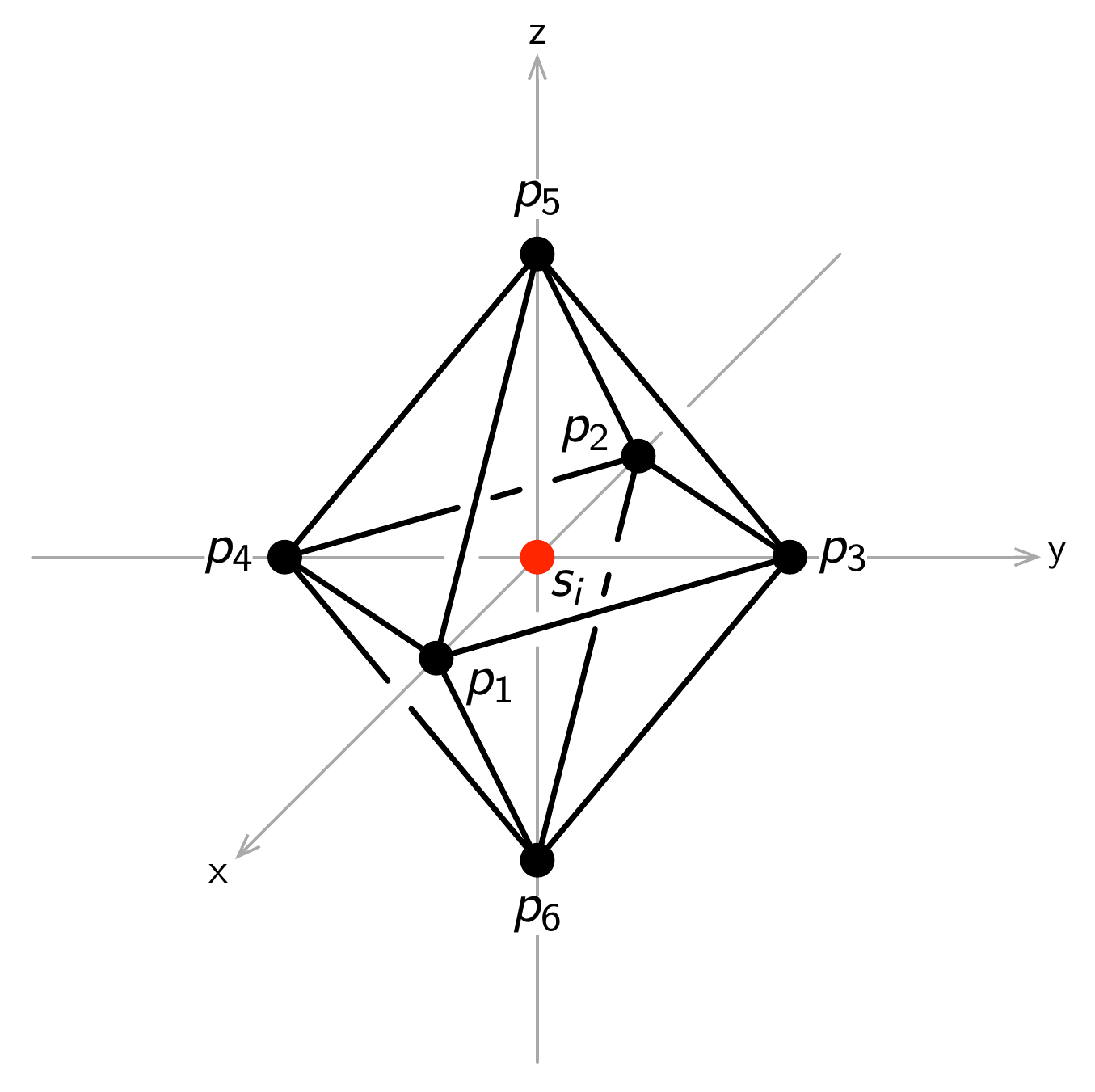}
}  
\caption{
Toric diagram for $Q^{1,1,1}/\mathbb{Z}_2$.
\label{q111z2-toric1}}
 \end{center}
 \end{figure} 

In order to understand the general reasons behind the invariance of the mesonic moduli space, we now pick a pair of these theories and discuss the connection between them in detail. Let us consider phases B and C of $Q^{1,1,1}/\mathbb{Z}_2$. For convenience, we reproduce the toric diagram for  $Q^{1,1,1}/\mathbb{Z}_2$ in \fref{q111z2-toric1}, where we indicate the positions of brick matchings.
The periodic quivers and brane brick models for both theories are given in appendix \sref{Q111-Z2-detail}.

Starting from phase B and performing a triality transformation on node 5 we obtain phase C. The $P$-matrix for phase B, summarizing the chiral field content of its brick matchings is:
\beal{q111z2b1-px}
\tiny
P^{(B)}=
\left(
\begin{array}{c|cccccc|ccccccccccc}
\; & p_1 & p_2 & p_3 & p_4 & p_5 & p_6 & s_1 & s_2 & s_3 & s_4 & s_5 & s_6 & s_7 & s_8 & s_9 & s_{10} & s_{11} \\
\hline
X_{37}^{+} & 1 & 0 & 0 & 0 & 0 & 0 & 1 & 0 & 0 & 0 & 0 & 0 & 0 & 0 &0 & 0 & 0 \\
X_{37}^{-} &  0 & 1 & 0 & 0 & 0 & 0 & 1 & 0 & 0 & 0 & 0 & 0 & 0 & 0 &0 & 0 & 0 \\
X_{62}^{+} &  1 & 0 & 0 & 0 & 0 & 0 & 0 & 0 & 0 & 0 & 0 & 1 & 1 & 0 &0 & 0 & 0 \\
X_{62}^{-} &  0 & 1 & 0 & 0 & 0 & 0 & 0 & 0 & 0 & 0 & 0 & 1 & 1 & 0 &0 & 0 & 0 \\
X_{84}^{+} &  1 & 0 & 0 & 0 & 0 & 0 & 0 & 0 & 0 & 0 & 0 & 0 & 1 & 0 &1 & 0 & 1 \\
X_{84}^{-} &  0 & 1 & 0 & 0 & 0 & 0 & 0 & 0 & 0 & 0 & 0 & 0 & 1 & 0 &1 & 0 & 1 \\
X_{24}^{+} &  0 & 0 & 1 & 0 & 0 & 0 & 0 & 0 & 0 & 0 & 0 & 0 & 0 & 1 &1 & 1 & 1 \\
X_{24}^{-} &  0 & 0 & 0 & 1 & 0 & 0 & 0 & 0 & 0 & 0 & 0 & 0 & 0 & 1 &1 & 1 & 1 \\
X_{68}^{+} &  0 & 0 & 1 & 0 & 0 & 0 & 0 & 0 & 0 & 0 & 0 & 1 & 0 & 1 &0 & 1 & 0 \\
X_{68}^{-} &  0 & 0 & 0 & 1 & 0 & 0 & 0 & 0 & 0 & 0 & 0 & 1 & 0 & 1 &0 & 1 & 0 \\
\rowcolor{cyan} X_{75}^{+} &  0 & 0 & 1 & 0 & 0 & 0 & 0 & 0 & 1 & 0 & 0 & 0 & 0 & 0 &0 & 0 & 0 \\
\rowcolor{cyan} X_{75}^{-} &  0 & 0 & 0 & 1 & 0 & 0 & 0 & 0 & 1 & 0 & 0 & 0 & 0 & 0 &0 & 0 & 0 \\
X_{43}^{+} &  0 & 0 & 0 & 0 & 1 & 0 & 0 & 0 & 0 & 1 & 1 & 0 & 0 & 0 &0 & 0 & 0 \\
X_{43}^{-} &  0 & 0 & 0 & 0 & 0 & 1 & 0 & 0 & 0 & 1 & 1 & 0 & 0 & 0 &0 & 0 & 0 \\
\rowcolor{cyan} X_{56}^{+} &  0 & 0 & 0 & 0 & 1 & 0 & 0 & 1 & 0 & 0 & 0 & 0 & 0 & 0 &0 & 0 & 0 \\
\rowcolor{cyan} X_{56}^{-} &  0 & 0 & 0 & 0 & 0 & 1 & 0 & 1 & 0 & 0 & 0 & 0 & 0 & 0 &0 & 0 & 0 \\
X_{21}^{+} &  0 & 0 & 0 & 0 & 1 & 0 & 0 & 0 & 0 & 0 & 1 & 0 & 0 & 0 &0 & 1 & 1 \\
X_{21}^{-} &  0 & 0 & 0 & 0 & 0 & 1 & 0 & 0 & 0 & 0 & 1 & 0 & 0 & 0 &0 & 1 & 1 \\
X_{13}^{+} &  0 & 0 & 1 & 0 & 0 & 0 & 0 & 0 & 0 & 1 & 0 & 0 & 0 & 1 &1 & 0 & 0 \\
X_{13}^{-} &  0 & 0 & 0 & 1 & 0 & 0 & 0 & 0 & 0 & 1 & 0 & 0 & 0 & 1 &1 & 0 & 0 \\
\end{array}
\right)~.~
\eea
In blue, we have highlighted the rows associated to chiral fields affected by triality, namely those charged under node 5. The $P$-matrix for phase C is given by:
\beal{q111z2c1-px}
\tiny
P^{(C)} =
\left(
\begin{array}{c|cccccc|cccccccccccccc}
\; & \tilde{p}_1 & \tilde{p}_2 & \tilde{p}_3 & \tilde{p}_4 & \tilde{p}_5 & \tilde{p}_6 & \tilde{s}_1 & \tilde{s}_2 & \tilde{s}_3 & \tilde{s}_4 & \tilde{s}_5 & \tilde{s}_6 & \tilde{s}_7 & \tilde{s}_8 & \tilde{s}_9 & \tilde{s}_{10} & \tilde{s}_{11} & \tilde{s}_{12} & \tilde{s}_{13} & \tilde{s}_{14}\\
\hline
\rowcolor{cyan} X_{15}^{+} & 1 & 0 & 0 & 0 & 0 & 0 & 1 & 0 & 0 & 0 & 0 & 0 & 1 & 0 & 0 & 0 & 1 & 1 & 0 & 0 \\
\rowcolor{cyan} X_{15}^{-} & 0 & 1 & 0 & 0 & 0 & 0 & 1 & 0 & 0 & 0 & 0 & 0 & 1 & 0 & 0 & 0 & 1 & 1 & 0 & 0 \\
X_{37}^{+} & 1 & 0 & 0 & 0 & 0 & 0 & 1 & 0 & 1 & 0 & 0 & 0 & 0 & 0 & 0 & 0 & 0 & 0 & 0 & 0 \\
X_{37}^{-} & 0 & 1 & 0 & 0 & 0 & 0 & 1 & 0 & 1 & 0 & 0 & 0 & 0 & 0 & 0 & 0 & 0 & 0 & 0 & 0 \\
X_{62}^{+} & 1 & 0 & 0 & 0 & 0 & 0 & 0 & 0 & 0 & 1 & 1 & 0 & 0 & 0 & 0 & 0 & 0 & 0 & 0 & 0 \\
X_{62}^{-} & 0 & 1 & 0 & 0 & 0 & 0 & 0 & 0 & 0 & 1 & 1 & 0 & 0 & 0 & 0 & 0 & 0 & 0 & 0 & 0 \\
X_{84}^{+} & 1 & 0 & 0 & 0 & 0 & 0 & 0 & 0 & 0 & 0 & 1 & 0 & 0 & 0 & 0 & 1 & 0 & 1 & 0 & 1 \\
X_{84}^{-} & 0 & 1 & 0 & 0 & 0 & 0 & 0 & 0 & 0 & 0 & 1 & 0 & 0 & 0 & 0 & 1 & 0 & 1 & 0 & 1 \\
X_{13}^{+} & 0 & 0 & 1 & 0 & 0 & 0 & 0 & 0 & 0 & 0 & 0 & 0 & 1 & 1 & 0 & 0 & 1 & 1 & 1 & 1 \\
X_{13}^{-} & 0 & 0 & 0 & 1 & 0 & 0 & 0 & 0 & 0 & 0 & 0 & 0 & 1 & 1 & 0 & 0 & 1 & 1 & 1 & 1 \\
X_{24}^{+} & 0 & 0 & 1 & 0 & 0 & 0 & 0 & 0 & 0 & 0 & 0 & 0 & 0 & 0 & 1 & 1 & 1 & 1 & 1 & 1 \\
X_{24}^{-} & 0 & 0 & 0 & 1 & 0 & 0 & 0 & 0 & 0 & 0 & 0 & 0 & 0 & 0 & 1 & 1 & 1 & 1 & 1 & 1 \\
\rowcolor{cyan} X_{57}^{+} & 0 & 0 & 1 & 0 & 0 & 0 & 0 & 0 & 1 & 0 & 0 & 0 & 0 & 1 & 0 & 0 & 0 & 0 & 1 & 1 \\
\rowcolor{cyan} X_{57}^{-} & 0 & 0 & 0 & 1 & 0 & 0 & 0 & 0 & 1 & 0 & 0 & 0 & 0 & 1 & 0 & 0 & 0 & 0 & 1 & 1 \\
X_{68}^{+} & 0 & 0 & 1 & 0 & 0 & 0 & 0 & 0 & 0 & 1 & 0 & 0 & 0 & 0 & 1 & 0 & 1 & 0 & 1 & 0 \\
X_{68}^{-} & 0 & 0 & 0 & 1 & 0 & 0 & 0 & 0 & 0 & 1 & 0 & 0 & 0 & 0 & 1 & 0 & 1 & 0 & 1 & 0 \\
X_{21}^{+} & 0 & 0 & 0 & 0 & 1 & 0 & 0 & 1 & 0 & 0 & 0 & 0 & 0 & 0 & 1 & 1 & 0 & 0 & 0 & 0 \\
X_{21}^{-} & 0 & 0 & 0 & 0 & 0 & 1 & 0 & 1 & 0 & 0 & 0 & 0 & 0 & 0 & 1 & 1 & 0 & 0 & 0 & 0 \\
X_{43}^{+} & 0 & 0 & 0 & 0 & 1 & 0 & 0 & 1 & 0 & 0 & 0 & 0 & 1 & 1 & 0 & 0 & 0 & 0 & 0 & 0 \\
X_{43}^{-} & 0 & 0 & 0 & 0 & 0 & 1 & 0 & 1 & 0 & 0 & 0 & 0 & 1 & 1 & 0 & 0 & 0 & 0 & 0 & 0 \\
\rowcolor{pink} X_{76}^{++} & 0 & 0 & 1 & 0 & 1 & 0 & 0 & 0 & 0 & 0 & 0 & 1 & 0 & 0 & 0 & 0 & 0 & 0 & 0 & 0 \\
\rowcolor{pink} X_{76}^{+-} & 0 & 0 & 1 & 0 & 0 & 1 & 0 & 0 & 0 & 0 & 0 & 1 & 0 & 0 & 0 & 0 & 0 & 0 & 0 & 0 \\
\rowcolor{pink} X_{76}^{-+} & 0 & 0 & 0 & 1 & 1 & 0 & 0 & 0 & 0 & 0 & 0 & 1 & 0 & 0 & 0 & 0 & 0 & 0 & 0 & 0 \\
\rowcolor{pink} X_{76}^{--} & 0 & 0 & 0 & 1 & 0 & 1 & 0 & 0 & 0 & 0 & 0 & 1 & 0 & 0 & 0 & 0 & 0 & 0 & 0 & 0 \\
\end{array}
\right)~.~
\eea
For clarity, we use tildes for the new brick matchings. Once again, we have highlighted the chiral fields participating in the triality transformation. In blue, we show the fields charged under node 5, i.e. the dual flavors. In pink, we show the chiral mesons, which extend between nodes 7 and 6.

Omitting the colored rows, both $P$-matrices are identical, up to column repetition. In other words, the brick matchings in the two theories only differ by the fields affected by triality. In particular, fields that intersect the boundaries of the unit cell remain unaltered. Hence, the fast forward algorithm implies that we obtain the same toric diagram.

Interestingly, the modification of the pieces of brick matchings affected by triality can relate different numbers of brick matchings in the dual theories. In this particular example, we have
\beq
\begin{array}{ccccccc}
s_{1} & \to & \{\tilde{s}_1,\tilde{s}_3 \} & \ \ \ \ \ \ &  s_{4} & \to & \{\tilde{s}_7,\tilde{s}_8 \} \\ 
s_{8} & \to & \{\tilde{s}_{11},\tilde{s}_{13} \} & \ \ \ \ \ \ &  s_{10} & \to & \{\tilde{s}_{12},\tilde{s}_{14} \}
\end{array}
\eeq
and
\beq
\{s_2,s_3\} \to \tilde{s}_6 \, .
\eeq

\subsection{General Invariance of the Mesonic Moduli Space}

\label{section_general_invariance_moduli_space}

\begin{figure}[ht!!]
\begin{center}
\resizebox{0.7\hsize}{!}{
\includegraphics[height=4.5cm]{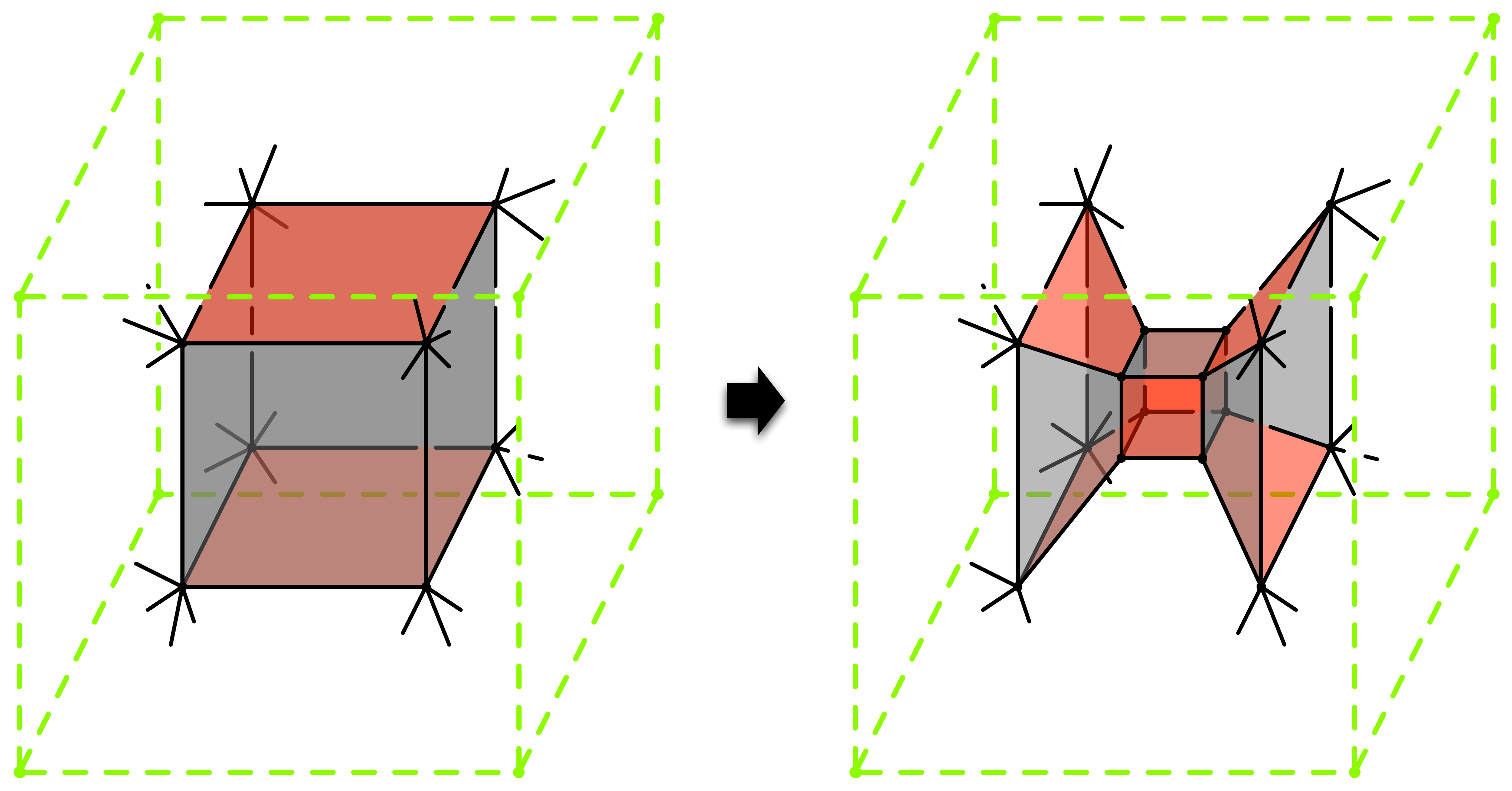}
}
\caption{It is possible to pick the unit cell of a brane brick model (here shown in green) such that all fields involved in the triality transformation are fully contained in its interior.}
\label{ftrialityunitcell}
 \end{center}
 \end{figure}

From the previous example, we can infer the general reasons underlying the invariance of the mesonic moduli space for brane brick models under triality. The key points are:
\begin{itemize}
\item[1.] Only fields affected by triality change in the chiral field content of brick matchings, i.e. fields charged under the dualized gauge group and chiral mesons. This highly non-trivial fact holds in all the explicit examples we have studied. An interesting consequence of it is that the map between brick matchings in dual theories is often not one-to-one.
\item[2.] It is always possible to pick the unit cell such that the region of the brane brick model modified by triality does not intersect the edges of the unit cell. This fact is schematically illustrated in \fref{ftrialityunitcell}.
\item[3.] The two previous points imply, via the fast forward algorithm, that the toric diagram of the mesonic moduli space is preserved.
\end{itemize}
These general arguments are rather compelling and we expect they underlie a rigorous proof of the invariance of the mesonic moduli space of brane brick models under triality.

\subsection{Connection to $(2,2)$ Duality} 

We now study another feature of our explicit examples of triality, their behavior under partial resolution of the corresponding CY$_4$. Partial resolution translates into classical higgsing at the level of the gauge theory. The connection between partial resolution and classical higgsing has been extensively studied in \cite{Franco:2015tna}, where it was exploited to generate gauge theories for arbitrary toric CY$_4$'s.

Verifying that the theories behave as expected under partial resolution is an additional check of the triality rules we used in this paper to derive them. As mentioned earlier, when multiple $(0,2)$ brane brick models are associated to the same toric CY$_4$, they are related by triality. When the CY$_4$ is of the special form CY$_3 \times \mathbb{C}$, we obtain $2d$ $(2,2)$ theories, which follow from dimensional reduction of the $4d$ $\mathcal{N}=1$ theories on the worldvolume of D3-branes probing the CY$_3$ factor of the geometry. Once again, there can also be multiple theories for a single geometry. In this case, the different theories are related by $2d$ $(2,2)$ duality \cite{Benini:2014mia}, which is the dimensional reduction of Seiberg duality for the $4d$ theories \cite{Seiberg:1994pq}.

The periodic quivers and brane brick models of the $2d$ $(2,2)$ theories can be systematically constructed from periodic quivers and brane tilings of the $4d$ theories by means of a {\it lifting algorithm} introduced in \cite{Franco:2015tna,Franco:2015tya}. Conversely, the objects encoding the $4d$ theories are obtained by a projection of those for the $2d$ theories.   

\begin{figure}[ht!!]
\begin{center}
\includegraphics[height=9cm]{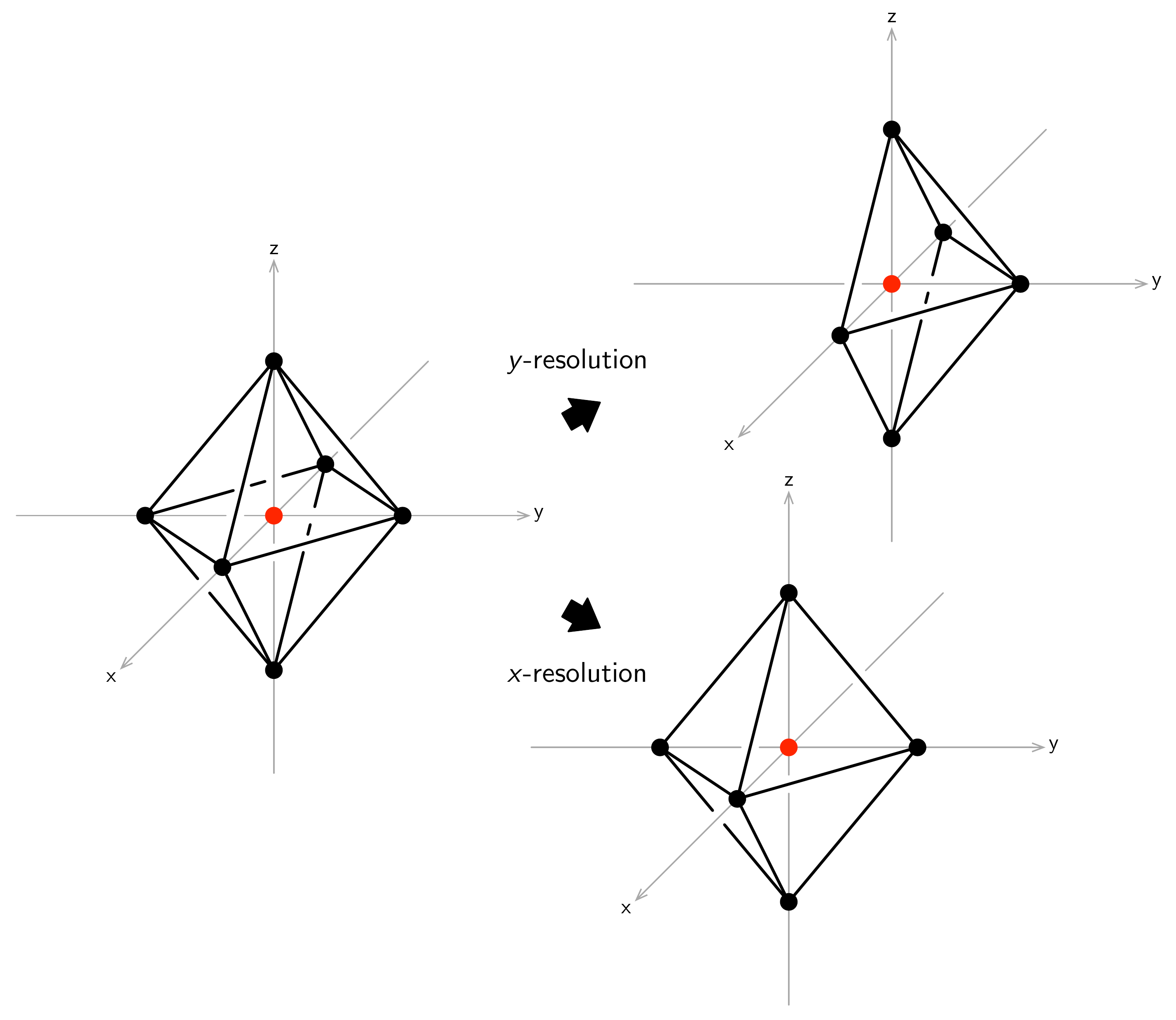}
\caption{Two different partial resolutions of $Q^{1,1,1}/\mathbb{Z}_2$ to 
$F_0 \times \mathbb{C}$. 
\label{Q11-Z2-F0xC-toric}}
 \end{center}
 \end{figure}

Let us consider $Q^{1,1,1}/\mathbb{Z}_2$. As shown in \fref{Q11-Z2-F0xC-toric}, this geometry is connected by partial resolution to $F_0 \times \mathbb{C}$. In fact, this can be achieved by removing any of the six corners of the toric diagram. The different choices map to different sets of chiral fields getting non-zero vevs. Performing partial resolution in different ways, it is thus possible to land on dual theories. 

\begin{figure}[ht!!]
\begin{center}
\includegraphics[height=10cm]{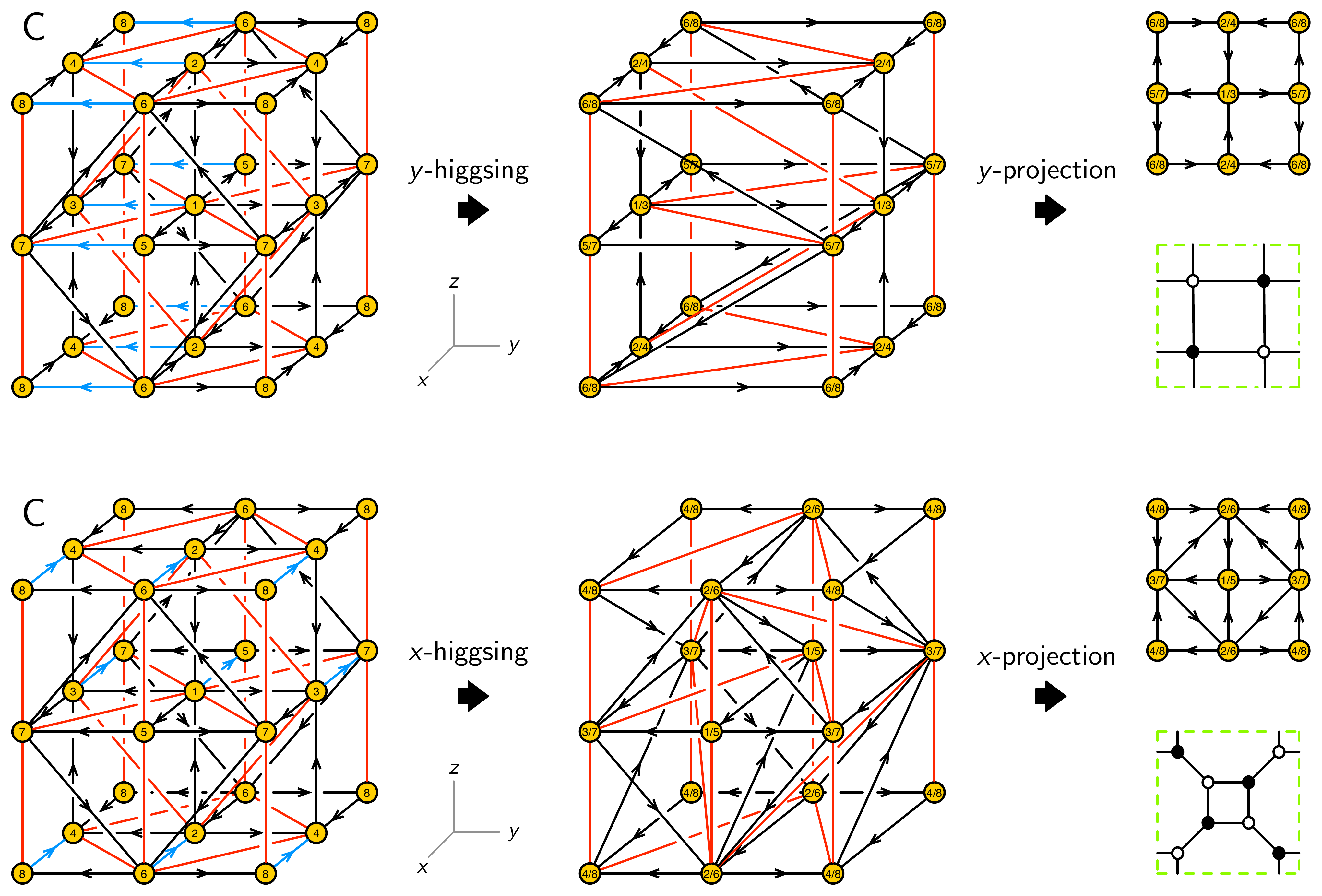}
\caption{Two different higgsings of $Q^{1,1,1}/\mathbb{Z}_2$ to 
$F_0 \times \mathbb{C}$. The chiral fields whose scalar components get non-zero vevs are shown in blue. The third column shows how they project down to the periodic quivers and brane tilings of the two toric phases of $F_0$ \cite{Feng:2002zw}, which are related by Seiberg duality.
\label{Q11-Z2-F0xC}}
 \end{center}
 \end{figure}

Let us consider for example phase C of $Q^{1,1,1}/\mathbb{Z}_2$. \fref{Q11-Z2-F0xC} presents two higgsings of this theory associated to the two ways of resolving the geometry down to $F_0 \times \mathbb{C}$ shown \fref{Q11-Z2-F0xC-toric}. They result in two different theories for $F_0 \times \mathbb{C}$, which are in fact related by $2d$ $(2,2)$ duality.

\section{Triality and Phase Boundaries \label{sec:phase-boundary}} 

So far, we have studied how triality is realized in terms of brane brick models and its effect on the geometry of the mesonic moduli space via brick matchings. In this section we continue investigating geometric aspects of triality, this time from the perspective of phase boundaries.

\subsection{Phase Boundaries for Cubic Nodes} 

Phase boundaries can be succinctly encoded in terms of the {\it phase boundary matrix} $H$ \cite{Franco:2015tya}. Columns in this matrix correspond to phase boundaries $\eta_\alpha$ and rows correspond to chiral and Fermi fields. An entry in $H_{i\alpha}$ is equal to $\pm 1$ if the face associated to the row $i$ is contained in the boundary represented by the column $\alpha$, with the sign determined by orientation, and $0$ otherwise. In other words, the $H$-matrix summarizes the net intersection numbers, counted with orientation, between phase boundaries and fields in the periodic quiver. The $H$-matrices for all the theories studied in this paper can be found in the appendices.

Let us consider phase A of $Q^{1,1,1}/\mathbb{Z}_2$. Its $H$-matrix is given in \eref{q111z2a-hx} and \eref{Q111Z2-Ha}. Let us focus on node 1, which is one of the four cubes in the brane brick model. It is useful to list the phase boundaries that intersect each of the fields charged under node 1. They are:
\be
\begin{array}{c|cccccccc}
X_{15}^+ & \ \textcolor{ablue}{\eta_{13}} \ & \ \textcolor{ablue}{\eta_{14}} \ & \ \eta_{15} \ &\ \eta_{16} \ & & & & \\
X_{15}^- & \ \textcolor{ablue}{\eta_{23}} \ & \ \textcolor{ablue}{\eta_{24}} \ & \ \eta_{25} \ &\ \eta_{26} \ & & & & \\
X_{31}^+ & \ \textcolor{ablue}{\eta_{13}} \ & \ \textcolor{ablue}{\eta_{23}} \ & \ \eta_{35} \ &\ \eta_{36} \ & & & & \\
X_{31}^- & \ \textcolor{ablue}{\eta_{14}} \ & \ \textcolor{ablue}{\eta_{24}} \ & \ \eta_{45} \ &\ \eta_{46} \ & & & & \\ \hline
\Lambda_{21}^+ & \ \eta_{15} \ & \ \eta_{25} \ & \ \eta_{36} \ &\ \eta_{46} \ & \ \textcolor{ablue}{\eta_{13}} \ & \ \textcolor{ablue}{\eta_{14}} \ & \ \textcolor{ablue}{\eta_{23}} \ & \ \textcolor{ablue}{\eta_{24}} \ \\
\Lambda_{21}^- & \ \eta_{16} \ & \ \eta_{26} \ & \ \eta_{35} \ &\ \eta_{45} \ & \ \textcolor{ablue}{\eta_{13}} \ & \ \textcolor{ablue}{\eta_{14}} \ & \ \textcolor{ablue}{\eta_{23}} \ & \ \textcolor{ablue}{\eta_{24}} \
\end{array}
\label{fields_phase_boundaries_cube}
\eeq
Here $\eta_{ij}$ indicates the phase boundary associated to the external edge between the points $p_i$ and $p_j$ of the toric diagram. In more detail:
\begin{itemize}
\item
The cube involves twelve different phase boundaries. This is equal to the total number of phase boundaries for both $Q^{1,1,1}$ and $Q^{1,1,1}/\mathbb{Z}_2$. However, it could be a subset of all the phase boundaries for more complicated geometries.
\item
Four phase boundaries intersect each chiral field.
\item
Eight phase boundaries intersect each Fermi field. Four of them, shown in blue in the previous table, are common to the two Fermis and are not part of the corresponding alternating cones. These phase boundaries also intersect the chiral fields and play an important role when phase boundaries get reorganized due to triality, as explained below. 
\end{itemize}
This configuration seems to be rather generic. Indeed, all the cubes in the phases $Q^{1,1,1}$ of $Q^{1,1,1}/\mathbb{Z}_2$ studied in this paper have the same general structure.

\begin{figure}[h]
\begin{center}
\includegraphics[width=15.5cm]{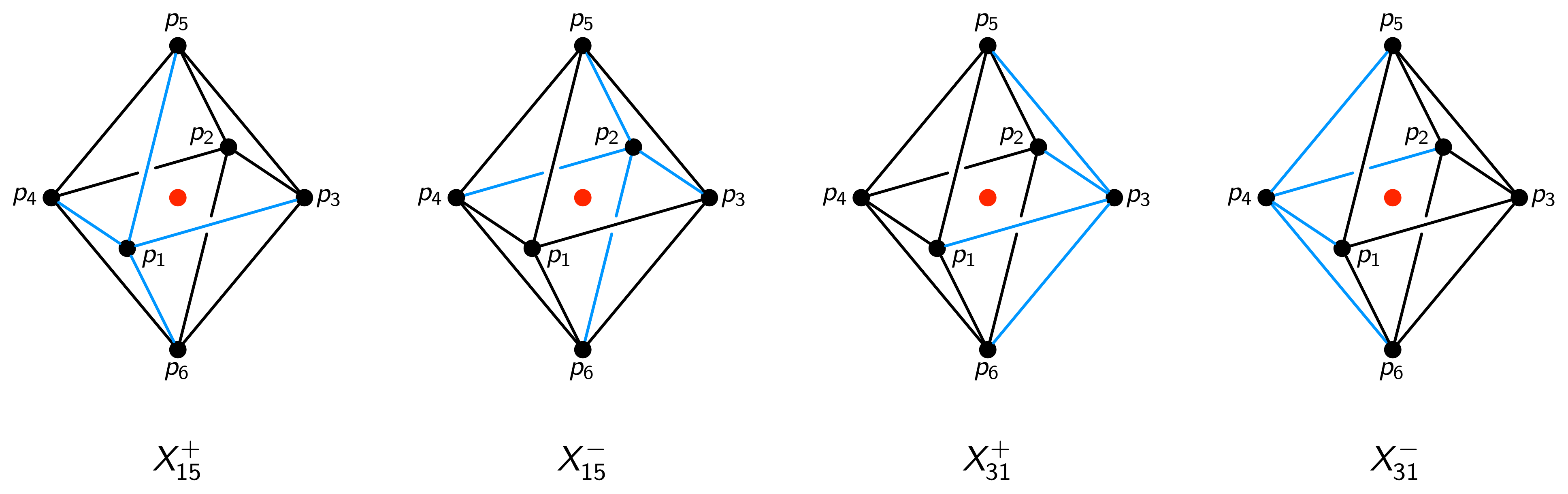}
\caption{Each chiral field involves four phase boundaries associated to edges in the toric diagram terminating on the same external point.
}
\label{Q111_Z2_toric_diagram_chirals}
\end{center}
\end{figure}

The four phase boundaries intersecting each of the chiral fields correspond to edges in the toric diagram connected to a common external point, as shown in \fref{Q111_Z2_toric_diagram_chirals}. It would be interesting to check whether this simple structure generalizes to cubes in theories for larger toric diagrams.

\begin{figure}[h]
\begin{center}
\includegraphics[width=12cm]{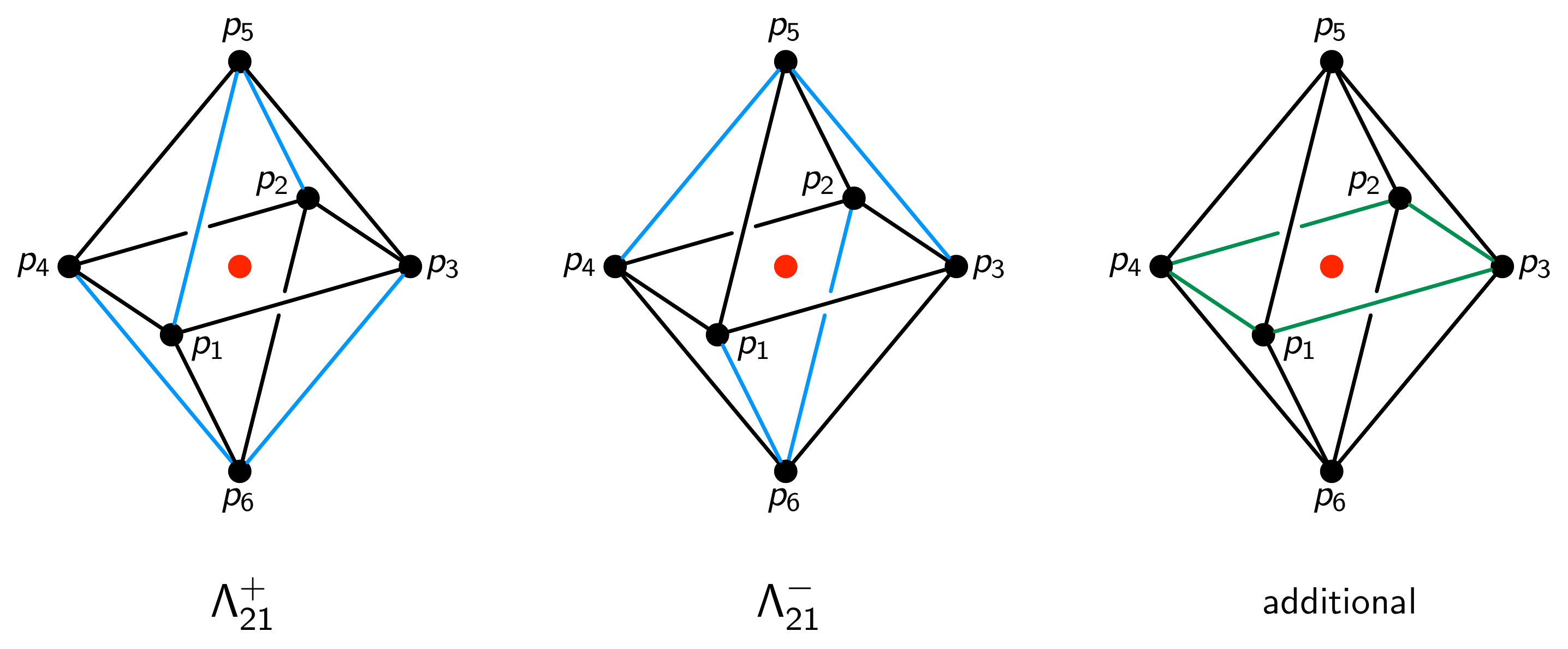}
\caption{Edges on the toric diagram for phase boundaries associated to Fermi fields. The ones that do not participate in the alternating cones are shown in green.
}
\label{Q111_Z2_toric_diagram_Fermis}
\end{center}
\end{figure}

For each Fermi field, the phase boundaries indicated in black in \eref{fields_phase_boundaries_cube}, namely those giving rise to the alternating cones, correspond to two pairs of edges connected to opposite external points in the toric diagram. We show these configurations in \fref{Q111_Z2_toric_diagram_Fermis}. The four additional phase boundaries, listed in blue in \eref{fields_phase_boundaries_cube}, are shown in green in the figure. We observe that their orientations are coplanar.

\begin{figure}[h]
\begin{center}
\includegraphics[height=4.5 cm]{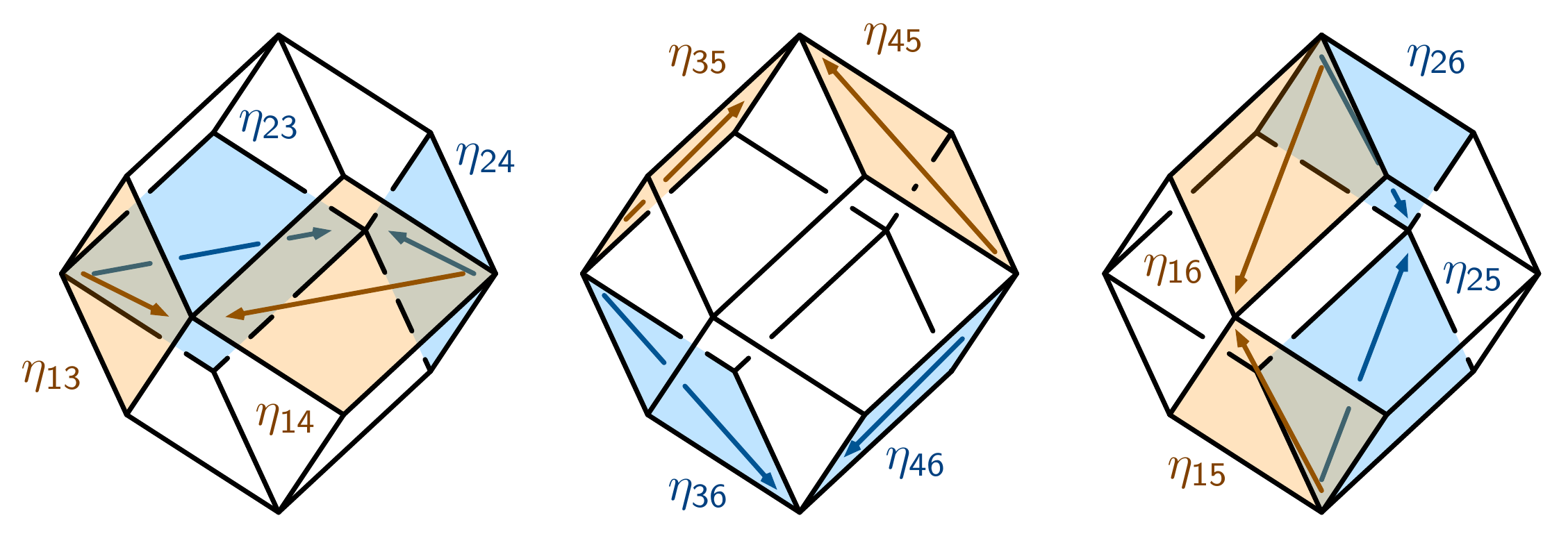} 
\caption{
The configuration of oriented phase boundaries forming the RD associated to cube 1 in the brane brick model. The orientations of phase boundaries are determined from the toric diagram with the prescription introduced in \cite{Franco:2015tya}.
\label{rd-boundary}}
 \end{center}
 \end{figure}

\begin{figure}[h]
\begin{center}
\includegraphics[width=5cm]{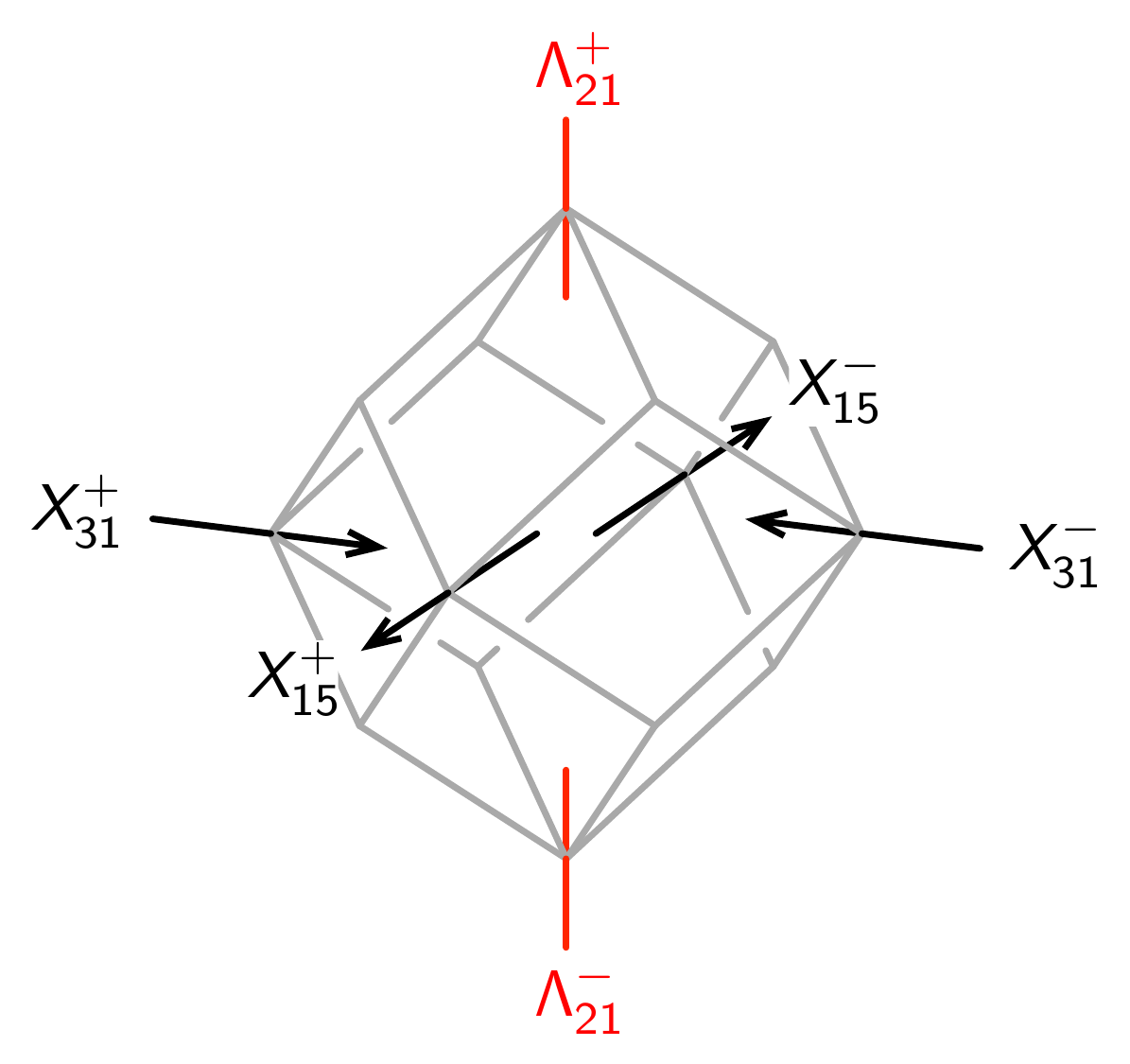}
\caption{Quiver fields charged under node 1 and their corresponding phase boundary intersections. 
}
\label{RD_quiver_fields}
\end{center}
\end{figure}

The twelve phase boundaries form a rhombic dodecahedron (RD) as shown in \fref{rd-boundary}. The relative positions of parallel phase boundaries are crucial, and are such that precisely the chiral and Fermi fields of the cubic node arise from oriented and alternating cones, as shown in \fref{RD_quiver_fields}. The cubic vertices of the RD are neither oriented nor alternating, hence they do not give rise to any field in the quiver or, equivalently, to a face in the brane brick model.\footnote{It is interesting to note that the phase boundaries in abelian orbifolds of $\mathbb{C}^4$ also form rhombic dodecahedra. However, the orientations of the phase boundaries in them are such that the twelve vertices of each RD correspond to fields.} 

\subsection{Phase Boundaries under Triality} 

\begin{figure}[H]
\begin{center}
\includegraphics[width=14cm]{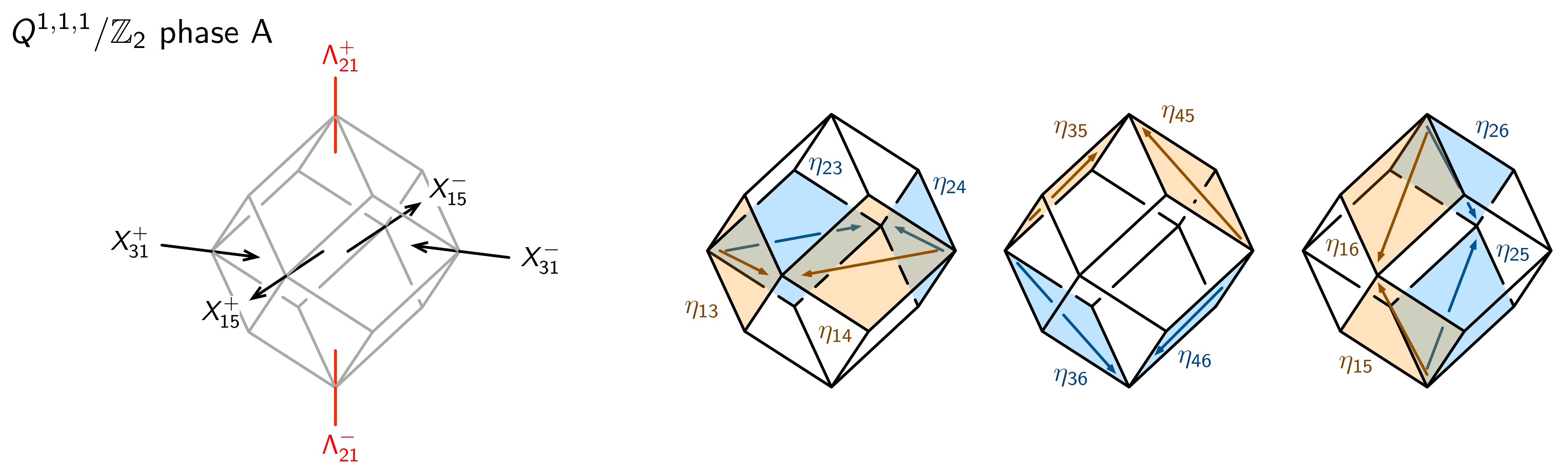} 
\\
\includegraphics[width=14cm]{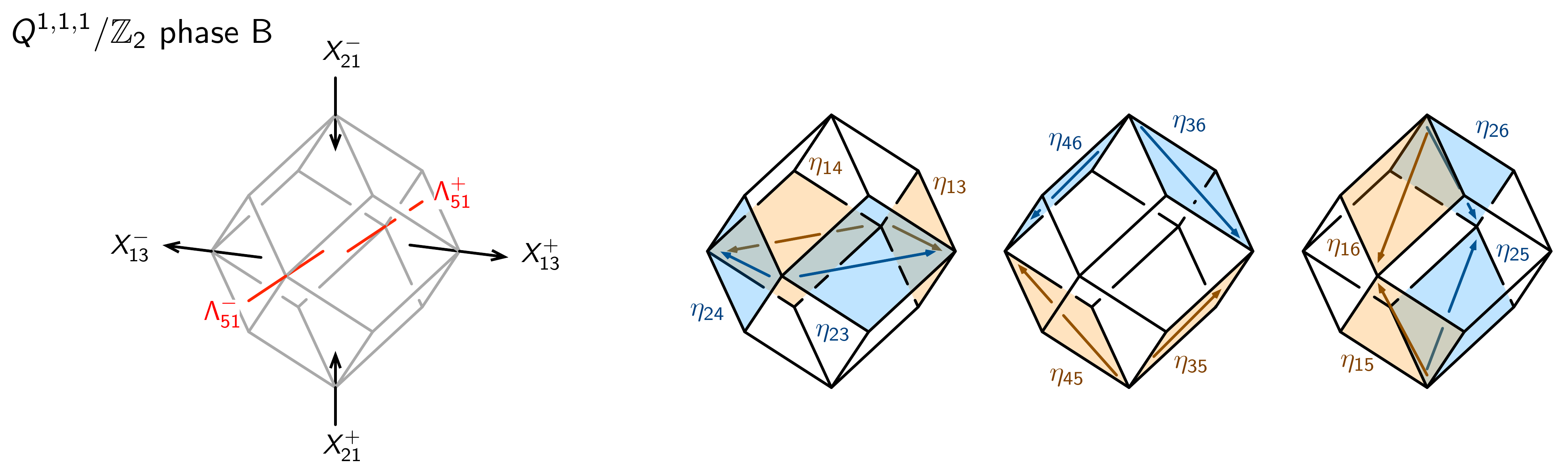}
\caption{Triality as a rearrangement of phase boundaries.
}
\label{rdAB}
\end{center}
\end{figure}

Triality corresponds to exchanging opposite phase boundaries in two of the ``squares" shown in \fref{rd-boundary}. At every dualization, the pair of squares whose phase boundaries are flipped are the one intersecting the four chirals and the one intersecting the incoming chirals and the Fermis. \fref{rdAB} shows the triality from phase A to phase B of $Q^{1,1,1}/\mathbb{Z}_2$. In this case, the phase boundaries are exchanged as follows: $\{\eta_{13} \leftrightarrow \eta_{24}, \eta_{14} \leftrightarrow \eta_{23} \}$ and $\{\eta_{35} \leftrightarrow \eta_{46}, \eta_{36} \leftrightarrow \eta_{46} \}$. The correspondence between oriented and alternating intersections and fields in the quiver implies that this operation precisely implements the modification of the periodic quiver under triality. Furthermore, as expected, iterating this transformation three times amounts to the identity. 

Phase boundaries should not be regarded as planes, but as $2d$ surfaces that get deformed while preserving their homology during triality. Below we provide a more detailed description of this process.

\subsection{Phase Boundaries as Collection of Faces} 

\begin{figure}[H]
\begin{center}
\includegraphics[height=11.5 cm]{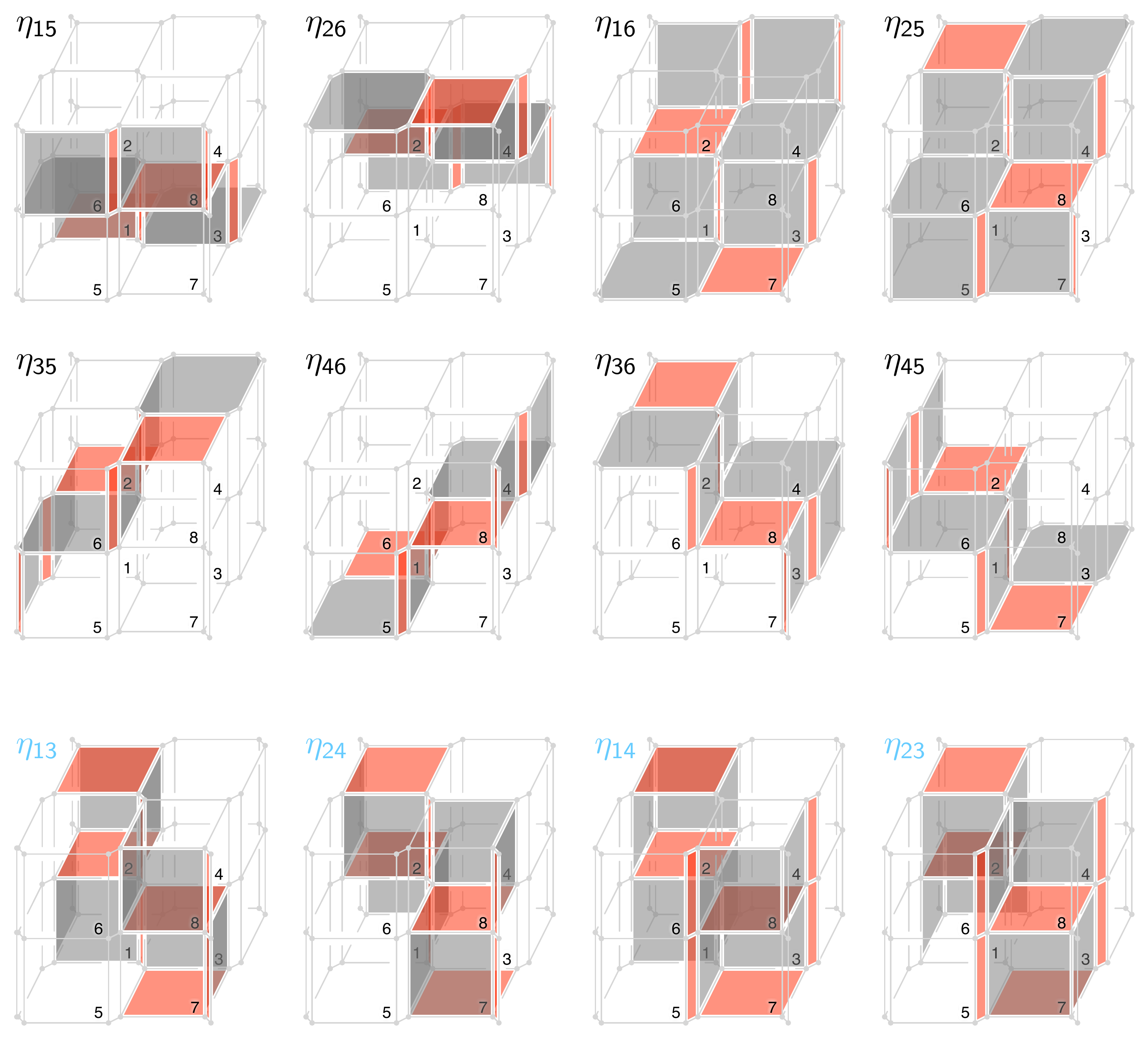}
\caption{
Phase boundaries for phase A of  $Q^{1,1,1}/\mathbb{Z}_2$ as collections of faces in the brane brick model. Faces that are repeated due to periodicity are colored only once.
\label{Q111-boundaries-global}}
 \end{center}
 \end{figure}

Let us now study the structure of phase boundaries in further detail, regarding them as collections of faces in the brane brick model. We can use the $H$-matrix in \eref{q111z2a-hx} and \eref{Q111Z2-Ha} to locate the twelve phase boundaries on the brane brick model. The result is shown in \fref{Q111-boundaries-global}. 

In order to simplify the representation of phase boundaries, we have used the same notation for the bricks as in \fref{fq111modelAbranebrick} (with the unit cell doubled as in the periodic quiver of \fref{fq111a1}, to account for the $\mathbb{Z}_2$ orbifold) except that we have shrunk the Fermi faces in the diagonal directions of octagonal cylinders until they become thin strips. As expected, the phase boundaries form 2-cycles whose homology vectors equal those of the corresponding edges in the toric diagram.  

\begin{figure}[h]
\begin{center}
\includegraphics[height=7 cm]{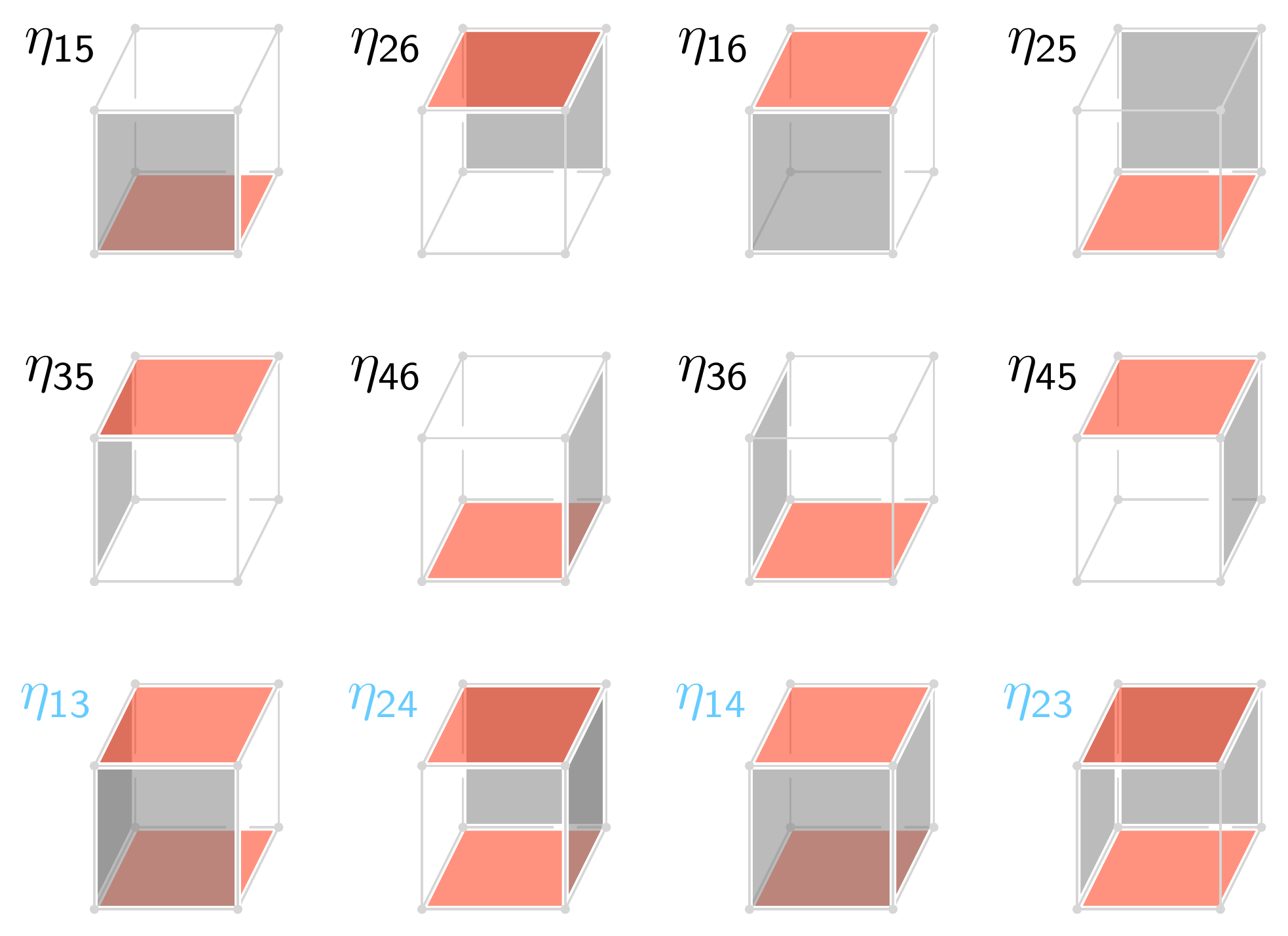}
\caption{
Local shape of the phase boundaries at the cubic node 1. There are two qualitatively different classes of phases boundaries: those involving two faces (top and middle rows) and those involving four faces (bottom row).
\label{Q111-boundaries-local}}
 \end{center}
 \end{figure}

\fref{Q111-boundaries-local} shows the local appearance of phase boundaries at node 1. We see a sharp distinction between the black and blue phase boundaries of \eref{fields_phase_boundaries_cube}, which involve two and four faces, respectively. 

\begin{figure}[ht]
\begin{center}
\includegraphics[height=7.5 cm]{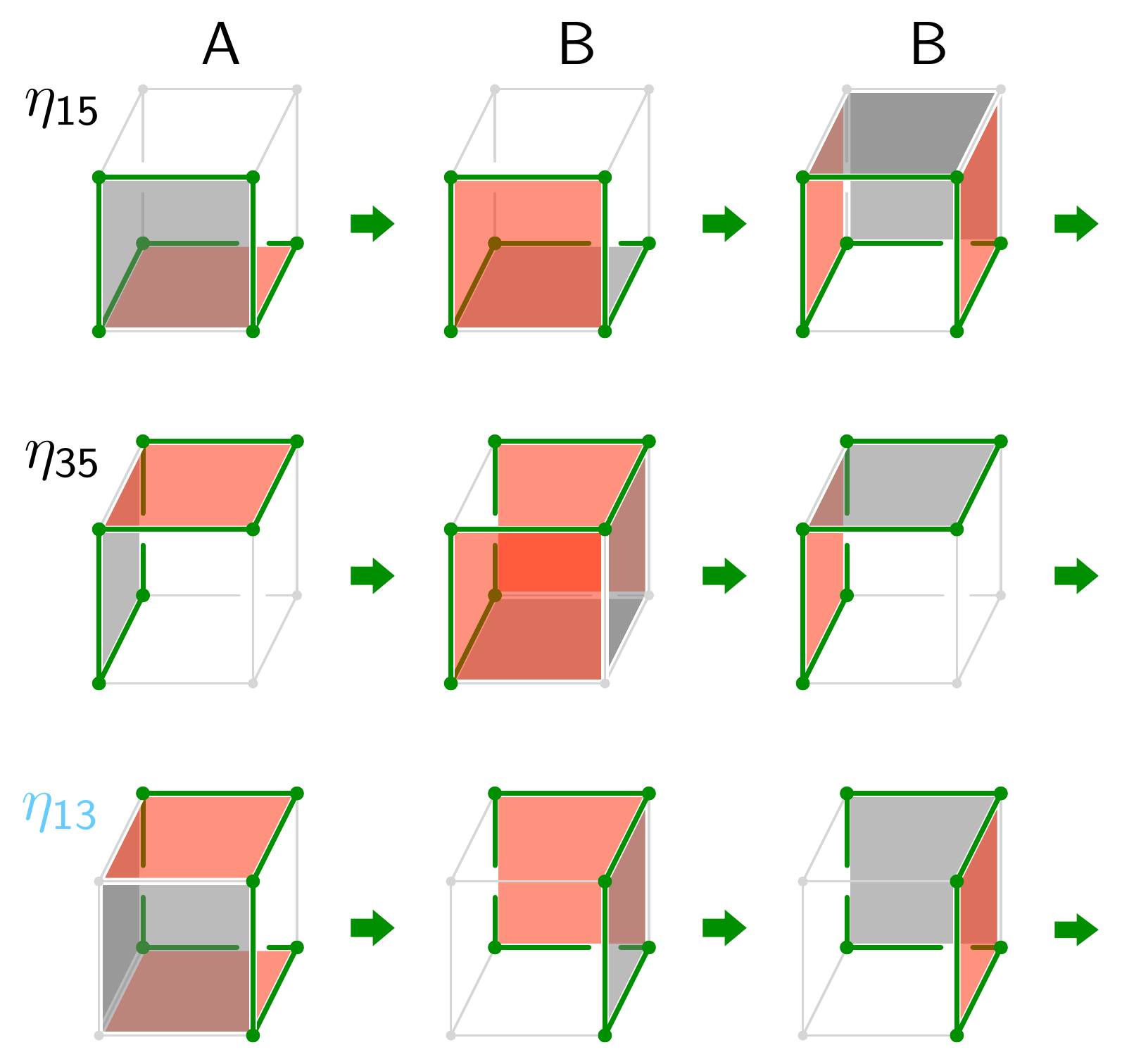}
\caption{
Local behavior of the phase boundaries under repeated triality on node 1.
\label{Q111-boundaries-triality}}
 \end{center}
 \end{figure}

Let us now consider how the discussion in the previous section translates into this language. To do so, we take the phase boundaries in the first column of \fref{Q111-boundaries-local} and examine how they transform under repeated triality on node 1. The result is summarized in \fref{Q111-boundaries-triality}. The phase boundary $\eta_{13}$ at the bottom row starts from a 4-face configuration. Triality transforms it into a 2-face configuration. The next move exchanges the types of the two faces. \fref{Q111-boundaries-triality} shows that the same rule applies to $\eta_{15}$ and $\eta_{35}$. It also works for all other phase boundaries in \fref{Q111-boundaries-local}.

A crucial feature of this transformation is that it is {\em local}. A local triality move should not affect the way the phase boundaries depart from node 1. In \fref{Q111-boundaries-triality}, the edges along which the phase boundaries depart from node 1 are colored in green. As it is clear from the figure, they are invariant under the triality moves. Another sign of the local nature of the triality move is that the transformation rule described in \fref{Q111-boundaries-triality} 
is universal for all cubes in all toric phases of $Q^{1,1,1}$ and $Q^{1,1,1}/\mathbb{Z}_2$.

\subsection{Possible Connections to Integrable Systems}

Phase boundaries are the brane brick model analogues of zig-zag paths for brane tilings \cite{Hanany:2005ss,Feng:2005gw}. Brane tilings encode the $4d$ $\mathcal{N}=1$ quiver gauge theories on the worldvolume of D3-branes probing toric CY$_3$ singularities \cite{Franco:2005rj,Franco:2005sm}. It was noted in \cite{Hanany:2005ss} that the square move that implements Seiberg duality leading to toric phases in brane tilings can be interpreted as a double Yang-Baxter transformation in terms of zig-zag paths. This turns out to be a manifestation of integrable structures underlying these $4d$ quivers and their dimensional reductions \cite{Goncharov:2011hp,Franco:2011sz,Eager:2011dp,Yamazaki:2012cp,Franco:2012hv,Yamazaki:2015voa,Franco:2015rnr}.

Zamolodchikov's {\it tetrahedron equation} is the $3d$ generalization of the Yang-Baxter equation \cite{Zamolodchikov:1981kf}. Geometrically, it is associated to inequivalent dissections of the rhombic dodecahedron into four hexahedra. As we explained above, triality of brane brick models corresponds to a flip of opposite phase boundaries in a rhombic dodecahedron configuration. It is natural to speculate that this fact hints to integrable structures related to triality. It would be interesting to explore this direction in further detail.

\section{Conclusions \label{sec:conclusion}}

We introduced the first brane realization of $2d$ $(0,2)$ triality. Our prescription applies to the infinite class of gauge theories arising on D1-branes probing toric CY$_4$ singularities, which are mapped to brane brick models by T-duality. 

We showed in explicit examples that triality preserves the classical mesonic moduli space of the theories, which corresponds to the probed CY$_4$. We also outlined a general proof of this invariance for arbitrary brane brick models. Conversely, brane brick models generically associate a class of $2d$ $(0,2)$ gauge theories to every toric CY$_4$ and, remarkably, the theories within this class turn out to be related by triality.

We studied brane brick models for $Q^{1,1,1}$ and $Q^{1,1,1}/\mathbb{Z}_2$, for which we derived several dual phases and explained how they are connected in the corresponding triality networks. 

Finally, we investigated triality from the perspective of the phase boundaries underlying the fast inverse algorithm. For cubic nodes, it simply amounts to an exchange of some of the opposite faces in a local rhombic dodecahedron configuration.

There are several directions for future investigation. The analysis of brane brick models in \cite{Franco:2015tna,Franco:2015tya} and the study of triality in this paper have been done mostly at a classical level. 
An important open question concerns the quantum dynamics of brane brick models. For the minimal triality quiver \cite{Gadde:2013lxa} we reviewed in section \sref{sec:triality-review}, the quantum dynamics was studied in depth in \cite{Gadde:2014ppa,Gadde:2015kda}.

It would also be desirable to obtain a deeper understanding of the RG flow of brane brick models. As explained in \cite{Gadde:2014ppa,Gadde:2015kda}, the RG flow of a $2d$ $(0,2)$ gauge theory proceeds in two steps. The original gauge theory is regarded as a non-abelian gauged linear sigma model (GLSM). In the first stage of the RG, the gauge coupling quickly becomes strong and the theory flows to a non-linear sigma model (NLSM), whose target space is the vacuum moduli space of the GLSM. In the second stage, the parameters of the theory, such as the complexified K\"ahler parameter of the target space geometry, undergo further RG running. 
At an RG fixed point, if it exists, the theory becomes a $2d$ $(0,2)$ superconformal field theory. In summary, we would like to follow the GLSM$\rightarrow$ NLSM $\rightarrow$ SCFT flow for brane brick models.

There are a handful of tools available to probe the SCFT directly from the GLSM. For example, we may compute the elliptic genus 
using localization techniques following \cite{Gadde:2013wq,Gadde:2013ftv,Benini:2013nda,Benini:2013xpa}. Among other things, the elliptic genus in the RR sector captures chiral operators of the theory. In $2d$ $(0,2)$ theories, the spectrum and the operator product expansions (OPE's) of chiral operators are described by quantum sheaf cohomology, first introduced in \cite{Katz:2004nn}.\footnote{See, {\it e.g.}, \cite{Donagi:2011uz,Donagi:2011va} and references therein for further developments in quantum sheaf cohomology, and \cite{Guo:2015gha} for a recent application to triality.}
In general, chiral operators of $2d$ $(0,2)$ theories do not form a ring, but under certain conditions they are closed under OPE's \cite{Adams:2005tc}.\footnote{We thank C. Beem for a discussion on this point.} In discussing such chiral operators, it is important to include the left-moving fermions that are annihilated by the right-moving supercharges. In our earlier works on brane brick models \cite{Franco:2015tna,Franco:2015tya}, we only enumerated gauge invariant operators consisting of chiral multiplets, as a way to determine the vacuum moduli space of the GLSM. It would be interesting to explore whether the combinatorial tools associated to brane brick models, such as brick matchings, are also useful for incorporating left-moving fermions. The computation of the elliptic genus and other SCFT observables from the GLSM will be the topic of a forthcoming paper \cite{topub1}.

\acknowledgments

We would like to thank A. Amariti, C. Beem, A. Gadde, P. Putrov, M. Romo, N. Seiberg, E. Sharpe and P. Yi for enjoyable and helpful discussions. We are also grateful to D. Ghim for collaboration on related topics. The work of S. F. is supported by the U.S. National Science Foundation grant PHY-1518967 and by a PSC-CUNY award. The work of S. L. is supported by Samsung Science and Technology Foundation under Project Number SSTF-BA1402-08 and by the IBM Einstein Fellowship of the Institute for Advanced Study.

\newpage

\appendix 

\section{Phases of $Q^{1,1,1}$ \label{Q111-detail}}

\begin{figure}[h]
\begin{center}
\resizebox{0.4\hsize}{!}{
\includegraphics[trim=0cm 0cm 0cm 0cm,totalheight=10 cm]{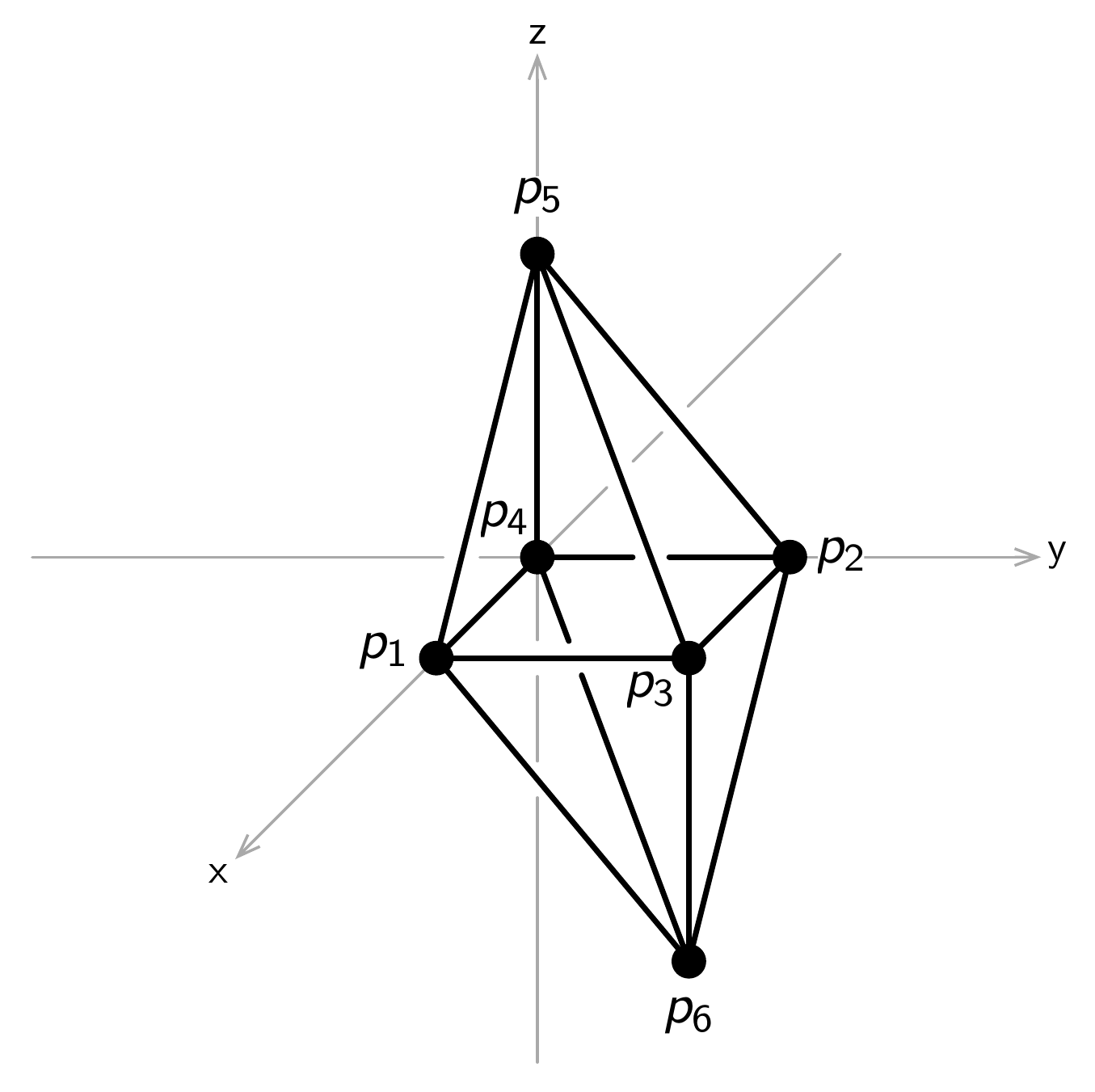}
}  
\caption{
Toric diagram for $Q^{1,1,1}$.
\label{quivertoricq111}}
 \end{center}
 \end{figure}

In the main text, we discussed three phases (A, S, NT) for $Q^{1,1,1}$ and four phases (A, B, C, D) for $Q^{1,1,1}/\mathbb{Z}_2$. In this appendix and the next one, we present detailed data for all these theories with the exception of NT, which is a non-toric phase. We start from $Q^{1,1,1}$, whose toric diagram is reproduced in \fref{quivertoricq111}. All the phases satisfy the vanishing trace condition.

\subsection*{Phase A}

\begin{figure}[h]
\begin{center}
\resizebox{0.6\hsize}{!}{
\includegraphics[trim=0cm 0cm 0cm 0cm,totalheight=10 cm]{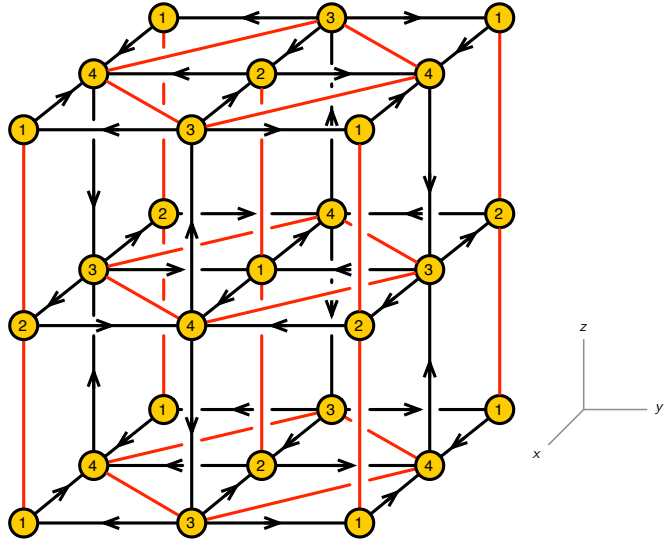}
}  
\caption{
Periodic quiver for phase A of $Q^{1,1,1}$. Notice that the region represented has twice the volume of the unit cell.  
\label{q111a-quiver-temp2}}
 \end{center}
 \end{figure}

This theory was originally introduced in \cite{Franco:2015tna,Franco:2015tya}. \fref{q111a-quiver-temp2} shows the periodic quiver for this phase. It has 4 gauge groups, 10 chiral and 6 Fermi fields. This is in agreement with $n^\chi=G+n^F$, with $n^\chi$, $G$ and $n^F$ the total numbers of chiral fields, gauge groups and Fermi fields, respectively. This general expression is valid for all brane brick models and is obtained by summing the anomaly cancellation condition \eref{anomaly-simple} over all gauge groups \cite{Franco:2015tya}. The $J$- and $E$-terms for the different Fermi fields are
{\small
\beq
\begin{array}{lcccc}
& &  J & &  E
\\
   \Lambda_{21}^{+} & :\ \ \ & X^{+}_{14} \cdot X^{-}_{43} \cdot X^{-}_{32} - X^{-}_{14} \cdot X^{-}_{43} \cdot X^{+}_{32}    & \ \ \ \ &  X^{+}_{24} \cdot X^{+}_{43} \cdot X^{-}_{31} -  X^{-}_{24} \cdot X^{+}_{43} \cdot X^{+}_{31}  
\\
   \Lambda_{21}^{-} & :\ \ \ & X^{-}_{14} \cdot X^{+}_{43} \cdot X^{+}_{32}  - X^{+}_{14} \cdot X^{+}_{43} \cdot X^{-}_{32}   & \ \ \ \ & X^{+}_{24} \cdot X^{-}_{43} \cdot X^{-}_{31} - X^{-}_{24} \cdot X^{-}_{43} \cdot X^{+}_{31}  
\\
   \Lambda_{34}^{++} & :\ \ \ & X^{+}_{43} \cdot X^{-}_{32} \cdot X^{-}_{24} \cdot X^{-}_{43} - X^{-}_{43} \cdot X^{-}_{31} \cdot X^{-}_{14} \cdot X^{+}_{43}   & \ \ \ \ & X^{+}_{32} \cdot X^{®†+}_{24} -  X^{+}_{31} \cdot X^{+}_{14}  
\\
   \Lambda_{34}^{--} & :\ \ \ & X^{+}_{43} \cdot X^{+}_{31} \cdot X^{+}_{14} \cdot X^{-}_{43} - X^{-}_{43} \cdot X^{+}_{32} \cdot X^{+}_{24} \cdot X^{+}_{43}  & \ \ \ \ &  X^{-}_{32} \cdot X^{-}_{24} - X^{-}_{31} \cdot X^{-}_{14}  
\\
    \Lambda_{34}^{+-} & :\ \ \ & X^{-}_{43} \cdot X^{+}_{31} \cdot X^{-}_{14} \cdot X^{+}_{43}  - X^{+}_{43} \cdot X^{-}_{32} \cdot X^{+}_{24} \cdot X^{-}_{43}   & \ \ \ \ &  X^{+}_{32} \cdot X^{-}_{24} - X^{-}_{31} \cdot X^{+}_{14}   
\\
   \Lambda_{34}^{-+} & :\ \ \ & X^{-}_{43} \cdot X^{+}_{32} \cdot X^{-}_{24} \cdot X^{+}_{43} - X^{+}_{43} \cdot X^{-}_{31} \cdot X^{+}_{14} \cdot X^{-}_{43}   & \ \ \ \ &   X^{-}_{32} \cdot X^{+}_{24} - X^{+}_{31} \cdot X^{-}_{14}  
\\ 
   \end{array}
\label{je-q111a}
\eeq
}

Given the large size of the brick matching matrices for the theories under consideration, it is convenient to split them into their chiral and Fermi parts, $P$ and $P_\Lambda$.\footnote{This notation differs slightly from the one used in \cite{Franco:2015tya}.} For phase A, they are given by
\beal{q111a-pxl}
\scriptsize
P =
\left(
\begin{array}{c|cccccc}
\; & p_1 & p_2 & p_3 & p_4  & p_5 & p_6 
\\
\hline 
X_{14}^{+} &  1 & 0 & 0 & 0 & 0 & 0
\\
X_{14}^{-} &  0 & 1 & 0 & 0 & 0 & 0
\\
X_{32}^{+} &  1 & 0 & 0 & 0 & 0 & 0
\\
X_{32}^{-} &  0 & 1 & 0 & 0 & 0 & 0
\\
X_{24}^{+} & 0 & 0 & 1 & 0 & 0 & 0
\\
X_{24}^{-} & 0 & 0 & 0 & 1 & 0 & 0
\\
X_{31}^{+} & 0 & 0 & 1 & 0 & 0 & 0
\\
X_{31}^{-} & 0 & 0 & 0 & 1 & 0 & 0
\\
X_{43}^{+} & 0 & 0 & 0 & 0 & 1 & 0
\\
X_{43}^{-} & 0 & 0 & 0 & 0 & 0 & 1
\end{array}
\right)~,~ \quad
P_\Lambda = 
\left(
\begin{array}{c|cccccc}
\; & p_1 & p_2 & p_3 & p_4  & p_5 & p_6 
\\
\hline
\Lambda_{21}^{+} & 0 & 0 & 1 & 1 & 1 & 0
\\
\Lambda_{21}^{-} & 0 & 0 & 1 & 1 & 0 & 1
\\
\Lambda_{34}^{++} & 1 & 0 & 1 & 0 & 0 & 0
\\
\Lambda_{34}^{+-} & 1 & 0 & 0 & 1 & 0 & 0
\\
\Lambda_{34}^{-+} & 0 & 1 & 1 & 0 & 0 & 0
\\
\Lambda_{34}^{--} & 0 & 1 & 0 & 1 & 0 & 0
\\
\end{array}
\right)~.~
\eea

\begin{figure}[ht]
\begin{center}
\resizebox{0.5\hsize}{!}{
\includegraphics[trim=0cm 0cm 0cm 0cm,totalheight=10 cm]{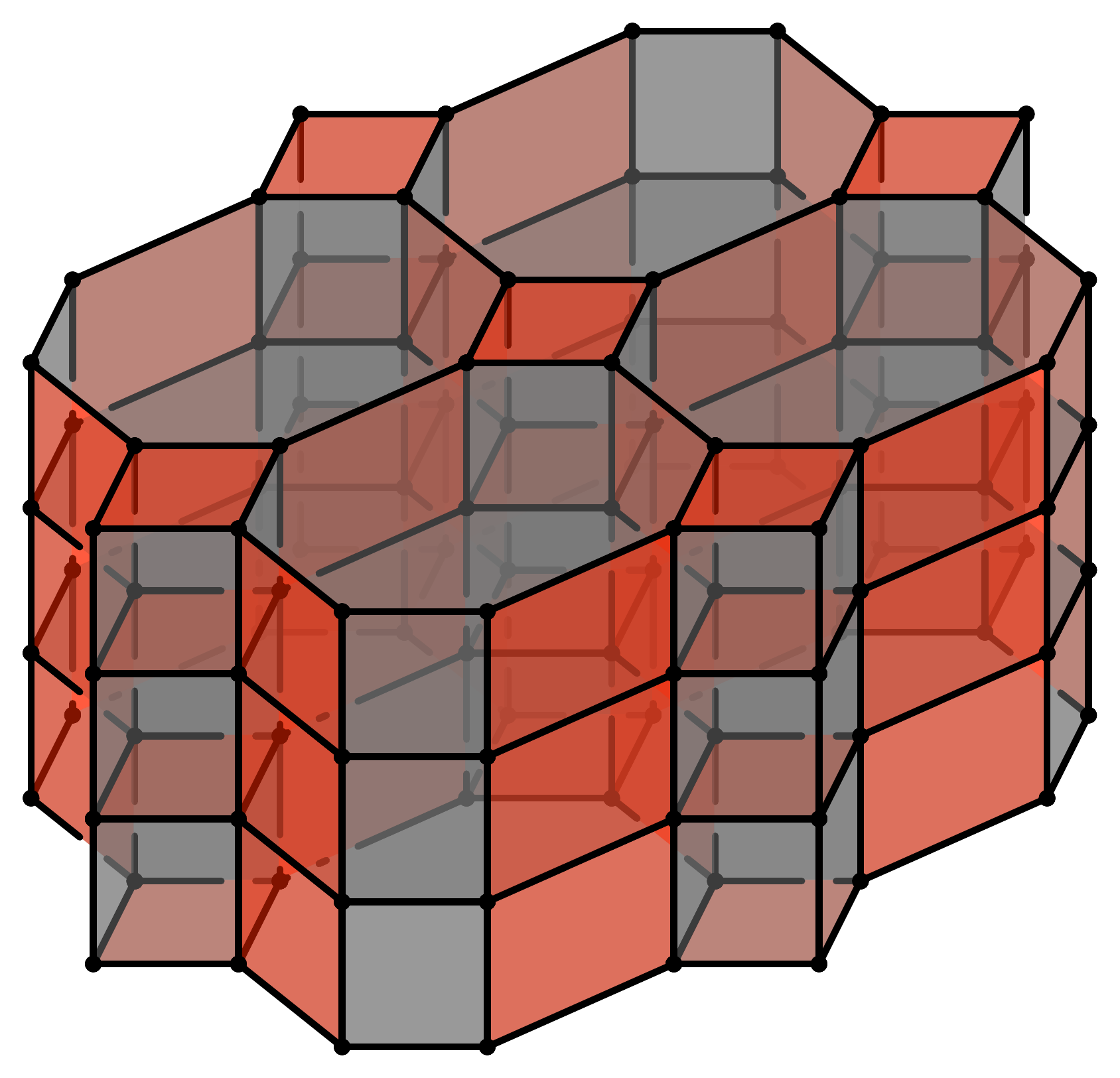}
}  
\caption{
Brane brick model for phase A of $Q^{1,1,1}$.
\label{fq111modelAbranebrick}}
 \end{center}
 \end{figure}

\begin{figure}[H]
\begin{center}
\resizebox{0.45\hsize}{!}{
\includegraphics[trim=0cm 0cm 0cm 0cm,totalheight=10 cm]{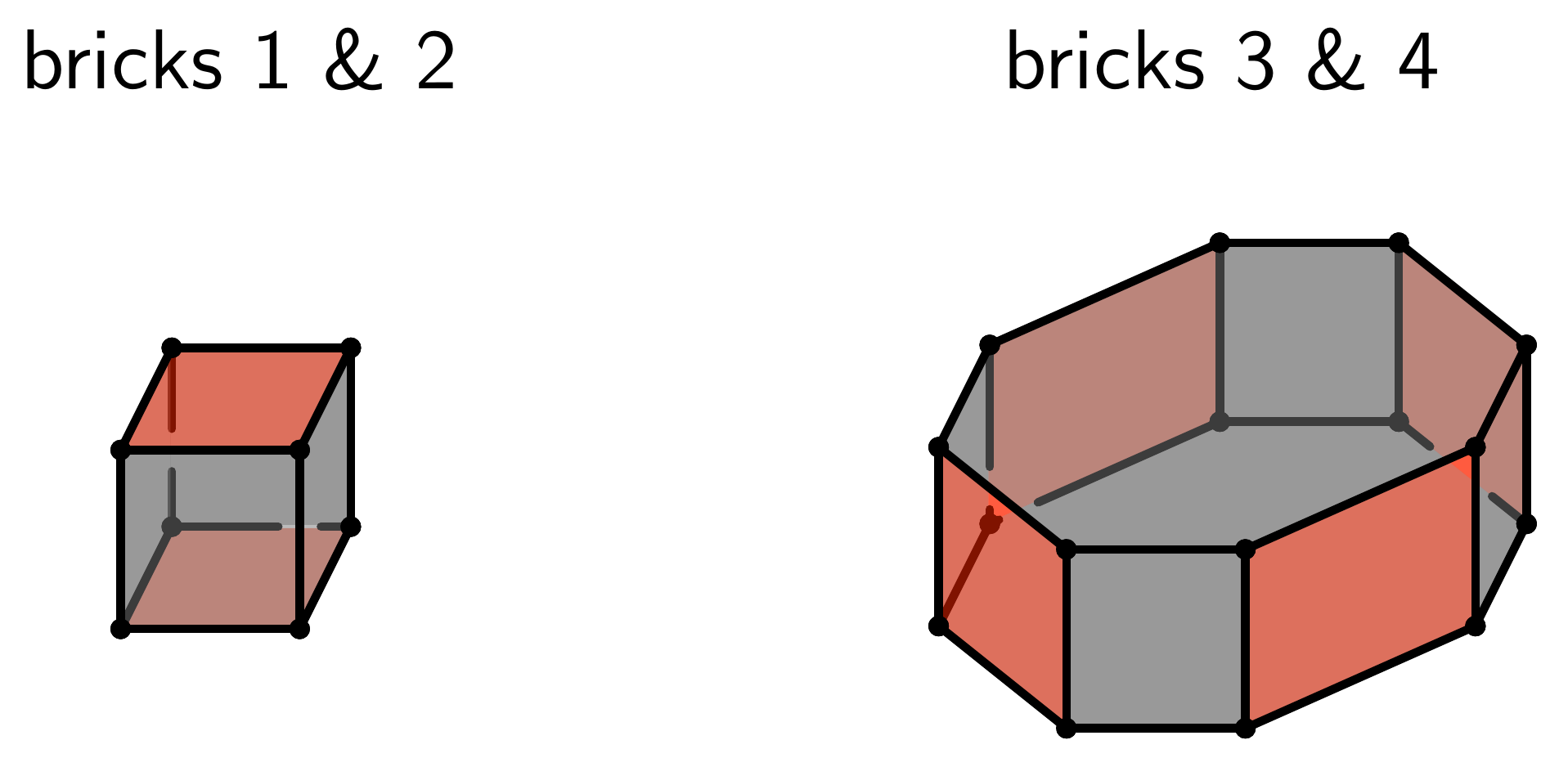}
}  
\caption{
Brane bricks for phase A of $Q^{1,1,1}$.
\label{fq111a-brick}}
 \end{center}
 \end{figure}

The brane brick model is obtained by dualizing the periodic quiver and it is shown in \fref{fq111modelAbranebrick}. It consists of four bricks, one per gauge group, which are independently listed in \fref{fq111a-brick}. 

As done with the $P$ and $P_\Lambda$ matrices, it is also convenient to split the chiral and Fermi parts of the $H$-matrix encoding the phase boundaries. They are
\beal{h-q111aX}
\scriptsize
H_X
=
\left(
\begin{array}{c|cccccccccccc}
\; &
\eta_{13} & \eta_{14} & \eta_{23} & \eta_{24} & \eta_{15} & \eta_{16} & \eta_{25} & \eta_{26} & \eta_{35} & \eta_{36} & \eta_{45} & \eta_{46} 
\\
\hline
X_{14}^{+} & 1 & 1 & 0 & 0 & 1 & 1 & 0 & 0 & 0 & 0 & 0 & 0 \\
X_{14}^{-} & 0 & 0 & 1 & 1 & 0 & 0 & 1 & 1 & 0 & 0 & 0 & 0 \\
X_{32}^{+} & 1 & 1 & 0 & 0 & 1 & 1 & 0 & 0 & 0 & 0 & 0 & 0 \\
X_{32}^{-} & 0 & 0 & 1 & 1 & 0 & 0 & 1 & 1 & 0 & 0 & 0 & 0 \\
X_{24}^{+} & -1 & 0 & -1 & 0 & 0 & 0 & 0 & 0 & 1 & 1 & 0 & 0 \\
X_{24}^{-} & 0 & -1 & 0 & -1 & 0 & 0 & 0 & 0 & 0 & 0 & 1 & 1 \\
X_{31}^{+} & -1 & 0 & -1 & 0 & 0 & 0 & 0 & 0 & 1 & 1 & 0 & 0 \\
X_{31}^{-} & 0 & -1 & 0 & -1 & 0 & 0 & 0 & 0 & 0 & 0 & 1 & 1 \\
X_{43}^{+} & 0 & 0 & 0 & 0 & -1 & 0 & -1 & 0 & -1 & 0 & -1 & 0 \\
X_{43}^{-} & 0 & 0 & 0 & 0 & 0 & -1 & 0 & -1 & 0 & -1 & 0 & -1 \\
\end{array}
\right)~,~
\eea

\beal{h-q111aLambda}
\scriptsize
H_\Lambda
=
\left(
\begin{array}{c|cccccccccccc}
\; &
\eta_{13} & \eta_{14} & \eta_{23} & \eta_{24} & \eta_{15} & \eta_{16} & \eta_{25} & \eta_{26} & \eta_{35} & \eta_{36} & \eta_{45} & \eta_{46} 
\\
\hline
\Lambda_{21}^{+} & -1 & -1 & -1 & -1 & 0 & -1 & 0 & -1 & 1 & 0 & 1 & 0 \\
\Lambda_{21}^{-} & -1 & -1 & -1 & -1 & -1 & 0 & -1 & 0 & 0 & 1 & 0 & 1 \\
\Lambda_{34}^{++} & 0 & 1 & -1 & 0 & 1 & 1 & 0 & 0 & 1 & 1 & 0 & 0 \\
\Lambda_{34}^{+-} & 1 & 0 & 0 & -1 & 1 & 1 & 0 & 0 & 0 & 0 & 1 & 1 \\
\Lambda_{34}^{-+} & -1 & 0 & 0 & 1 & 0 & 0 & 1 & 1 & 1 & 1 & 0 & 0 \\
\Lambda_{34}^{--} & 0 & -1 & 1 & 0 & 0 & 0 & 1 & 1 & 0 & 0 & 1 & 1 \\
\end{array}
\right)~.~
\eea

We choose the unit cell of the brane brick model such that the only non-trivial intersections between chiral faces and the unit cell edges
\beq
\begin{array}{cclcccl}
\vec{n}(X^{+}_{14}) & = &  (1,0,0) & \ \ \ \ \ \ &
\vec{n}(X^{-}_{14}) & = & (0,1,0) \\[.1cm]
\vec{n}(X^{+}_{31}) & = & (1,1,0) & \ \ \ \ \ \ &
\vec{n}(X^{-}_{31}) & = &(0,0,0) \\[.1cm]
\vec{n}(X^{+}_{43}) & = & (0,0,1) & \ \ \ \ \ \ &
\vec{n}(X^{-}_{43}) & = & (1,1,-1) 
\end{array}
\label{q111a-toric-data}
\eeq
where $\vec{n}(X)$ indicates the individual contribution of a chiral field $X$ to \eref{es200a1}. It is now possible to determine the mesonic moduli space using the fast forward algorithm explained in section \sref{section_review_fast_forward_algorithm}. Combining \eref{q111a-toric-data} with the chiral field content of brick matchings summarized by the $P$-matrix \eref{q111a-pxl}, we indeed obtain the $Q^{1,1,1}$ toric diagram shown in \fref{quivertoricq111}.

\subsection*{Phase S}

\begin{figure}[H]
\begin{center}
\resizebox{0.6\hsize}{!}{
\includegraphics[height=8 cm]{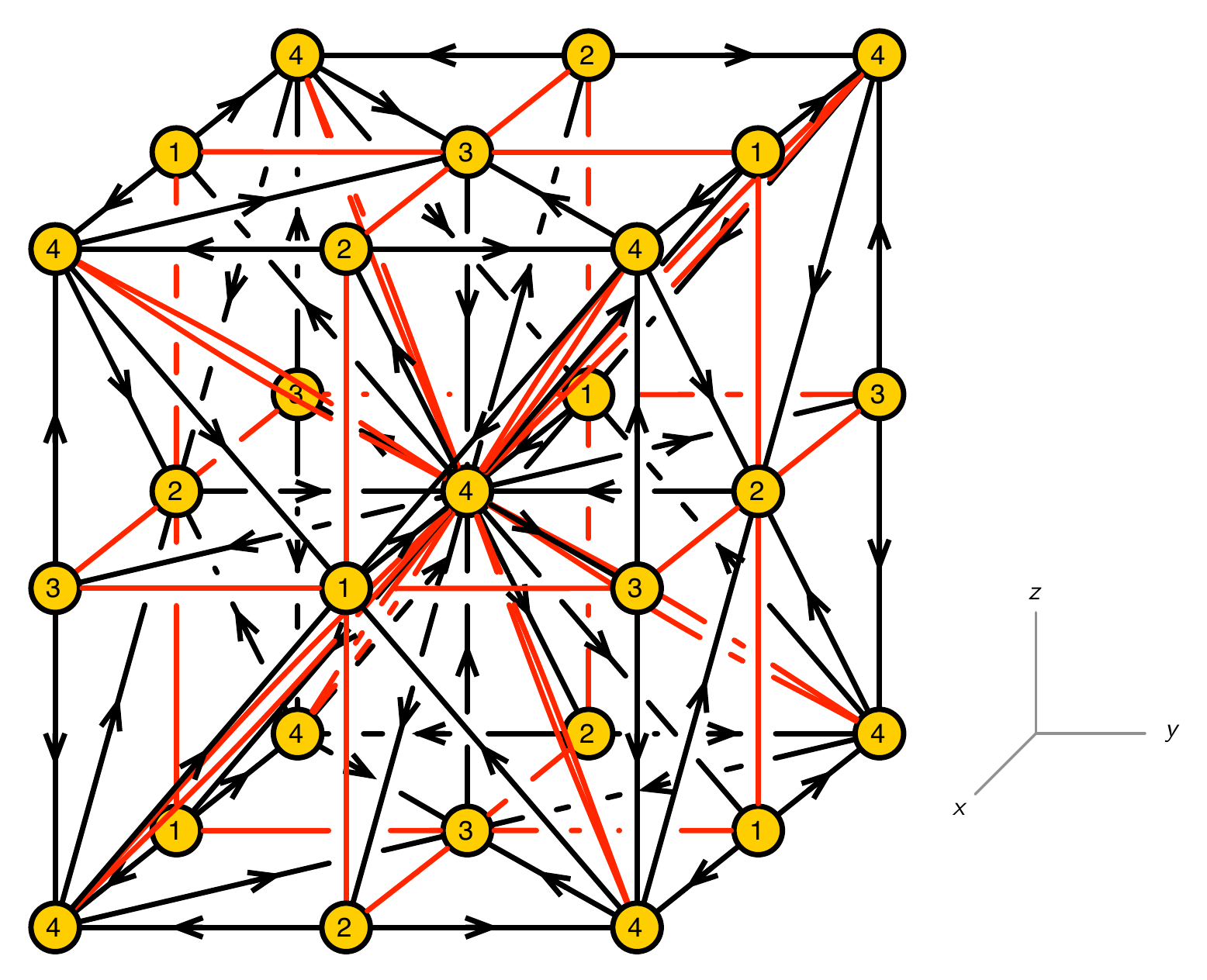}
}  
\caption{
Periodic quiver for phase S of $Q^{1,1,1}$. Notice that the region represented has twice the volume of the unit cell. The radial Fermi lines have multiplicity 2.
\label{q111s-quiver}}
 \end{center}
 \end{figure}

The periodic quiver for phase S of $Q^{1,1,1}$ is shown in \fref{q111s-quiver}. The theory has 4 gauge groups, 18 chiral fields and 14 Fermi fields. 

\vspace{-.04cm} 

In contrast to phase A, the periodic quiver of phase S exhibits a manifest octahedral symmetry. In order to simplify the discussion of the symmetries of the theory, we have shifted the periodic quiver with respect to \fref{Q111-oct}, placing node 4 at the center. It is also convenient to split the nodes into a {\it central} node (4) and three {\it satellite} nodes (1, 2, 3). To emphasize how the octahedral symmetry is realized, we will use the following notation for the fields: 
\begin{align}
\mbox{Outgoing chiral (from center to satellite)} &: \;\; X_{s,0,0}, X_{0,s,0}, X_{0,0,s} \,,
\nn \\
\mbox{Incoming chiral (to center from satellite)} &: \;\; Y_{0,s,s'}, Y_{s',0,s}, Y_{s,s',0} \,,
\nn \\
\mbox{Orbiting Fermi (from satellite to another)} &: \;\; \Lambda_{s,0,0}, \Lambda_{0,s,0}, \Lambda_{0,0,s} \,,
\nn \\
\mbox{Radial Fermi (center adjoint)} &: \;\; \Psi_{s,s',s''} \,.
\label{four_types_of_fields}
\end{align}
The three subindices, with $s,s',s''=\pm$, indicate the directions of the fields in the periodic quiver with respect to the three coordinate axes. It is interesting to point out a rather special feature of this periodic quiver: the multiplicity of the lines for radial Fermis is two. More specifically, the pair of Fermis on each of these lines is $\Psi_{s,s',s''}$ and $\Psi_{-s,-s',-s''}$. We will elaborate on this fact below. These fields transform in the adjoint representation of node 4.  

The $J$- and $E$-terms for the orbiting Fermi fields are
\begin{align}
\begin{array}{rcrc}
& J & & E 
\\
\Lambda_{+00} : \ \ \  &  
Y_{-0+} \cdot X_{00-} - Y_{-0-} \cdot X_{00+}   & \ \ \ \ & 
Y_{++0} \cdot X_{0-0} - Y_{+-0} \cdot X_{0+0}   
\\ 
\Lambda_{-00} : \ \ \  &  
Y_{+0+} \cdot X_{00-} - Y_{+0-} \cdot X_{00+}   & \ \ \ \ & 
Y_{--0} \cdot X_{0+0} - Y_{-+0} \cdot X_{0-0}   
\\
\Lambda_{0+0} : \ \ \  &  
Y_{+-0} \cdot X_{-00} - Y_{--0} \cdot X_{+00}   & \ \ \ \ & 
Y_{0++} \cdot X_{00-} - Y_{0+-} \cdot X_{00+}   
\\ 
\Lambda_{0-0} : \ \ \  &  
Y_{++0} \cdot X_{-00} - Y_{-+0} \cdot X_{+00}   & \ \ \ \ & 
Y_{0--} \cdot X_{00+} - Y_{0-+} \cdot X_{00-}   
\\
\Lambda_{00+} : \ \ \  &  
Y_{0+-} \cdot X_{0-0} - Y_{0--} \cdot X_{0+0}   & \ \ \ \ & 
Y_{+0+} \cdot X_{-00} - Y_{-0+} \cdot X_{+00}   
\\ 
\Lambda_{00-} : \ \ \  &  
Y_{0++} \cdot X_{0-0} - Y_{0-+} \cdot X_{0+0}   & \ \ \ \ & 
Y_{-0-} \cdot X_{+00} - Y_{+0-} \cdot X_{-00}   
\\
\end{array}
\label{q111s-je1}
\end{align}
The $J$- and $E$-terms for the radial Fermi fields are 
\begin{align}
\begin{array}{rcrc}
& J & & E  
\\
\Psi_{+--} : \ \ \  &  
X_{00+} \cdot Y_{-+0} - X_{-00} \cdot Y_{0++}   & \ \ \ \ & 
X_{+00} \cdot Y_{0--} - X_{0-0} \cdot Y_{+0-}   
\\ 
\Psi_{-++} : \ \ \  &  
X_{+00} \cdot Y_{0--} - X_{00-} \cdot Y_{+-0}   & \ \ \ \ & 
X_{-00} \cdot Y_{0++} - X_{0+0} \cdot Y_{-0+}   
\\
\Psi_{-+-} : \ \ \  &  
X_{+00} \cdot Y_{0-+} - X_{0-0} \cdot Y_{+0+}   & \ \ \ \ & 
X_{0+0} \cdot Y_{-0-} - X_{00-} \cdot Y_{-+0}   
\\ 
\Psi_{+-+} : \ \ \  &  
X_{0+0} \cdot Y_{-0-} - X_{-00} \cdot Y_{0+-}   & \ \ \ \ & 
X_{0-0} \cdot Y_{+0+} - X_{00+} \cdot Y_{+-0}   
\\
\Psi_{--+} : \ \ \  &  
X_{0+0} \cdot Y_{+0-} - X_{00-} \cdot Y_{++0}   & \ \ \ \ & 
X_{00+} \cdot Y_{--0} - X_{-00} \cdot Y_{0-+}   
\\ 
\Psi_{++-} : \ \ \  &  
X_{00+} \cdot Y_{--0} - X_{0-0} \cdot Y_{-0+}   & \ \ \ \ & 
X_{00-} \cdot Y_{++0} - X_{+00} \cdot Y_{0+-}   
\\
\Psi_{+++} : \ \ \  &  
X_{00-} \cdot Y_{--0} - X_{-00} \cdot Y_{0--}   & \ \ \ \ & 
X_{00+} \cdot Y_{++0} - X_{0+0} \cdot Y_{+0+}   
\\ 
\Psi_{---} : \ \ \  &  
X_{+00} \cdot Y_{0++} - X_{00+} \cdot Y_{++0}   & \ \ \ \ & 
X_{00-} \cdot Y_{--0} - X_{0-0} \cdot Y_{-0-}   
\end{array}
\label{q111s-je2}
\end{align}

The four types of fields in \eref{four_types_of_fields} independently form representations of 
the octahedral symmetry of $Q^{1,1,1}$. In particular, the different $J$- and $E$-terms transform into each other under cyclic permutations of the three indices of the corresponding Fermi fields, with the exception of those for $\Psi_{+++}$ and $\Psi_{---}$.

Vanishing of the $J$- and $E$-terms for $\Psi_{+++}$ and $\Psi_{---}$ requires that
\beq
\begin{array}{ccccc}
X_{+00} \cdot Y_{0++} & = & X_{0+0} \cdot Y_{+0+} & = & X_{00+} \cdot Y_{++0}  \\[.1cm]
X_{-00} \cdot Y_{0--} & = & X_{0-0} \cdot Y_{-0-} & = & X_{00-} \cdot Y_{--0}  
\end{array}
\label{Q111JEs}
\eeq
It means that the three paths in the chiral ring connecting the center and the corner of \fref{q111s-quiver}  on each line are equivalent. This statement is invariant under the cyclic subgroup of the octahedral symmetry. As a result, the chiral ring constructed based on these relations 
will be invariant under the full octahedral symmetry. However, at the level of the periodic quiver, realizing these relations in terms of toric $J$- and $E$-terms, i.e. in terms of plaquettes, inevitably leads to a spontaneous breaking of the symmetry. Below we will address this problem in terms of brane brick models.  It is important to emphasize that $\Psi_{+++}$ and $\Psi_{---}$ are by no means special among the radial Fermi fields. They are singled out in our analysis simply due to our choice of notation and the realization of part of the symmetry as cyclic permutation of indices.

The brick matchings of phase S are summarized by the matrices
\begin{align}
\scriptsize
P= \left(
\begin{array}{c|cccccc}
\; & p_1 & p_2 & p_3 & p_4 & p_5 & p_6 
\\
\hline
X_{+00} & 1 & 0 & 0 & 0 & 0 & 0
\\
X_{-00} & 0 & 1 & 0 & 0 & 0 & 0
\\
X_{0+0} & 0 & 0 & 1 & 0 & 0 & 0
\\
X_{0-0} & 0 & 0 & 0 & 1 & 0 & 0
\\
X_{00+} & 0 & 0 & 0 & 0 & 1 & 0
\\
X_{00-} & 0 & 0 & 0 & 0 & 0 & 1
\\
\hline
Y_{0++} & 0 & 0 & 1 & 0 & 1 & 0
\\
Y_{0--} & 0 & 0 & 0 & 1 & 0 & 1
\\
Y_{0+-} & 0 & 0 & 1 & 0 & 0 & 1
\\
Y_{0-+} & 0 & 0 & 0 & 1 & 1 & 0
\\
Y_{+0+} & 1 & 0 & 0 & 0 & 1 & 0
\\
Y_{-0-} & 0 & 1 & 0 & 0 & 0 & 1
\\
Y_{-0+} & 0 & 1 & 0 & 0 & 1 & 0
\\
Y_{+0-} & 1 & 0 & 0 & 0 & 0 & 1
\\
Y_{++0} & 1 & 0 & 1 & 0 & 0 & 0
\\
Y_{--0} & 0 & 1 & 0 & 1 & 0 & 0
\\
Y_{+-0} & 1 & 0 & 0 & 1 & 0 & 0
\\
Y_{-+0} & 0 & 1 & 1 & 0 & 0 & 0
\end{array}
\right) ~,~
\qquad 
P_\Lambda= \left(
\begin{array}{c|cccccc}
\; & p_1 & p_2 & p_3 & p_4 & p_5 & p_6 
\\
\hline
\Lambda_{+00} & 1 & 0 & 1 & 1 & 0 & 0
\\
\Lambda_{-00} & 0 & 1 & 1 & 1 & 0 & 0
\\
\Lambda_{0+0} & 0 & 0 & 1 & 0 & 1 & 1
\\
\Lambda_{0-0} & 0 & 0 & 0 & 1 & 1 & 1
\\
\Lambda_{00+} & 1 & 1 & 0 & 0 & 1 & 0
\\
\Lambda_{00-} & 1 & 1 & 0 & 0 & 0 & 1
\\
\hline 
\Psi_{+--} & 1 & 0 & 0 & 1 & 0 & 1
\\
\Psi_{-++} & 0 & 1 & 1 & 0 & 1 & 0
\\
\Psi_{-+-} & 0 & 1 & 1 & 0 & 0 & 1
\\
\Psi_{+-+} & 1 & 0 & 0 & 1 & 1 & 0
\\
\Psi_{--+} & 0 & 1 & 0 & 1 & 1 & 0
\\
\Psi_{++-} & 1 & 0 & 1 & 0 & 0 & 1
\\
\Psi_{+++} & 1 & 0 & 1 & 0 & 1 & 0
\\
\Psi_{---} & 0 & 1 & 0 & 1 & 0 & 1
\end{array}
\right) ~.~
\label{q111s-pxl}
\end{align}

\begin{figure}[H]
\begin{center}
\resizebox{0.5\hsize}{!}{
\includegraphics[trim=0cm 0cm 0cm 0cm,totalheight=10 cm]{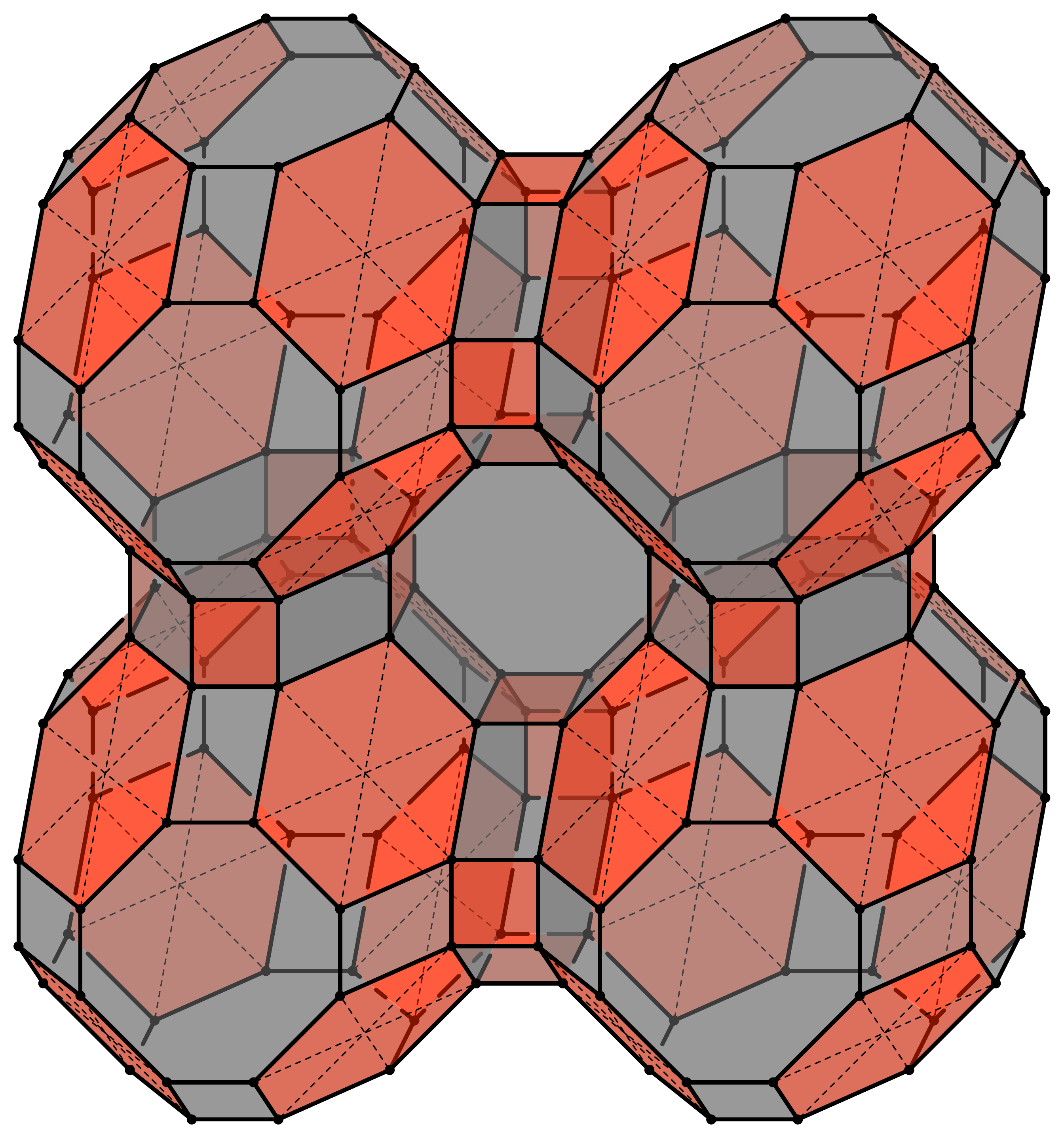}
}  
\caption{
Brane brick model for phase S of $Q^{1,1,1}$.
\label{fq111modelSbranebrick}}
 \end{center}
 \end{figure}

\begin{figure}[H]
\begin{center}
\resizebox{0.9\hsize}{!}{
\includegraphics[trim=0cm 0cm 0cm 0cm,totalheight=10 cm]{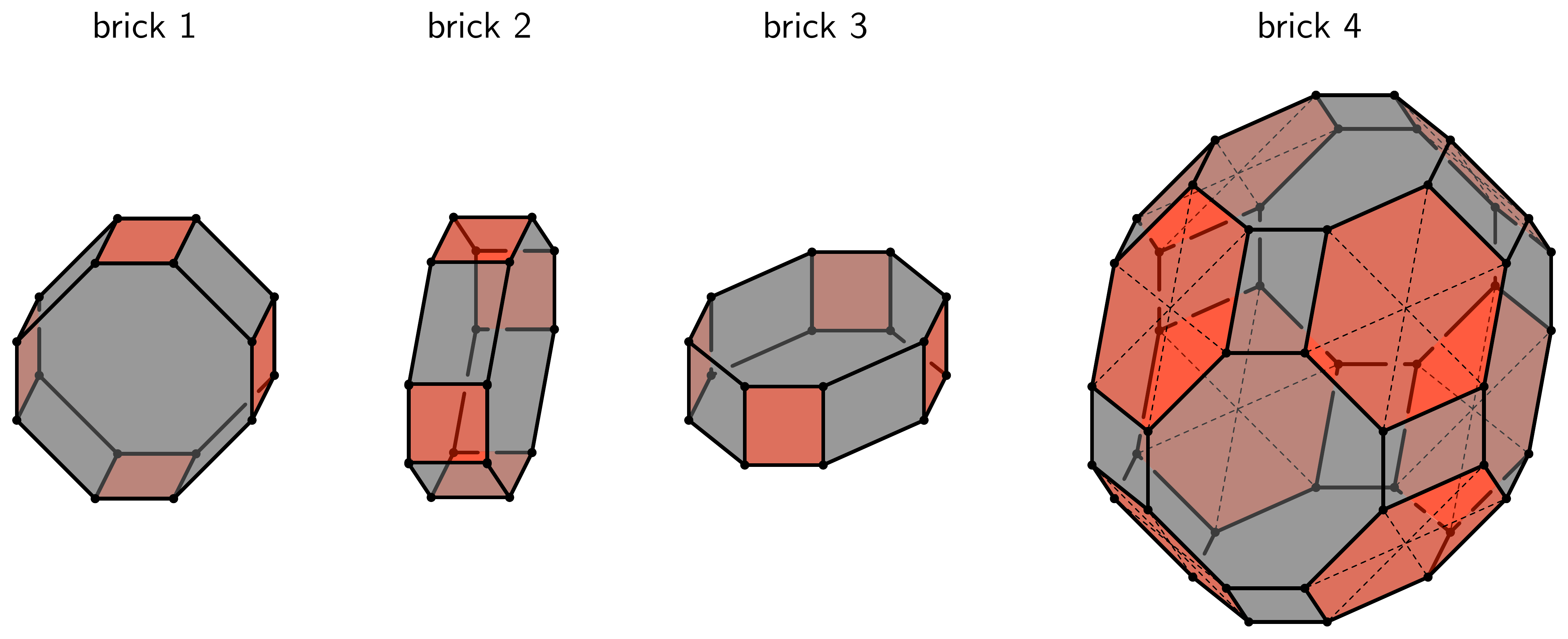}
}  
\caption{
Brane bricks for phase S of $Q^{1,1,1}$.
\label{fq111s-brick}}
 \end{center}
 \end{figure}

The brane brick model for phase $S$ is shown in  \fref{fq111modelSbranebrick}. \fref{fq111s-brick} lists the individual bricks.

Let us revisit the spontaneous breaking of the octahedral symmetry from the perspective of the brane brick model. The source of symmetry breaking can be traced to the hexagonal faces associated to the radial Fermi fields.  Due to the toric condition on $J$- and $E$-terms, Fermi faces must always be quadrilaterals. Every hexagonal face should be understood as two quadrilaterals, representing a $\Psi_{s,s',s''}$ and $\Psi_{-s,-s',-s''}$ pair, joined by a common edge. There are three different ways to draw a diagonal line splitting each of the hexagons into two trapezoids. This fact nicely matches the three ways in which we can pick toric $J$- and $E$-terms for every $\Psi_{s,s',s''}$ and $\Psi_{-s,-s',-s''}$ pair such that they give rise to the same relations when they vanish.

The chiral and Fermi field parts of the phase boundary matrix are
\begin{align}
\scriptsize
H_X = \left(
\begin{array}{c|cccc|cccc|cccc}
\; & \eta_{13} & \eta_{14} & \eta_{23} & \eta_{24} & \eta_{15} & \eta_{16} & \eta_{25} & \eta_{26} & \eta_{35} & \eta_{36} & \eta_{45} & \eta_{46} 
\\
\hline
X_{+00} & 1 & 1 & 0 & 0 & 1 & 1 & 0 & 0 & 0 & 0 & 0 & 0
\\
X_{-00} & 0 & 0 & 1 & 1 & 0 & 0 & 1 & 1 & 0 & 0 & 0 & 0
\\
X_{0+0} & -1 & 0 & -1 & 0 & 0 & 0 & 0 & 0 & 1 & 1 & 0 & 0
\\
X_{0-0} & 0 & -1 & 0 & -1 & 0 & 0 & 0 & 0 & 0 & 0 & 1 & 1
\\
X_{00+} & 0 & 0 & 0 & 0 & -1 & 0 & -1 & 0 & -1 & 0 & -1 & 0
\\
X_{00-} & 0 & 0 & 0 & 0 & 0 & -1 & 0 & -1 & 0 & -1 & 0 & -1
\\
\hline
Y_{0++} & -1 & 0 & -1 & 0 & -1 & 0 & -1 & 0 & 0 & 1 & -1 & 0
\\
Y_{0--} & 0 & -1 & 0 & -1 & 0 & -1 & 0 & -1 & 0 & -1 & 1 & 0
\\
Y_{0+-} & -1 & 0 & -1 & 0 & 0 & -1 & 0 & -1 & 1 & 0 & 0 & -1
\\
Y_{0-+} & 0 & -1 & 0 & -1 & -1 & 0 & -1 & 0 & -1 & 0 & 0 & 1
\\
Y_{+0+} & 1 & 1 & 0 & 0 & 0 & 1 & -1 & 0 & -1 & 0 & -1 & 0
\\
Y_{-0-} & 0 & 0 & 1 & 1 & 0 & -1 & 1 & 0 & 0 & -1 & 0 & -1
\\
Y_{-0+} & 0 & 0 & 1 & 1 & -1 & 0 & 0 & 1 & -1 & 0 & -1 & 0
\\
Y_{+0-} & 1 & 1 & 0 & 0 & 1 & 0 & 0 & -1 & 0 & -1 & 0 & -1
\\
Y_{++0} & 0 & 1 & -1 & 0 & 1 & 1 & 0 & 0 & 1 & 1 & 0 & 0
\\
Y_{--0} & 0 & -1 & 1 & 0 & 0 & 0 & 1 & 1 & 0 & 0 & 1 & 1
\\
Y_{+-0} & 1 & 0 & 0 & -1 & 1 & 1 & 0 & 0 & 0 & 0 & 1 & 1
\\
Y_{-+0} & -1 & 0 & 0 & 1 & 0 & 0 & 1 & 1 & 1 & 1 & 0 & 0
\end{array}
\right) ~,~
\end{align}

\begin{align}
\scriptsize
H_\Lambda = \left(
\begin{array}{c|cccc|cccc|cccc}
\; & \eta_{13} & \eta_{14} & \eta_{23} & \eta_{24} & \eta_{15} & \eta_{16} & \eta_{25} & \eta_{26} & \eta_{35} & \eta_{36} & \eta_{45} & \eta_{46} 
\\
\hline 
\Lambda_{+00} & 0 & 0 & -1 & -1 & 1 & 1 & 0 & 0 & 1 & 1 & 1 & 1
\\
\Lambda_{-00} & -1 & -1 & 0 & 0 & 0 & 0 & 1 & 1 & 1 & 1 & 1 & 1
\\
\Lambda_{0+0} & -1 & 0 & -1 & 0 & -1 & -1 & -1 & -1 & 0 & 0 & -1 & -1
\\
\Lambda_{0-0} & 0 & -1 & 0 & -1 & -1 & -1 & -1 & -1 & -1 & -1 & 0 & 0
\\
\Lambda_{00+} & 1 & 1 & 1 & 1 & 0 & 1 & 0 & 1 & -1 & 0 & -1 & 0
\\
\Lambda_{00-} & 1 & 1 & 1 & 1 & 1 & 0 & 1 & 0 & 0 & -1 & 0 & -1
\\
\hline 
\Psi_{+--} & 1 & 0 & 0 & -1 & 1 & 0 & 0 & -1 & 0 & -1 & 1 & 0
\\
\Psi_{-++} & -1 & 0 & 0 & 1 & -1 & 0 & 0 & 1 & 0 & 1 & -1 & 0
\\
\Psi_{-+-} & -1 & 0 & 0 & 1 & 0 & -1 & 1 & 0 & 1 & 0 & 0 & -1
\\
\Psi_{+-+} & 1 & 0 & 0 & -1 & 0 & 1 & -1 & 0 & -1 & 0 & 0 & 1
\\
\Psi_{--+} & 0 & -1 & 1 & 0 & -1 & 0 & 0 & 1 & -1 & 0 & 0 & 1
\\
\Psi_{++-} & 0 & 1 & -1 & 0 & 1 & 0 & 0 & -1 & 1 & 0 & 0 & -1
\\
\Psi_{+++} & 0 & 1 & -1 & 0 & 0 & 1 & -1 & 0 & 0 & 1 & -1 & 0
\\
\Psi_{---} & 0 & -1 & 1 & 0 & 0 & -1 & 1 & 0 & 0 & -1 & 1 & 0
\end{array}
\right) ~.~
\end{align}

We choose the unit cell of the brane brick model such that the non-zero intersection between chiral field faces and the unit cell edges are 
\beq
\begin{array}{cclcccl}
\vec{n}(Y_{++0}) & = & (1,0,0) & \ \ \ \ \ \ &
\vec{n}(Y_{-+0}) & = & (0,1,0) \\[.1cm]
\vec{n}(X_{00+}) & = & (0,0,1) & \ \ \ \ \ \ &
\vec{n}(X_{00-}) & = & (1,1,-1)
\end{array}
\label{q111s-toric-data}
\eeq
Once again, combining \eref{q111s-pxl} and \eref{q111s-toric-data}, the fast forward algorithm implies that the toric diagram for the mesonic moduli space is the one for $Q^{1,1,1}$.

\newpage
\section{Phases of $Q^{1,1,1}/\mathbb{Z}_2$ \label{Q111-Z2-detail}}

\begin{figure}[H]
\begin{center}
\resizebox{0.4\hsize}{!}{
\includegraphics[trim=0cm 0cm 0cm 0cm,totalheight=10 cm]{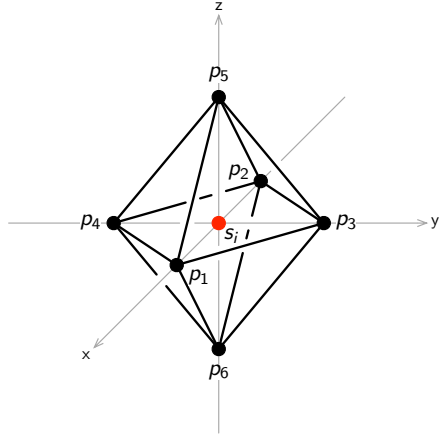}
}  
\vspace{-.3cm}\caption{
Toric diagram for $Q^{1,1,1}/\mathbb{Z}_2$.
\label{q111z2-toric}}
 \end{center}
 \end{figure}

We now discuss the A, B, C and D toric phases of $Q^{1,1,1}/\mathbb{Z}_2$, whose toric diagram is reproduced in \fref{q111z2-toric}. All these theories satisfy the vanishing trace condition.

\vspace{-.3cm} 

\vspace{-.4cm}\subsection*{Phase A}

\begin{figure}[H]
\begin{center}
\resizebox{0.6\hsize}{!}{
\includegraphics[trim=0cm 0cm 0cm 0cm,totalheight=10 cm]{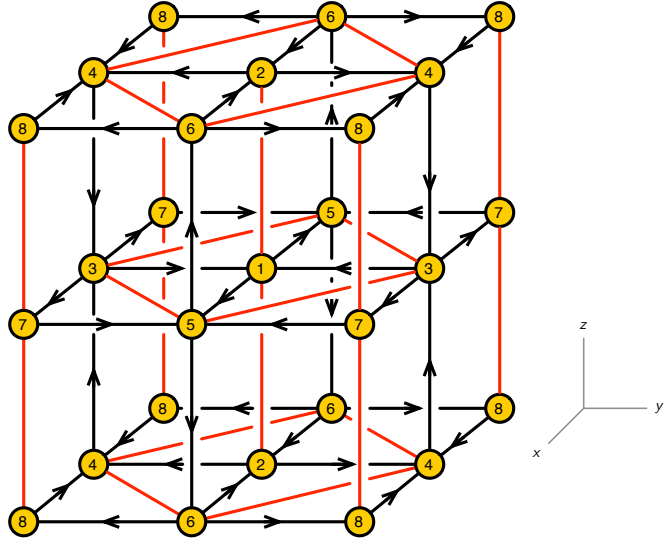}
}  
\vspace{-.3cm}\caption{
Periodic quiver for phase A of $Q^{1,1,1}/\mathbb{Z}_2$.
\label{fq111z2a-quiver}}
 \end{center}
 \end{figure}

Phase A of the $Q^{1,1,1}/\mathbb{Z}_2$ model is defined by the periodic quiver in \fref{fq111z2a-quiver}. It has 8 gauge groups, 20 chiral fields and 12 Fermi fields. It is identical to \fref{q111a-quiver-temp2} up to a doubling of the unit cell, which accounts for the $\mathbb{Z}_2$ orbifold.

The $J$- and $E$-terms are
{\small
\beq
\begin{array}{lcccc}
& &  J & &  E 
\\
   \Lambda_{21}^{+} & :\ \ \ & 
X^{+}_{15} \cdot X^{-}_{56} \cdot X^{-}_{62} - 
X^{-}_{15} \cdot X^{-}_{56} \cdot X^{+}_{62} 
& \ \ \ \ &  
X^{+}_{24} \cdot X^{+}_{43} \cdot X^{-}_{31} - 
X^{-}_{24} \cdot X^{+}_{43} \cdot X^{+}_{31} 
   \\
   \Lambda_{21}^{-} & :\ \ \ & 
X^{-}_{15} \cdot X^{+}_{56} \cdot X^{+}_{62} -
X^{+}_{15} \cdot X^{+}_{56} \cdot X^{-}_{62} 
& \ \ \ \ & 
X^{+}_{24} \cdot X^{-}_{43} \cdot X^{-}_{31}  -
X^{-}_{24} \cdot X^{-}_{43} \cdot X^{+}_{31}  
\\ 
  \Lambda_{78}^{+} & :\ \ \ &  
  X^{+}_{84} \cdot X^{-}_{43} \cdot X^{-}_{37}  -
  X^{-}_{84} \cdot X^{-}_{43} \cdot X^{+}_{37}   
  & \ \ \ \ &   
  X^{+}_{75} \cdot X^{+}_{56} \cdot X^{-}_{68} -
  X^{-}_{75} \cdot X^{+}_{56} \cdot X^{+}_{68}     
  \\
   \Lambda_{78}^{-} 
   & :\ \ \ &  
   X^{-}_{84} \cdot X^{+}_{43} \cdot X^{+}_{37} -
    X^{+}_{84} \cdot X^{+}_{43} \cdot X^{-}_{37}    
    & \ \ \ \ &  
    X^{+}_{75} \cdot X^{-}_{56} \cdot X^{-}_{68} - 
    X^{-}_{75} \cdot X^{-}_{56} \cdot X^{+}_{68}    
 \\
   \Lambda_{64}^{++} & :\ \ \ & X^{+}_{43} \cdot X^{-}_{37} \cdot X^{-}_{75} \cdot X^{-}_{56} - X^{-}_{43} \cdot X^{-}_{31} \cdot X^{-}_{15} \cdot X^{+}_{56}    & \ \ \ \ &  
X^{+}_{62} \cdot X^{+}_{24} - X^{+}_{68} \cdot X^{+}_{84} 
\\
    \Lambda_{64}^{--} & :\ \ \ & 
    X^{+}_{43} \cdot X^{+}_{31} \cdot X^{+}_{15} \cdot X^{-}_{56} -
    X^{-}_{43} \cdot X^{+}_{37} \cdot X^{+}_{75} \cdot X^{+}_{56} 
    & \ \ \ \ &  
X^{-}_{62} \cdot X^{-}_{24} - X^{-}_{68} \cdot X^{-}_{84}    
\\
    \Lambda_{64}^{+-} & :\ \ \ & 
    X^{-}_{43} \cdot X^{+}_{31} \cdot X^{-}_{15} \cdot X^{+}_{56} -
    X^{+}_{43} \cdot X^{-}_{37} \cdot X^{+}_{75} \cdot X^{-}_{56} 
    & \ \ \ \ &  
X^{+}_{62} \cdot X^{-}_{24} - X^{-}_{68} \cdot X^{+}_{84}    
\\
   \Lambda_{64}^{-+} & :\ \ \ & X^{-}_{43} \cdot X^{+}_{37} \cdot X^{-}_{75} \cdot X^{+}_{56} - X^{+}_{43} \cdot X^{-}_{31} \cdot X^{+}_{15} \cdot X^{-}_{56}    & \ \ \ \ &  
X^{-}_{62} \cdot X^{+}_{24} - X^{+}_{68} \cdot X^{-}_{84} 
\\ 
   \Lambda_{35}^{++} & :\ \ \ & X^{+}_{56} \cdot X^{-}_{62} \cdot X^{-}_{24} \cdot X^{-}_{43} - X^{-}_{56} \cdot X^{-}_{68} \cdot X^{-}_{84} \cdot X^{+}_{43}    & \ \ \ \ &  
X^{+}_{37} \cdot X^{+}_{75} - X^{+}_{31} \cdot X^{+}_{15}    
\\
    \Lambda_{35}^{--} & :\ \ \ & X^{+}_{56} \cdot X^{+}_{68} \cdot X^{+}_{84} \cdot X^{-}_{43}    
 - 
 X^{-}_{56} \cdot X^{+}_{62} \cdot X^{+}_{24} \cdot X^{+}_{43}     
 & \ \ \ \ &  
X^{-}_{37} \cdot X^{-}_{75} - X^{-}_{31} \cdot X^{-}_{15}   
\\ 
   \Lambda_{35}^{+-} & :\ \ \ & 
   X^{-}_{56} \cdot X^{+}_{68} \cdot X^{-}_{84} \cdot X^{+}_{43} -
   X^{+}_{56} \cdot X^{-}_{62} \cdot X^{+}_{24} \cdot X^{-}_{43} 
   & \ \ \ \ &  
X^{+}_{37} \cdot X^{-}_{75} - X^{-}_{31} \cdot X^{+}_{15}   
\\
    \Lambda_{35}^{-+} & :\ \ \ & X^{-}_{56} \cdot X^{+}_{62} \cdot X^{-}_{24} \cdot X^{+}_{43} - X^{+}_{56} \cdot X^{-}_{68} \cdot X^{+}_{84} \cdot X^{-}_{43}    & \ \ \ \ &  
X^{-}_{37} \cdot X^{+}_{75} - X^{+}_{31} \cdot X^{-}_{15}  
\\ 
\end{array}
\label{es1a1}
\eeq
}

The brick matchings are given by 
\beal{q111z2a-px}
\tiny
P =
\left(
\begin{array}{c|cccccc|cccccccccc}
\; & p_1 & p_2 & p_3 & p_4  & p_5 & p_6 & s_1 & s_2 & s_3 & s_4 & s_5 & s_6 & s_7 & s_8 & s_9 & s_{10} \\
\hline 
X_{37}^{+} &  1 & 0 & 0 & 0 & 0 & 0 & 0 & 0 & 1 & 0 & 0 & 0 & 1 & 0 & 0
   & 0 \\
X_{37}^{-} &  0 & 1 & 0 & 0 & 0 & 0 & 0 & 0 & 1 & 0 & 0 & 0 & 1 & 0 & 0
   & 0 \\
X_{62}^{+} & 1 & 0 & 0 & 0 & 0 & 0 & 0 & 0 & 0 & 0 & 0 & 0 & 0 & 0 & 1
   & 1 \\
X_{62}^{-} & 0 & 1 & 0 & 0 & 0 & 0 & 0 & 0 & 0 & 0 & 0 & 0 & 0 & 0 & 1
   & 1 \\
X_{84}^{+} & 1 & 0 & 0 & 0 & 0 & 0 & 0 & 0 & 0 & 0 & 0 & 1 & 0 & 0 & 0
   & 1 \\
X_{84}^{-} & 0 & 1 & 0 & 0 & 0 & 0 & 0 & 0 & 0 & 0 & 0 & 1 & 0 & 0 & 0
   & 1 \\
X_{24}^{+} & 0 & 0 & 1 & 0 & 0 & 0 & 0 & 0 & 0 & 0 & 1 & 1 & 0 & 0 & 0
   & 0 \\
X_{24}^{-} & 0 & 0 & 0 & 1 & 0 & 0 & 0 & 0 & 0 & 0 & 1 & 1 & 0 & 0 & 0
   & 0 \\
X_{68}^{+} & 0 & 0 & 1 & 0 & 0 & 0 & 0 & 0 & 0 & 0 & 1 & 0 & 0 & 0 & 1
   & 0 \\
X_{68}^{-} & 0 & 0 & 0 & 1 & 0 & 0 & 0 & 0 & 0 & 0 & 1 & 0 & 0 & 0 & 1
   & 0 \\
X_{75}^{+} & 0 & 0 & 1 & 0 & 0 & 0 & 0 & 0 & 0 & 1 & 0 & 0 & 0 & 1 & 0
   & 0 \\
X_{75}^{-} & 0 & 0 & 0 & 1 & 0 & 0 & 0 & 0 & 0 & 1 & 0 & 0 & 0 & 1 & 0
   & 0 \\
X_{43}^{+} & 0 & 0 & 0 & 0 & 1 & 0 & 1 & 0 & 0 & 0 & 0 & 0 & 0 & 0 & 0
   & 0 \\
X_{43}^{-} & 0 & 0 & 0 & 0 & 0 & 1 & 1 & 0 & 0 & 0 & 0 & 0 & 0 & 0 & 0
   & 0 \\
X_{56}^{+} & 0 & 0 & 0 & 0 & 1 & 0 & 0 & 1 & 0 & 0 & 0 & 0 & 0 & 0 & 0
   & 0 \\
X_{56}^{-} & 0 & 0 & 0 & 0 & 0 & 1 & 0 & 1 & 0 & 0 & 0 & 0 & 0 & 0 & 0
   & 0 \\
X_{31}^{+} & 0 & 0 & 1 & 0 & 0 & 0 & 0 & 0 & 0 & 0 & 0 & 0 & 1 & 1 & 0
   & 0 \\
X_{31}^{-} & 0 & 0 & 0 & 1 & 0 & 0 & 0 & 0 & 0 & 0 & 0 & 0 & 1 & 1 & 0
   & 0 \\   
X_{15}^{+} & 1 & 0 & 0 & 0 & 0 & 0 & 0 & 0 & 1 & 1 & 0 & 0 & 0 & 0 & 0
   & 0 \\
X_{15}^{-} &  0 & 1 & 0 & 0 & 0 & 0 & 0 & 0 & 1 & 1 & 0 & 0 & 0 & 0 & 0
   & 0 \\   
\end{array}
\right)~,~ \nonumber
\eea
\beal{q111z2a-pl}
\tiny
P_\Lambda = 
\left(
\begin{array}{c|cccccc|cccccccccc}
\; & p_1 & p_2 & p_3 & p_4  & p_5 & p_6 & s_1 & s_2 & s_3 & s_4 & s_5 & s_6 & s_7 & s_8 & s_9 & s_{10} \\
\hline
\Lambda_{21}^{+} & 0 & 0 & 1 & 1 & 1 & 0 & 1 & 0 & 0 & 0 & 1 & 1 & 1 & 1 & 0 & 0 \\
\Lambda_{21}^{-} & 0 & 0 & 1 & 1 & 0 & 1 & 1 & 0 & 0 & 0 & 1 & 1 & 1 & 1 & 0 & 0 \\
\Lambda_{78}^{+} & 0 & 0 & 1 & 1 & 1 & 0 & 0 & 1 & 0 & 1 & 1 & 0 & 0 & 1 & 1 & 0 \\
\Lambda_{78}^{-} & 0 & 0 & 1 & 1 & 0 & 1 & 0 & 1 & 0 & 1 & 1 & 0 & 0 & 1 & 1 & 0 \\
\Lambda_{64}^{++} & 1 & 0 & 1 & 0 & 0 & 0 & 0 & 0 & 0 & 0 & 1 & 1 & 0 & 0 & 1 & 1 \\
\Lambda_{64}^{+-} & 1 & 0 & 0 & 1 & 0 & 0 & 0 & 0 & 0 & 0 & 1 & 1 & 0 & 0 & 1 & 1 \\
\Lambda_{64}^{-+} & 0 & 1 & 1 & 0 & 0 & 0 & 0 & 0 & 0 & 0 & 1 & 1 & 0 & 0 & 1 & 1 \\
\Lambda_{64}^{--} & 0 & 1 & 0 & 1 & 0 & 0 & 0 & 0 & 0 & 0 & 1 & 1 & 0 & 0 & 1 & 1 \\
\Lambda_{35}^{++} & 1 & 0 & 1 & 0 & 0 & 0 & 0 & 0 & 1 & 1 & 0 & 0 & 1 & 1 & 0 & 0 \\
\Lambda_{35}^{+-} & 1 & 0 & 0 & 1 & 0 & 0 & 0 & 0 & 1 & 1 & 0 & 0 & 1 & 1 & 0 & 0 \\
\Lambda_{35}^{-+} & 0 & 1 & 1 & 0 & 0 & 0 & 0 & 0 & 1 & 1 & 0 & 0 & 1 & 1 & 0 & 0 \\
\Lambda_{35}^{--} & 0 & 1 & 0 & 1 & 0 & 0 & 0 & 0 & 1 & 1 & 0 & 0 & 1 & 1 & 0 & 0 \\
\end{array}
\right)~.~
\eea

The brane brick model for phase A is identical to the one in \fref{fq111modelAbranebrick} for $Q^{1,1,1}$, with the appropriate doubling of the size of the unit cell corresponding to the periodic quiver in \fref{fq111z2a-quiver}.

The chiral and Fermi field pieces of the phase boundary matrix are
\beal{q111z2a-hx}
\scriptsize
H_X
=
\left(
\begin{array}{c|cccccccccccc}
\; &
\eta_{13} & \eta_{14} & \eta_{23} & \eta_{24} & \eta_{15} & \eta_{16} & \eta_{25} & \eta_{26} & \eta_{35} & \eta_{36} & \eta_{45} & \eta_{46} 
\\
\hline
X_{37}^{+} & 1 & 1 & 0 & 0 & 1 & 1 & 0 & 0 & 0 & 0 & 0 & 0 \\
X_{37}^{-} & 0 & 0 & 1 & 1 & 0 & 0 & 1 & 1 & 0 & 0 & 0 & 0 \\
X_{62}^{+} & 1 & 1 & 0 & 0 & 1 & 1 & 0 & 0 & 0 & 0 & 0 & 0 \\
X_{62}^{-} & 0 & 0 & 1 & 1 & 0 & 0 & 1 & 1 & 0 & 0 & 0 & 0 \\
X_{84}^{+} & 1 & 1 & 0 & 0 & 1 & 1 & 0 & 0 & 0 & 0 & 0 & 0 \\
X_{84}^{-} & 0 & 0 & 1 & 1 & 0 & 0 & 1 & 1 & 0 & 0 & 0 & 0 \\
X_{24}^{+} & -1 & 0 & -1 & 0 & 0 & 0 & 0 & 0 & 1 & 1 & 0 & 0 \\
X_{24}^{-} & 0 & -1 & 0 & -1 & 0 & 0 & 0 & 0 & 0 & 0 & 1 & 1 \\
X_{68}^{+} & -1 & 0 & -1 & 0 & 0 & 0 & 0 & 0 & 1 & 1 & 0 & 0 \\
X_{68}^{-} & 0 & -1 & 0 & -1 & 0 & 0 & 0 & 0 & 0 & 0 & 1 & 1 \\
X_{75}^{+} & -1 & 0 & -1 & 0 & 0 & 0 & 0 & 0 & 1 & 1 & 0 & 0 \\
X_{75}^{-} & 0 & -1 & 0 & -1 & 0 & 0 & 0 & 0 & 0 & 0 & 1 & 1 \\
X_{43}^{+} & 0 & 0 & 0 & 0 & -1 & 0 & -1 & 0 & -1 & 0 & -1 & 0 \\
X_{43}^{-} & 0 & 0 & 0 & 0 & 0 & -1 & 0 & -1 & 0 & -1 & 0 & -1 \\
X_{56}^{+} & 0 & 0 & 0 & 0 & -1 & 0 & -1 & 0 & -1 & 0 & -1 & 0 \\
X_{56}^{-} & 0 & 0 & 0 & 0 & 0 & -1 & 0 & -1 & 0 & -1 & 0 & -1 \\
X_{15}^{+} & 1 & 1 & 0 & 0 & 1 & 1 & 0 & 0 & 0 & 0 & 0 & 0 \\
X_{15}^{-} & 0 & 0 & 1 & 1 & 0 & 0 & 1 & 1 & 0 & 0 & 0 & 0 \\
X_{31}^{+} & -1 & 0 & -1 & 0 & 0 & 0 & 0 & 0 & 1 & 1 & 0 & 0 \\
X_{31}^{-} & 0 & -1 & 0 & -1 & 0 & 0 & 0 & 0 & 0 & 0 & 1 & 1 \\
\end{array}
\right) ~,~
\eea
\beal{q111z2a-hl}
\scriptsize
H_\Lambda = 
\left(
\begin{array}{c|cccccccccccc}
\; &
\eta_{13} & \eta_{14} & \eta_{23} & \eta_{24} & \eta_{15} & \eta_{16} & \eta_{25} & \eta_{26} & \eta_{35} & \eta_{36} & \eta_{45} & \eta_{46} 
\\
\hline
\Lambda_{21}^{+} & -1 & -1 & -1 & -1 & -1 & 0 & -1 & 0 & 0 & 1 & 0 & 1 \\
\Lambda_{21}^{-} & -1 & -1 & -1 & -1 & 0 & -1 & 0 & -1 & 1 & 0 & 1 & 0 \\
\Lambda_{78}^{+} & -1 & -1 & -1 & -1 & -1 & 0 & -1 & 0 & 0 & 1 & 0 & 1 \\
\Lambda_{78}^{-} & -1 & -1 & -1 & -1 & 0 & -1 & 0 & -1 & 1 & 0 & 1 & 0 \\
\Lambda_{64}^{++} & 0 & 1 & -1 & 0 & 1 & 1 & 0 & 0 & 1 & 1 & 0 & 0 \\
\Lambda_{64}^{+-} & 1 & 0 & 0 & -1 & 1 & 1 & 0 & 0 & 0 & 0 & 1 & 1 \\
\Lambda_{64}^{-+} & -1 & 0 & 0 & 1 & 0 & 0 & 1 & 1 & 1 & 1 & 0 & 0 \\
\Lambda_{64}^{--} & 0 & -1 & 1 & 0 & 0 & 0 & 1 & 1 & 0 & 0 & 1 & 1 \\
\Lambda_{35}^{++} & 0 & 1 & -1 & 0 & 1 & 1 & 0 & 0 & 1 & 1 & 0 & 0 \\
\Lambda_{35}^{+-} & 1 & 0 & 0 & -1 & 1 & 1 & 0 & 0 & 0 & 0 & 1 & 1 \\
\Lambda_{35}^{-+} & -1 & 0 & 0 & 1 & 0 & 0 & 1 & 1 & 1 & 1 & 0 & 0 \\
\Lambda_{35}^{--} & 0 & -1 & 1 & 0 & 0 & 0 & 1 & 1 & 0 & 0 & 1 & 1 \\
\end{array}
\right)~.~
\label{Q111Z2-Ha}
\eea

We choose the unit cell to have the origin at node 4 and its three axes along the $x$, $y$ and $z$ directions of \fref{fq111z2a-quiver}. The non-zero intersection numbers of chiral field faces with the edges of the unit cell are
\beal{q111z2a-unit}
\vec{n}(X^{\pm}_{84}) = (\pm 1,0,0) ~,~
\vec{n}(X^{\pm}_{24}) = (0,\pm 1,0) ~,~
\vec{n}(X^{\pm}_{43}) = (0,0,\pm 1) ~.
&
\eea
Combining this with \eref{q111z2a-pl}, the fast forward algorithm produces the toric diagram for $Q^{1,1,1}/\mathbb{Z}_2$ shown in \fref{q111z2-toric}.
Each chiral field in \eqref{q111z2a-unit} contributes to a single external brick matching. For internal brick matchings, the contributions 
from $X_{ij}^+$ and $X_{ij}^-$ cancel out. 

A nice feature or our choice of unit cell is that the chiral fields with non-zero intersection numbers remain intact under the triality transformations we consider. As a result, following the general discussion in section \sref{section_general_invariance_moduli_space}, the moduli space of all other phases (B, C, D) is the same. The number of brick matchings associated to the internal point in the toric diagram may however vary from phase to phase.

\subsection*{Phase B}

\begin{figure}[H]
\begin{center}
\resizebox{0.6\hsize}{!}{
\includegraphics[trim=0cm 0cm 0cm 0cm,totalheight=10 cm]{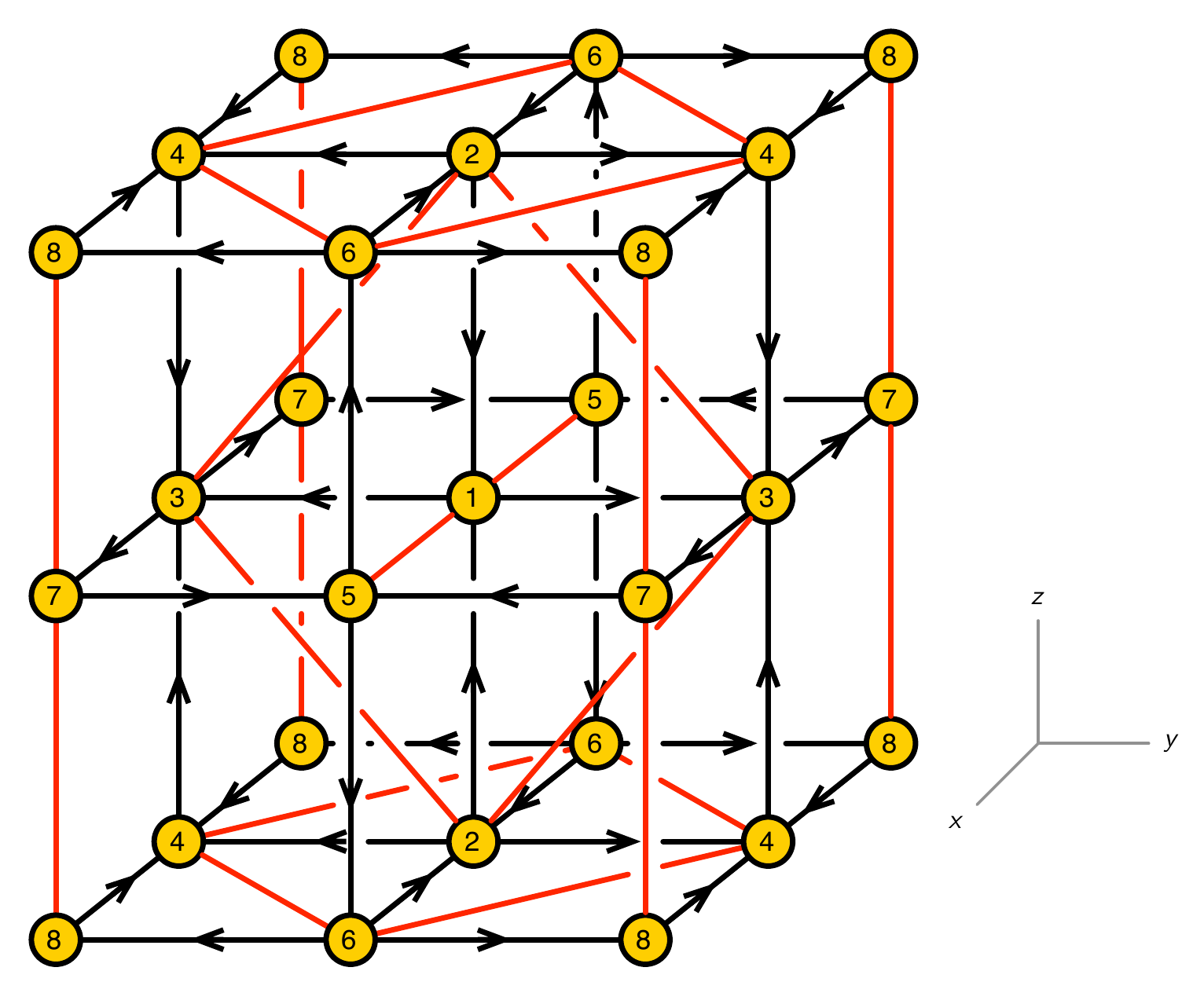}
}  
\caption{
Periodic quiver for phase B of $Q^{1,1,1}/\mathbb{Z}_2$.
\label{fq111z2b-quiver}}
 \end{center}
 \end{figure}

Phase B of $Q^{1,1,1}/\mathbb{Z}_2$ is defined by the periodic quiver in \fref{fq111z2b-quiver}. It has 8 gauge groups, 20 chiral fields and 12 Fermi fields.

The $J$- and $E$-terms are
{\small
\beq
\begin{array}{lcccc}
& &  J  & &  E  
\\  
   \Lambda_{51}^{+} 
 & :\ \ \ &  
  X^{+}_{13} \cdot X^{-}_{37} \cdot X^{-}_{75} -
    X^{-}_{13} \cdot X^{-}_{37} \cdot X^{+}_{75}    
  & \ \ \ \ &   
   X^{+}_{56} \cdot X^{+}_{62} \cdot X^{-}_{21} - 
       X^{-}_{56} \cdot X^{+}_{62} \cdot X^{+}_{21} 
  \\
 \Lambda_{51}^{-} & :\ \ \ &  
 X^{-}_{13} \cdot X^{+}_{37} \cdot X^{+}_{75} -
 X^{+}_{13} \cdot X^{+}_{37} \cdot X^{-}_{75}    
 & \ \ \ \ &   
   X^{+}_{56} \cdot X^{-}_{62} \cdot X^{-}_{21} -
   X^{-}_{56} \cdot X^{-}_{62} \cdot X^{+}_{21}
   \\  
  \Lambda_{78}^{+} & :\ \ \ &  
  X^{+}_{84} \cdot X^{-}_{43} \cdot X^{-}_{37}  -
  X^{-}_{84} \cdot X^{-}_{43} \cdot X^{+}_{37}   
  & \ \ \ \ & 
  X^{+}_{75} \cdot X^{+}_{56} \cdot X^{-}_{68} -    
  X^{-}_{75} \cdot X^{+}_{56} \cdot X^{+}_{68}  
  \\
   \Lambda_{78}^{-} 
   & :\ \ \ &  
   X^{-}_{84} \cdot X^{+}_{43} \cdot X^{+}_{37} -
    X^{+}_{84} \cdot X^{+}_{43} \cdot X^{-}_{37}    
    & \ \ \ \ &  
X^{+}_{75} \cdot X^{-}_{56} \cdot X^{-}_{68}  - 
X^{-}_{75} \cdot X^{-}_{56} \cdot X^{+}_{68}  
   \\
   \Lambda_{64}^{++} & :\ \ \ &  
   X^{+}_{43} \cdot X^{-}_{37} \cdot X^{-}_{75} \cdot X^{-}_{56}  -
   X^{-}_{43} \cdot X^{-}_{37} \cdot X^{-}_{75} \cdot X^{+}_{56}     
   & \ \ \ \ &   
X_{62}^{+} \cdot X_{24}^{+} - X_{68}^{+} \cdot X_{84}^{+}   
\\
         \Lambda_{64}^{--} & :\ \ \ &  X^{+}_{43} \cdot X^{+}_{37} \cdot X^{+}_{75} \cdot X^{-}_{56} - X^{-}_{43} \cdot X^{+}_{37} \cdot X^{+}_{75} \cdot X^{+}_{56}   
& \ \ \ \ &   
X_{62}^{-} \cdot X_{24}^{-} - X_{68}^{-} \cdot X_{84}^{-}   
\\
   \Lambda_{64}^{+-} & :\ \ \ &  
X^{-}_{43} \cdot X^{-}_{37} \cdot X^{+}_{75} \cdot X^{+}_{56}   -
   X^{+}_{43} \cdot X^{-}_{37} \cdot X^{+}_{75} \cdot X^{-}_{56}    
& \ \ \ \ &   
X_{62}^{+} \cdot X_{24}^{-} - X_{68}^{-} \cdot X_{84}^{+}      
\\
   \Lambda_{64}^{-+} & :\ \ \ &  X^{-}_{43} \cdot X^{+}_{37} \cdot X^{-}_{75} \cdot X^{+}_{56} - X^{+}_{43} \cdot X^{+}_{37} \cdot X^{-}_{75} \cdot X^{-}_{56}   & \ \ \ \ &   
X_{62}^{-} \cdot X_{24}^{+} - X_{68}^{+} \cdot X_{84}^{-}    
\\
\nonumber
\end{array}
\label{q111z2b-je}
\eeq

{\small
\beq
\begin{array}{lcccc}
& &  J  & &  E  
\\  
\Lambda_{23}^{++} & :\ \ \ &  
X^{+}_{37} \cdot X^{-}_{75} \cdot X^{-}_{56} \cdot X^{-}_{62}   -
X^{-}_{37} \cdot X^{-}_{75} \cdot X^{-}_{56} \cdot X^{+}_{62} 
& \ \ \ \ &   X^{+}_{24} \cdot X^{+}_{43} - X^{+}_{21} \cdot X^{+}_{13}   \\
\Lambda_{23}^{--} & :\ \ \ &  X^{+}_{37} \cdot X^{+}_{75} \cdot X^{+}_{56} \cdot X^{-}_{62} - X^{-}_{37} \cdot X^{+}_{75} \cdot X^{+}_{56} \cdot X^{+}_{62}   & \ \ \ \ &   X^{-}_{24} \cdot X^{-}_{43} - X^{-}_{21} \cdot X^{-}_{13}   \\
   \Lambda_{23}^{+-} 
   & :\ \ \ &  
   X^{-}_{37} \cdot X^{-}_{75} \cdot X^{+}_{56} \cdot X^{+}_{62}   -
   X^{+}_{37} \cdot X^{-}_{75} \cdot X^{+}_{56} \cdot X^{-}_{62}    & \ \ \ \ &   
   X^{+}_{24} \cdot X^{-}_{43} - X^{-}_{21} \cdot X^{+}_{13}   
   \\
   \Lambda_{23}^{-+} & :\ \ \ &  X^{-}_{37} \cdot X^{+}_{75} \cdot X^{-}_{56} \cdot X^{+}_{62} - X^{+}_{37} \cdot X^{+}_{75} \cdot X^{-}_{56} \cdot X^{-}_{62}   & \ \ \ \ &   X^{-}_{24} \cdot X^{+}_{43} - X^{+}_{21} \cdot X^{-}_{13}   \\
\end{array}
\label{q111z2b-je}
\eeq
}
The brick matchings are given by
\beal{q111z2b-px}
\scriptsize
P=
\left(
\begin{array}{c|cccccc|ccccccccccc}
\; & p_1 & p_2 & p_3 & p_4 & p_5 & p_6 & s_1 & s_2 & s_3 & s_4 & s_5 & s_6 & s_7 & s_8 & s_9 & s_{10} & s_{11} \\
\hline
X_{37}^{+} & 1 & 0 & 0 & 0 & 0 & 0 & 1 & 0 & 0 & 0 & 0 & 0 & 0 & 0 &0 & 0 & 0 \\
X_{37}^{-} &  0 & 1 & 0 & 0 & 0 & 0 & 1 & 0 & 0 & 0 & 0 & 0 & 0 & 0 &0 & 0 & 0 \\
X_{62}^{+} &  1 & 0 & 0 & 0 & 0 & 0 & 0 & 0 & 0 & 0 & 0 & 1 & 1 & 0 &0 & 0 & 0 \\
X_{62}^{-} &  0 & 1 & 0 & 0 & 0 & 0 & 0 & 0 & 0 & 0 & 0 & 1 & 1 & 0 &0 & 0 & 0 \\
X_{84}^{+} &  1 & 0 & 0 & 0 & 0 & 0 & 0 & 0 & 0 & 0 & 0 & 0 & 1 & 0 &1 & 0 & 1 \\
X_{84}^{-} &  0 & 1 & 0 & 0 & 0 & 0 & 0 & 0 & 0 & 0 & 0 & 0 & 1 & 0 &1 & 0 & 1 \\
X_{24}^{+} &  0 & 0 & 1 & 0 & 0 & 0 & 0 & 0 & 0 & 0 & 0 & 0 & 0 & 1 &1 & 1 & 1 \\
X_{24}^{-} &  0 & 0 & 0 & 1 & 0 & 0 & 0 & 0 & 0 & 0 & 0 & 0 & 0 & 1 &1 & 1 & 1 \\
X_{68}^{+} &  0 & 0 & 1 & 0 & 0 & 0 & 0 & 0 & 0 & 0 & 0 & 1 & 0 & 1 &0 & 1 & 0 \\
X_{68}^{-} &  0 & 0 & 0 & 1 & 0 & 0 & 0 & 0 & 0 & 0 & 0 & 1 & 0 & 1 &0 & 1 & 0 \\
X_{75}^{+} &  0 & 0 & 1 & 0 & 0 & 0 & 0 & 0 & 1 & 0 & 0 & 0 & 0 & 0 &0 & 0 & 0 \\
X_{75}^{-} &  0 & 0 & 0 & 1 & 0 & 0 & 0 & 0 & 1 & 0 & 0 & 0 & 0 & 0 &0 & 0 & 0 \\
X_{43}^{+} &  0 & 0 & 0 & 0 & 1 & 0 & 0 & 0 & 0 & 1 & 1 & 0 & 0 & 0 &0 & 0 & 0 \\
X_{43}^{-} &  0 & 0 & 0 & 0 & 0 & 1 & 0 & 0 & 0 & 1 & 1 & 0 & 0 & 0 &0 & 0 & 0 \\
X_{56}^{+} &  0 & 0 & 0 & 0 & 1 & 0 & 0 & 1 & 0 & 0 & 0 & 0 & 0 & 0 &0 & 0 & 0 \\
X_{56}^{-} &  0 & 0 & 0 & 0 & 0 & 1 & 0 & 1 & 0 & 0 & 0 & 0 & 0 & 0 &0 & 0 & 0 \\
X_{21}^{+} &  0 & 0 & 0 & 0 & 1 & 0 & 0 & 0 & 0 & 0 & 1 & 0 & 0 & 0 &0 & 1 & 1 \\
X_{21}^{-} &  0 & 0 & 0 & 0 & 0 & 1 & 0 & 0 & 0 & 0 & 1 & 0 & 0 & 0 &0 & 1 & 1 \\
X_{13}^{+} &  0 & 0 & 1 & 0 & 0 & 0 & 0 & 0 & 0 & 1 & 0 & 0 & 0 & 1 &1 & 0 & 0 \\
X_{13}^{-} &  0 & 0 & 0 & 1 & 0 & 0 & 0 & 0 & 0 & 1 & 0 & 0 & 0 & 1 &1 & 0 & 0 \\
\end{array}
\right)~,~
\nonumber
\eea
\beal{q111z2b-pl}
\scriptsize
P_\Lambda =
\left(
\begin{array}{c|cccccc|ccccccccccc}
\; & p_1 & p_2 & p_3 & p_4 & p_5 & p_6 & s_1 & s_2 & s_3 & s_4 & s_5 & s_6 & s_7 & s_8 & s_9 & s_{10} & s_{11} \\
\hline
\Lambda_{51}^{+} & 1 & 0 & 0 & 0 & 1 & 1 & 0 & 1 & 0 & 0 & 1 & 1 & 1 & 0 & 0 & 1 & 1 \\
\Lambda_{51}^{-} & 0 & 1 & 0 & 0 & 1 & 1 & 0 & 1 & 0 & 0 & 1 & 1 & 1 & 0 & 0 & 1 & 1 \\
\Lambda_{78}^{+} & 0 & 0 & 1 & 1 & 1 & 0 & 0 & 1 & 1 & 0 & 0 & 1 & 0 & 1 & 0 & 1 & 0 \\
\Lambda_{78}^{-} & 0 & 0 & 1 & 1 & 0 & 1 & 0 & 1 & 1 & 0 & 0 & 1 & 0 & 1 & 0 & 1 & 0 \\
\Lambda_{64}^{++} & 1 & 0 & 1 & 0 & 0 & 0 & 0 & 0 & 0 & 0 & 0 & 1 & 1 & 1 & 1 & 1 & 1 \\
\Lambda_{64}^{+-} & 1 & 0 & 0 & 1 & 0 & 0 & 0 & 0 & 0 & 0 & 0 & 1 & 1 & 1 & 1 & 1 & 1 \\
\Lambda_{64}^{-+} & 0 & 1 & 1 & 0 & 0 & 0 & 0 & 0 & 0 & 0 & 0 & 1 & 1 & 1 & 1 & 1 & 1 \\
\Lambda_{64}^{--} & 0 & 1 & 0 & 1 & 0 & 0 & 0 & 0 & 0 & 0 & 0 & 1 & 1 & 1 & 1 & 1 & 1 \\
\Lambda_{23}^{++} & 0 & 0 & 1 & 0 & 1 & 0 & 0 & 0 & 0 & 1 & 1 & 0 & 0 & 1 & 1 & 1 & 1 \\
\Lambda_{23}^{+-} & 0 & 0 & 1 & 0 & 0 & 1 & 0 & 0 & 0 & 1 & 1 & 0 & 0 & 1 & 1 & 1 & 1 \\
\Lambda_{23}^{-+} & 0 & 0 & 0 & 1 & 1 & 0 & 0 & 0 & 0 & 1 & 1 & 0 & 0 & 1 & 1 & 1 & 1 \\
\Lambda_{23}^{--} & 0 & 0 & 0 & 1 & 0 & 1 & 0 & 0 & 0 & 1 & 1 & 0 & 0 & 1 & 1 & 1 & 1 \\
\end{array}
\right)~.~
\eea

As explained above, the fast forward algorithm leads to the same mesonic moduli space for all phases of $Q^{1,1,1}/\mathbb{Z}_2$. For this theory and the ones that follow, the column labels in the $P$-matrix indicate the corresponding point in the toric diagram.

Figures \ref{fq111z2b-brick2} and \ref{fq111z2b-brick1} show the brane brick model and individual bricks for phase B. 

\begin{figure}[H]
\begin{center}
\resizebox{0.6\hsize}{!}{
\includegraphics[trim=0cm 0cm 0cm 0cm,totalheight=10 cm]{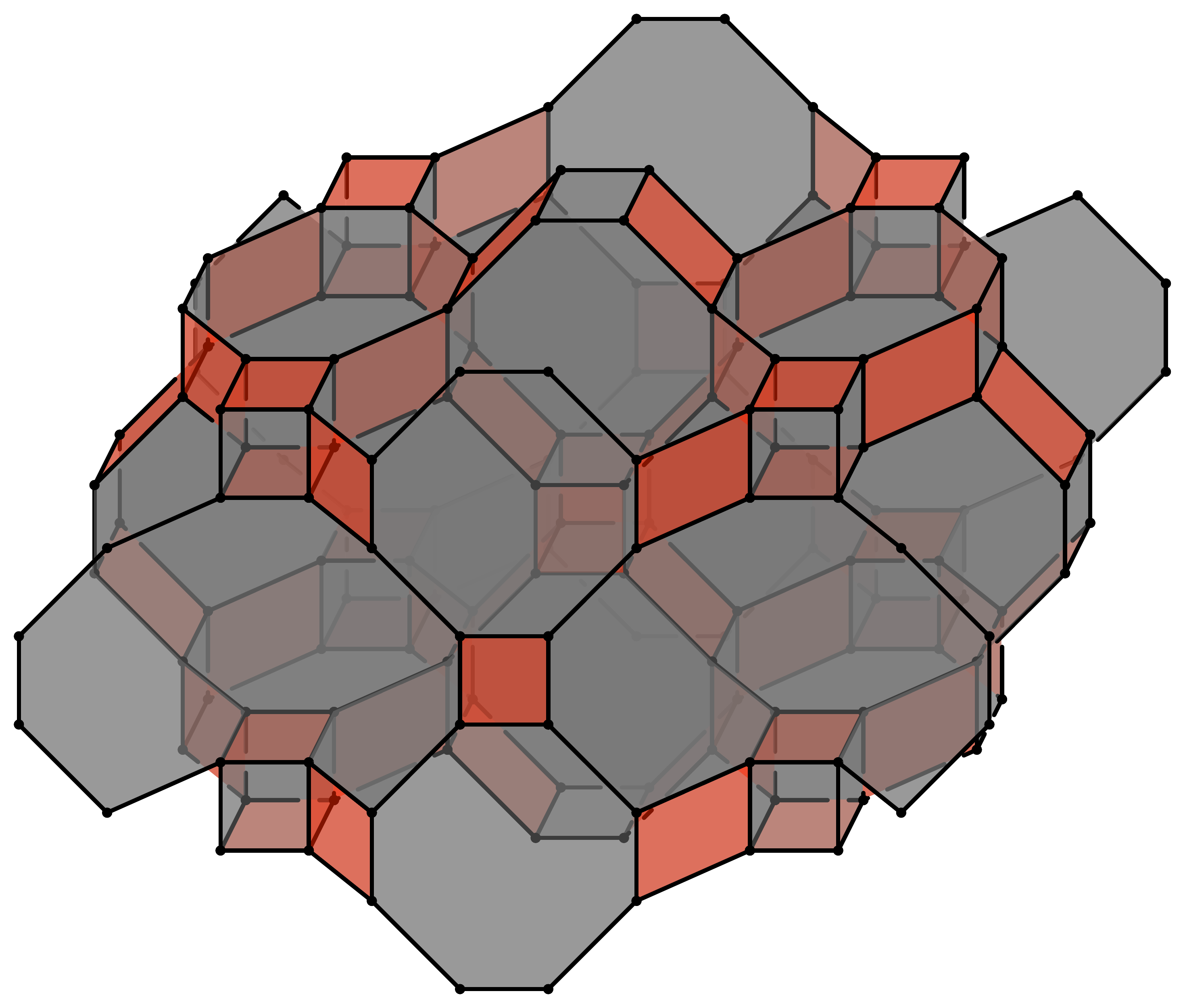}
}  
\caption{
Brane brick model for phase B of $Q^{1,1,1}/\mathbb{Z}_2$.
\label{fq111z2b-brick2}}
 \end{center}
 \end{figure} 

\begin{figure}[H]
\begin{center}
\resizebox{0.8\hsize}{!}{
\includegraphics[trim=0cm 0cm 0cm 0cm,totalheight=10 cm]{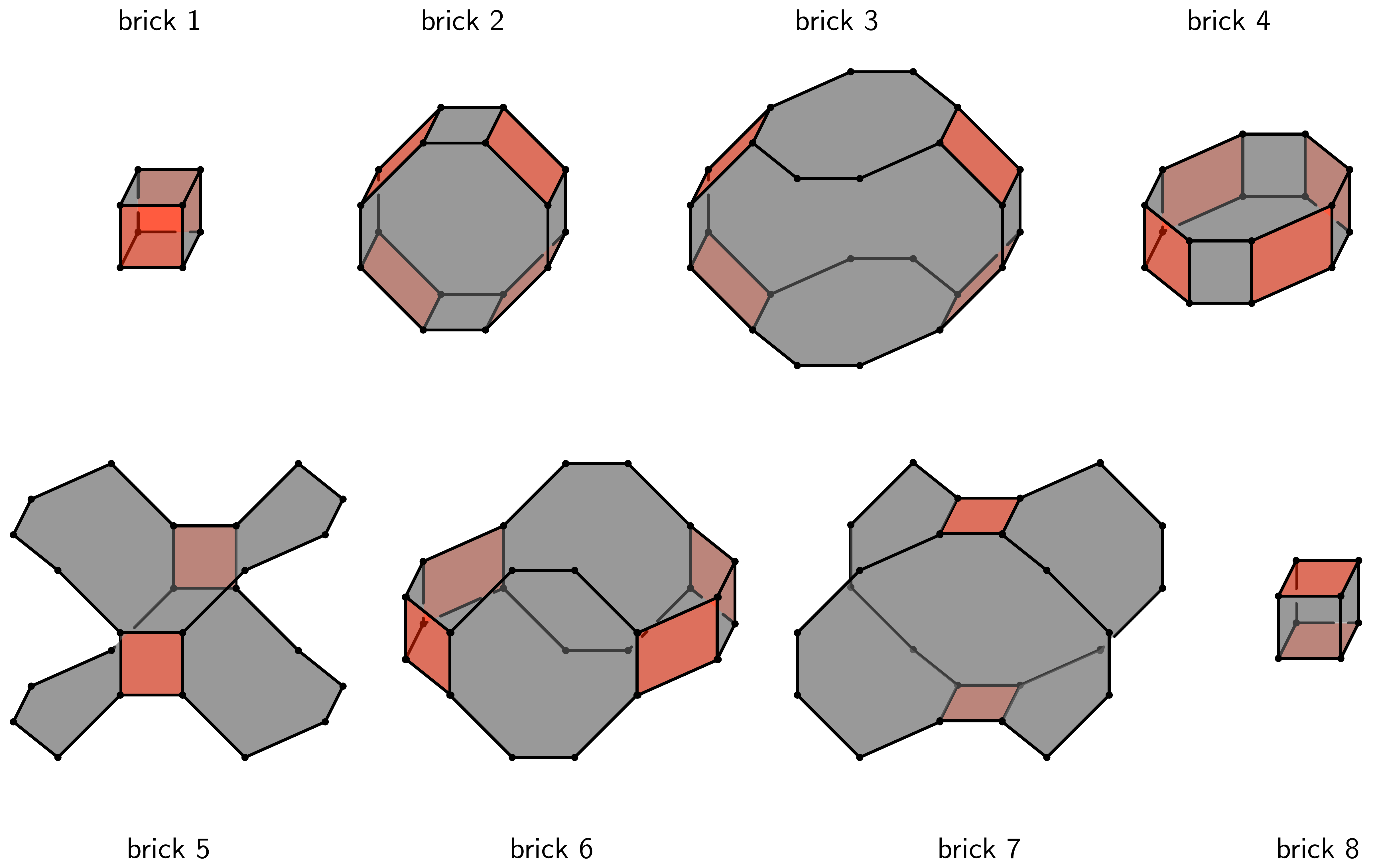}
}  
\caption{
Brane bricks for phase B of $Q^{1,1,1}/\mathbb{Z}_2$.
\label{fq111z2b-brick1}}
 \end{center}
 \end{figure} 

The chiral and Fermi field pieces of the phase boundary matrix are
\beal{q111z2b-hx}
\scriptsize
H_X =
\left(
\begin{array}{c|cccccccccccc}
\; &
\eta_{13} & \eta_{14} & \eta_{23} & \eta_{24} & \eta_{15} & \eta_{16} & \eta_{25} & \eta_{26} & \eta_{35} & \eta_{36} & \eta_{45} & \eta_{46} 
\\
\hline
X_{37}^{+} & 1 & 1 & 0 & 0 & 1 & 1 & 0 & 0 & 0 & 0 & 0 & 0 \\
X_{37}^{-} & 0 & 0 & 1 & 1 & 0 & 0 & 1 & 1 & 0 & 0 & 0 & 0 \\
X_{62}^{+} & 1 & 1 & 0 & 0 & 1 & 1 & 0 & 0 & 0 & 0 & 0 & 0 \\
X_{62}^{-} & 0 & 0 & 1 & 1 & 0 & 0 & 1 & 1 & 0 & 0 & 0 & 0 \\
X_{84}^{+} & 1 & 1 & 0 & 0 & 1 & 1 & 0 & 0 & 0 & 0 & 0 & 0 \\
X_{84}^{-} & 0 & 0 & 1 & 1 & 0 & 0 & 1 & 1 & 0 & 0 & 0 & 0 \\
X_{24}^{+} & -1 & 0 & -1 & 0 & 0 & 0 & 0 & 0 & 1 & 1 & 0 & 0 \\
X_{24}^{-} & 0 & -1 & 0 & -1 & 0 & 0 & 0 & 0 & 0 & 0 & 1 & 1 \\
X_{68}^{+} & -1 & 0 & -1 & 0 & 0 & 0 & 0 & 0 & 1 & 1 & 0 & 0 \\
X_{68}^{-} & 0 & -1 & 0 & -1 & 0 & 0 & 0 & 0 & 0 & 0 & 1 & 1 \\
X_{75}^{+} & -1 & 0 & -1 & 0 & 0 & 0 & 0 & 0 & 1 & 1 & 0 & 0 \\
X_{75}^{-} & 0 & -1 & 0 & -1 & 0 & 0 & 0 & 0 & 0 & 0 & 1 & 1 \\
X_{43}^{+} & 0 & 0 & 0 & 0 & -1 & 0 & -1 & 0 & -1 & 0 & -1 & 0 \\
X_{43}^{-} & 0 & 0 & 0 & 0 & 0 & -1 & 0 & -1 & 0 & -1 & 0 & -1 \\
X_{56}^{+} & 0 & 0 & 0 & 0 & -1 & 0 & -1 & 0 & -1 & 0 & -1 & 0 \\
X_{56}^{-} & 0 & 0 & 0 & 0 & 0 & -1 & 0 & -1 & 0 & -1 & 0 & -1 \\
X_{21}^{+} & 0 & 0 & 0 & 0 & -1 & 0 & -1 & 0 & -1 & 0 & -1 & 0 \\
X_{21}^{-} & 0 & 0 & 0 & 0 & 0 & -1 & 0 & -1 & 0 & -1 & 0 & -1 \\
X_{13}^{+} & -1 & 0 & -1 & 0 & 0 & 0 & 0 & 0 & 1 & 1 & 0 & 0 \\
X_{13}^{-} & 0 & -1 & 0 & -1 & 0 & 0 & 0 & 0 & 0 & 0 & 1 & 1 \\
\end{array}
\right) ~,~
\eea
\beal{q111z2b-hl}
\scriptsize
H_\Lambda = 
\left(
\begin{array}{c|cccccccccccc}
\; &
\eta_{13} & \eta_{14} & \eta_{23} & \eta_{24} & \eta_{15} & \eta_{16} & \eta_{25} & \eta_{26} & \eta_{35} & \eta_{36} & \eta_{45} & \eta_{46} 
\\
\hline
\Lambda_{51}^{+} & 1 & 1 & 0 & 0 & 0 & 0 & -1 & -1 & -1 & -1 & -1 & -1 \\
\Lambda_{51}^{-} & 0 & 0 & 1 & 1 & -1 & -1 & 0 & 0 & -1 & -1 & -1 & -1 \\
\Lambda_{78}^{+} & -1 & -1 & -1 & -1 & -1 & 0 & -1 & 0 & 0 & 1 & 0 & 1 \\
\Lambda_{78}^{-} & -1 & -1 & -1 & -1 & 0 & -1 & 0 & -1 & 1 & 0 & 1 & 0 \\
\Lambda_{64}^{++} & 0 & 1 & -1 & 0 & 1 & 1 & 0 & 0 & 1 & 1 & 0 & 0 \\
\Lambda_{64}^{+-} & 1 & 0 & 0 & -1 & 1 & 1 & 0 & 0 & 0 & 0 & 1 & 1 \\
\Lambda_{64}^{-+} & -1 & 0 & 0 & 1 & 0 & 0 & 1 & 1 & 1 & 1 & 0 & 0 \\
\Lambda_{64}^{--} & 0 & -1 & 1 & 0 & 0 & 0 & 1 & 1 & 0 & 0 & 1 & 1 \\
\Lambda_{23}^{++} & -1 & 0 & -1 & 0 & -1 & 0 & -1 & 0 & 0 & 1 & -1 & 0 \\
\Lambda_{23}^{+-} & -1 & 0 & -1 & 0 & 0 & -1 & 0 & -1 & 1 & 0 & 0 & -1 \\
\Lambda_{23}^{-+} & 0 & -1 & 0 & -1 & -1 & 0 & -1 & 0 & -1 & 0 & 0 & 1 \\
\Lambda_{23}^{--} & 0 & -1 & 0 & -1 & 0 & -1 & 0 & -1 & 0 & -1 & 1 & 0 \\
\end{array}
\right)~.~
\eea

\newpage

\subsection*{Phase C}

The periodic quiver for phase C is given in \fref{fq111z2c-quiver}. It has 8 gauge groups, 24 chiral fields and 16 Fermi fields.

\begin{figure}[H]
\begin{center}
\resizebox{0.6\hsize}{!}{
\includegraphics[trim=0cm 0cm 0cm 0cm,totalheight=10 cm]{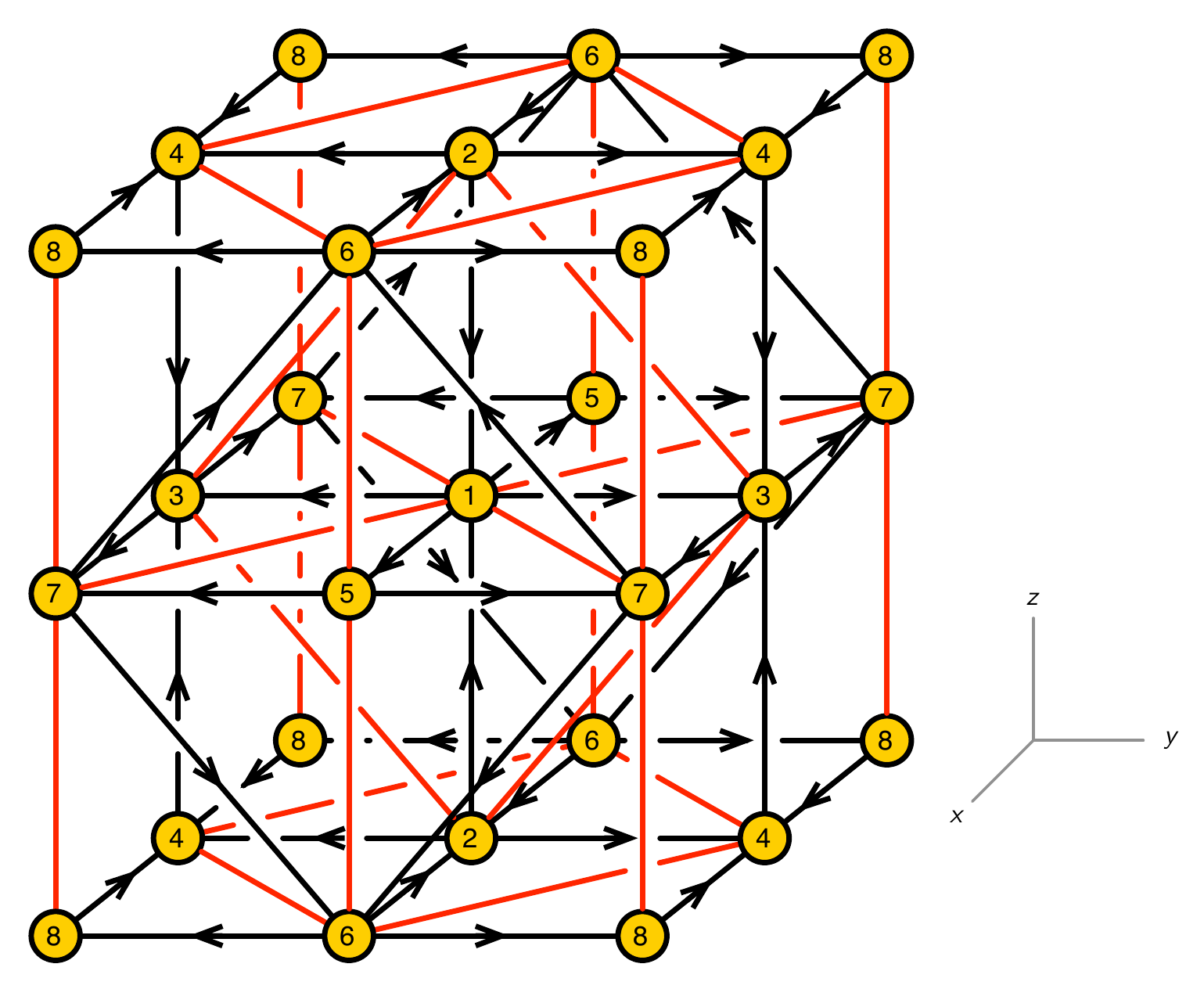}
}  
\caption{
Periodic quiver for phase C of $Q^{1,1,1}/\mathbb{Z}_2$.
\label{fq111z2c-quiver}}
 \end{center}
 \end{figure} 
 
The and $J$- and $E$-terms are
{\small
\beq
\begin{array}{lcccc}
& &  J  & &  E 
\\  
 \Lambda_{56}^{+} 
 & :\ \ \ &  
X_{62}^{+} \cdot X_{21}^{-} \cdot X_{15}^{-} -
X_{62}^{-} \cdot X_{21}^{-} \cdot X_{15}^{+} & \ \ \ \ &    
X_{57}^{+} \cdot X_{76}^{-+} - X_{57}^{-} \cdot X_{76}^{++}     
\\
 \Lambda_{56}^{-} 
& :\ \ \ &  
X_{62}^{-} \cdot X_{21}^{+} \cdot X_{15}^{+} -
X_{62}^{+} \cdot X_{21}^{+} \cdot X_{15}^{-} 
& \ \ \ \ &    
X_{57}^{+} \cdot X_{76}^{--} - X_{57}^{-} \cdot X_{76}^{+-}  
\\ 
\Lambda_{78}^{+} & :\ \ \ &  
X_{84}^{+} \cdot X_{43}^{-} \cdot X_{37}^{-}   
-
X_{84}^{-} \cdot X_{43}^{-} \cdot X_{37}^{+}
& \ \ \ \ &
X_{76}^{++} \cdot X_{68}^{-} - X_{76}^{-+} \cdot X_{68}^{+}  
\\ 
\Lambda_{78}^{-} 
& :\ \ \ &  
X_{84}^{-} \cdot X_{43}^{+} \cdot X_{37}^{+} -
X_{84}^{+} \cdot X_{43}^{+} \cdot X_{37}^{-} 
& \ \ \ \ &     
X_{76}^{+-} \cdot X_{68}^{-} - X_{76}^{--} \cdot X_{68}^{+}  
\\  
\Lambda_{64}^{++} 
& :\ \ \ &  
X_{43}^{+} \cdot X_{37}^{-} \cdot X_{76}^{--} - X_{43}^{-} \cdot X_{37}^{-} \cdot X_{76}^{-+}    
& \ \ \ \ &     
X_{62}^{+} \cdot X_{24}^{+} - X_{68}^{+} \cdot X_{84}^{+}    
\\  
\Lambda_{64}^{--} 
& :\ \ \ &  
X_{43}^{+} \cdot X_{37}^{+} \cdot X_{76}^{+-}    
-
X_{43}^{-} \cdot X_{37}^{+} \cdot X_{76}^{++} & \ \ \ \ &     
X_{62}^{-} \cdot X_{24}^{-} - X_{68}^{-} \cdot X_{84}^{-}    
\\  
\Lambda_{64}^{+-} 
& :\ \ \ &  
X_{43}^{-} \cdot X_{37}^{-} \cdot X_{76}^{++} - X_{43}^{+} \cdot X_{37}^{-} \cdot X_{76}^{+-}    
& \ \ \ \ &     
X_{62}^{+} \cdot X_{24}^{-} - X_{68}^{-} \cdot X_{84}^{+}    
\\  
\Lambda_{64}^{-+} 
& :\ \ \ &  
X_{43}^{-} \cdot X_{37}^{+} \cdot X_{76}^{-+}    
-
X_{43}^{+} \cdot X_{37}^{+} \cdot X_{76}^{--} & \ \ \ \ &     
X_{62}^{-} \cdot X_{24}^{+} - X_{68}^{+} \cdot X_{84}^{-}    
\\ 
\Lambda_{17}^{++} 
& :\ \ \ &  
X_{76}^{-+} \cdot X_{62}^{-} \cdot X_{21}^{-} - X_{76}^{--} \cdot X_{62}^{-} \cdot X_{21}^{+}    
& \ \ \ \ &    
X_{15}^{+} \cdot X_{57}^{+} - X_{13}^{+} \cdot X_{37}^{+}    
\\ 
\Lambda_{17}^{--} 
& :\ \ \ &  
X_{76}^{++} \cdot X_{62}^{+} \cdot X_{21}^{-} - X_{76}^{+-} \cdot X_{62}^{+} \cdot X_{21}^{+}    
& \ \ \ \ &    
X_{15}^{-} \cdot X_{57}^{-} - X_{13}^{-} \cdot X_{37}^{-}    
\\ 
\Lambda_{17}^{+-} 
& :\ \ \ &  
X_{76}^{+-} \cdot X_{62}^{-} \cdot X_{21}^{+} -
X_{76}^{++} \cdot X_{62}^{-} \cdot X_{21}^{-} 
& \ \ \ \ &    
X_{15}^{+} \cdot X_{57}^{-} - X_{13}^{-} \cdot X_{37}^{+}    
\\ 
\Lambda_{17}^{-+} 
& :\ \ \ &  
X_{76}^{--} \cdot X_{62}^{+} \cdot X_{21}^{+}-
X_{76}^{-+} \cdot X_{62}^{+} \cdot X_{21}^{-} 
& \ \ \ \ &    
X_{15}^{-} \cdot X_{57}^{+} - X_{13}^{+} \cdot X_{37}^{-}    
\\ 
\Lambda_{23}^{++} 
& :\ \ \ &  
X_{37}^{+} \cdot X_{76}^{--} \cdot X_{62}^{-} - X_{37}^{-} \cdot X_{76}^{--} 
\cdot X_{62}^{+}
& \ \ \ \ &     
X_{24}^{+} \cdot X_{43}^{+} - X_{21}^{+} \cdot X_{13}^{+}   
\\ 
\Lambda_{23}^{--} 
& :\ \ \ &  
X_{37}^{+} \cdot X_{76}^{++} \cdot X_{62}^{-} - X_{37}^{-} \cdot X_{76}^{++} \cdot X_{62}^{+}    
& \ \ \ \ &     
X_{24}^{-} \cdot X_{43}^{-} - X_{21}^{-} \cdot X_{13}^{-} 
\\ 
\Lambda_{23}^{+-} 
& :\ \ \ &  
X_{37}^{-} \cdot X_{76}^{-+} \cdot X_{62}^{+}    
-
X_{37}^{+} \cdot X_{76}^{-+} \cdot X_{62}^{-} & \ \ \ \ &     
X_{24}^{+} \cdot X_{43}^{-} - X_{21}^{-} \cdot X_{13}^{+}   
\\ 
\Lambda_{23}^{-+} 
& :\ \ \ &  
X_{37}^{-} \cdot X_{76}^{+-} \cdot X_{62}^{+}   
-
X_{37}^{+} \cdot X_{76}^{+-} \cdot X_{62}^{-} & \ \ \ \ &     
X_{24}^{-} \cdot X_{43}^{+} - X_{21}^{+} \cdot X_{13}^{-}  
\\  
\end{array}
\label{q111z2c-je}
\eeq
}
The brick matchings are given by
\beal{q111z2c-px}
\scriptsize
P =
\left(
\begin{array}{c|cccccc|cccccccccccccc}
\; & p_1 & p_2 & p_3 & p_4 & p_5 & p_6 & s_1 & s_2 & s_3 & s_4 & s_5 & s_6 & s_7 & s_8 & s_9 & s_{10} & s_{11} & s_{12} & s_{13} & s_{14}\\
\hline
X_{15}^{+} & 1 & 0 & 0 & 0 & 0 & 0 & 1 & 0 & 0 & 0 & 0 & 0 & 1 & 0 & 0 & 0 & 1 & 1 & 0 & 0 \\
X_{15}^{-} & 0 & 1 & 0 & 0 & 0 & 0 & 1 & 0 & 0 & 0 & 0 & 0 & 1 & 0 & 0 & 0 & 1 & 1 & 0 & 0 \\
X_{37}^{+} & 1 & 0 & 0 & 0 & 0 & 0 & 1 & 0 & 1 & 0 & 0 & 0 & 0 & 0 & 0 & 0 & 0 & 0 & 0 & 0 \\
X_{37}^{-} & 0 & 1 & 0 & 0 & 0 & 0 & 1 & 0 & 1 & 0 & 0 & 0 & 0 & 0 & 0 & 0 & 0 & 0 & 0 & 0 \\
X_{62}^{+} & 1 & 0 & 0 & 0 & 0 & 0 & 0 & 0 & 0 & 1 & 1 & 0 & 0 & 0 & 0 & 0 & 0 & 0 & 0 & 0 \\
X_{62}^{-} & 0 & 1 & 0 & 0 & 0 & 0 & 0 & 0 & 0 & 1 & 1 & 0 & 0 & 0 & 0 & 0 & 0 & 0 & 0 & 0 \\
X_{84}^{+} & 1 & 0 & 0 & 0 & 0 & 0 & 0 & 0 & 0 & 0 & 1 & 0 & 0 & 0 & 0 & 1 & 0 & 1 & 0 & 1 \\
X_{84}^{-} & 0 & 1 & 0 & 0 & 0 & 0 & 0 & 0 & 0 & 0 & 1 & 0 & 0 & 0 & 0 & 1 & 0 & 1 & 0 & 1 \\
X_{13}^{+} & 0 & 0 & 1 & 0 & 0 & 0 & 0 & 0 & 0 & 0 & 0 & 0 & 1 & 1 & 0 & 0 & 1 & 1 & 1 & 1 \\
X_{13}^{-} & 0 & 0 & 0 & 1 & 0 & 0 & 0 & 0 & 0 & 0 & 0 & 0 & 1 & 1 & 0 & 0 & 1 & 1 & 1 & 1 \\
X_{24}^{+} & 0 & 0 & 1 & 0 & 0 & 0 & 0 & 0 & 0 & 0 & 0 & 0 & 0 & 0 & 1 & 1 & 1 & 1 & 1 & 1 \\
X_{24}^{-} & 0 & 0 & 0 & 1 & 0 & 0 & 0 & 0 & 0 & 0 & 0 & 0 & 0 & 0 & 1 & 1 & 1 & 1 & 1 & 1 \\
X_{57}^{+} & 0 & 0 & 1 & 0 & 0 & 0 & 0 & 0 & 1 & 0 & 0 & 0 & 0 & 1 & 0 & 0 & 0 & 0 & 1 & 1 \\
X_{57}^{-} & 0 & 0 & 0 & 1 & 0 & 0 & 0 & 0 & 1 & 0 & 0 & 0 & 0 & 1 & 0 & 0 & 0 & 0 & 1 & 1 \\
X_{68}^{+} & 0 & 0 & 1 & 0 & 0 & 0 & 0 & 0 & 0 & 1 & 0 & 0 & 0 & 0 & 1 & 0 & 1 & 0 & 1 & 0 \\
X_{68}^{-} & 0 & 0 & 0 & 1 & 0 & 0 & 0 & 0 & 0 & 1 & 0 & 0 & 0 & 0 & 1 & 0 & 1 & 0 & 1 & 0 \\
X_{21}^{+} & 0 & 0 & 0 & 0 & 1 & 0 & 0 & 1 & 0 & 0 & 0 & 0 & 0 & 0 & 1 & 1 & 0 & 0 & 0 & 0 \\
X_{21}^{-} & 0 & 0 & 0 & 0 & 0 & 1 & 0 & 1 & 0 & 0 & 0 & 0 & 0 & 0 & 1 & 1 & 0 & 0 & 0 & 0 \\
X_{43}^{+} & 0 & 0 & 0 & 0 & 1 & 0 & 0 & 1 & 0 & 0 & 0 & 0 & 1 & 1 & 0 & 0 & 0 & 0 & 0 & 0 \\
X_{43}^{-} & 0 & 0 & 0 & 0 & 0 & 1 & 0 & 1 & 0 & 0 & 0 & 0 & 1 & 1 & 0 & 0 & 0 & 0 & 0 & 0 \\
X_{76}^{++} & 0 & 0 & 1 & 0 & 1 & 0 & 0 & 0 & 0 & 0 & 0 & 1 & 0 & 0 & 0 & 0 & 0 & 0 & 0 & 0 \\
X_{76}^{+-} & 0 & 0 & 1 & 0 & 0 & 1 & 0 & 0 & 0 & 0 & 0 & 1 & 0 & 0 & 0 & 0 & 0 & 0 & 0 & 0 \\
X_{76}^{-+} & 0 & 0 & 0 & 1 & 1 & 0 & 0 & 0 & 0 & 0 & 0 & 1 & 0 & 0 & 0 & 0 & 0 & 0 & 0 & 0 \\
X_{76}^{--} & 0 & 0 & 0 & 1 & 0 & 1 & 0 & 0 & 0 & 0 & 0 & 1 & 0 & 0 & 0 & 0 & 0 & 0 & 0 & 0 \\
\end{array}
\right)~,~
\nonumber
\eea
\beal{q111z2c-pl}
\scriptsize
P_\Lambda =
\left(
\begin{array}{c|cccccc|cccccccccccccc}
\; & p_1 & p_2 & p_3 & p_4 & p_5 & p_6 & s_1 & s_2 & s_3 & s_4 & s_5 & s_6 & s_7 & s_8 & s_9 & s_{10} & s_{11} & s_{12} & s_{13} & s_{14}\\
\hline
\Lambda_{56}^{+} & 0 & 0 & 1 & 1 & 1 & 0 & 0 & 0 & 1 & 0 & 0 & 1 & 0 & 1 & 0 & 0 & 0 & 0 & 1 & 1 \\
\Lambda_{56}^{-} & 0 & 0 & 1 & 1 & 0 & 1 & 0 & 0 & 1 & 0 & 0 & 1 & 0 & 1 & 0 & 0 & 0 & 0 & 1 & 1 \\
\Lambda_{78}^{+} & 0 & 0 & 1 & 1 & 1 & 0 & 0 & 0 & 0 & 1 & 0 & 1 & 0 & 0 & 1 & 0 & 1 & 0 & 1 & 0 \\
\Lambda_{78}^{-} & 0 & 0 & 1 & 1 & 0 & 1 & 0 & 0 & 0 & 1 & 0 & 1 & 0 & 0 & 1 & 0 & 1 & 0 & 1 & 0 \\
\Lambda_{64}^{++} & 1 & 0 & 1 & 0 & 0 & 0 & 0 & 0 & 0 & 1 & 1 & 0 & 0 & 0 & 1 & 1 & 1 & 1 & 1 & 1 \\
\Lambda_{64}^{+-} & 1 & 0 & 0 & 1 & 0 & 0 & 0 & 0 & 0 & 1 & 1 & 0 & 0 & 0 & 1 & 1 & 1 & 1 & 1 & 1 \\
\Lambda_{64}^{-+} & 0 & 1 & 1 & 0 & 0 & 0 & 0 & 0 & 0 & 1 & 1 & 0 & 0 & 0 & 1 & 1 & 1 & 1 & 1 & 1 \\
\Lambda_{64}^{-} & 0 & 1 & 0 & 1 & 0 & 0 & 0 & 0 & 0 & 1 & 1 & 0 & 0 & 0 & 1 & 1 & 1 & 1 & 1 & 1 \\
\Lambda_{17}^{++} & 1 & 0 & 1 & 0 & 0 & 0 & 1 & 0 & 1 & 0 & 0 & 0 & 1 & 1 & 0 & 0 & 1 & 1 & 1 & 1 \\
\Lambda_{17}^{+-} & 1 & 0 & 0 & 1 & 0 & 0 & 1 & 0 & 1 & 0 & 0 & 0 & 1 & 1 & 0 & 0 & 1 & 1 & 1 & 1 \\
\Lambda_{17}^{-+} & 0 & 1 & 1 & 0 & 0 & 0 & 1 & 0 & 1 & 0 & 0 & 0 & 1 & 1 & 0 & 0 & 1 & 1 & 1 & 1 \\
\Lambda_{17}^{--} & 0 & 1 & 0 & 1 & 0 & 0 & 1 & 0 & 1 & 0 & 0 & 0 & 1 & 1 & 0 & 0 & 1 & 1 & 1 & 1 \\
\Lambda_{23}^{++} & 0 & 0 & 1 & 0 & 1 & 0 & 0 & 1 & 0 & 0 & 0 & 0 & 1 & 1 & 1 & 1 & 1 & 1 & 1 & 1 \\
\Lambda_{23}^{+-} & 0 & 0 & 1 & 0 & 0 & 1 & 0 & 1 & 0 & 0 & 0 & 0 & 1 & 1 & 1 & 1 & 1 & 1 & 1 & 1 \\
\Lambda_{23}^{-+} & 0 & 0 & 0 & 1 & 1 & 0 & 0 & 1 & 0 & 0 & 0 & 0 & 1 & 1 & 1 & 1 & 1 & 1 & 1 & 1 \\
\Lambda_{23}^{--} & 0 & 0 & 0 & 1 & 0 & 1 & 0 & 1 & 0 & 0 & 0 & 0 & 1 & 1 & 1 & 1 & 1 & 1 & 1 & 1 \\
\end{array}
\right)~.~
\eea

The brane brick model for phase C and its individual bricks are shown in Figures \ref{fq111z2c-brick2} and \ref{fq111z2c-brick1}.

\begin{figure}[H]
\begin{center}
\resizebox{0.6\hsize}{!}{
\includegraphics[trim=0cm 0cm 0cm 0cm,totalheight=10 cm]{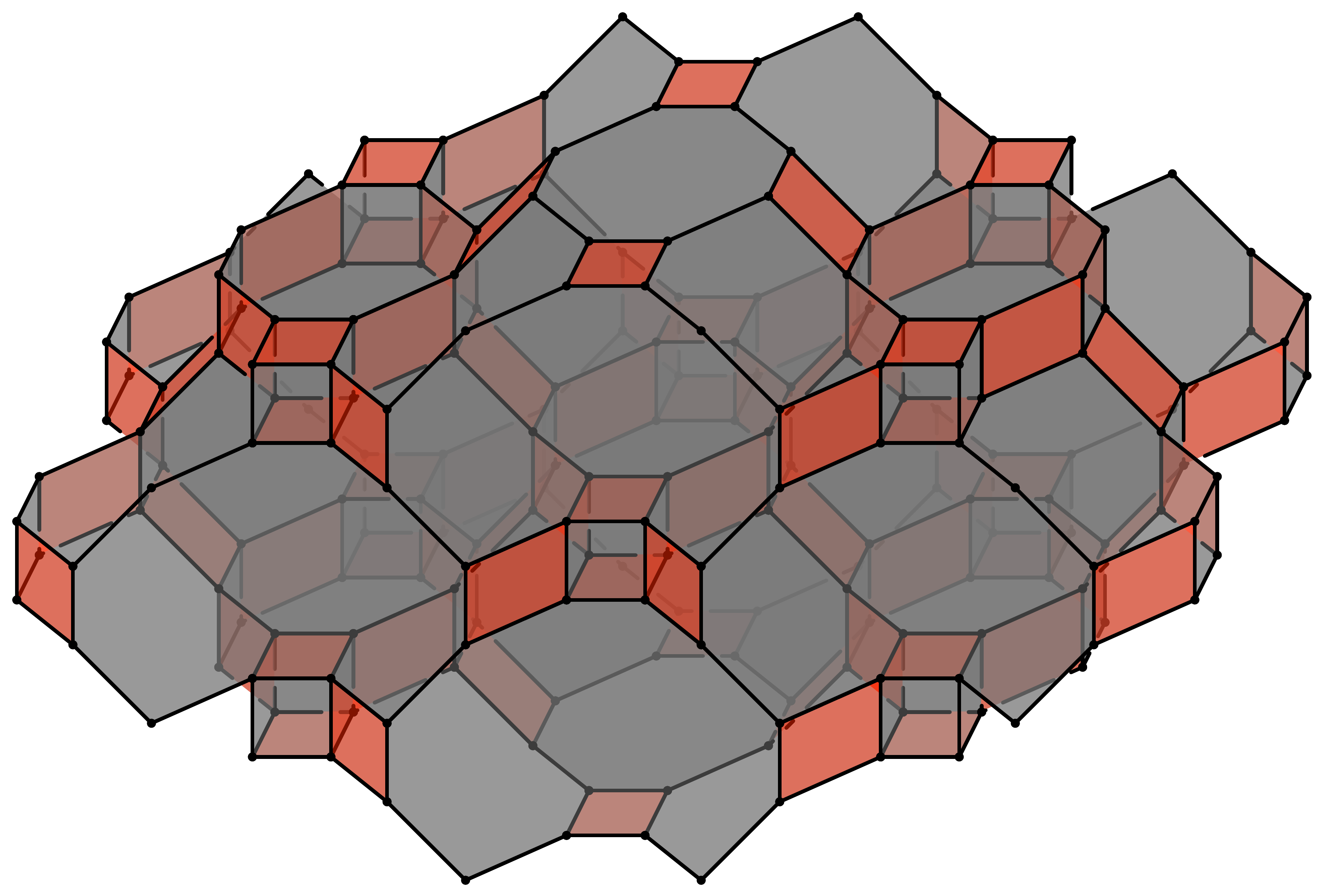}
}  
\caption{
Brane brick model for phase C of $Q^{1,1,1}/\mathbb{Z}_2$.
\label{fq111z2c-brick2}}
 \end{center}
 \end{figure} 

\begin{figure}[H]
\begin{center}
\resizebox{0.8\hsize}{!}{
\includegraphics[trim=0cm 0cm 0cm 0cm,totalheight=10 cm]{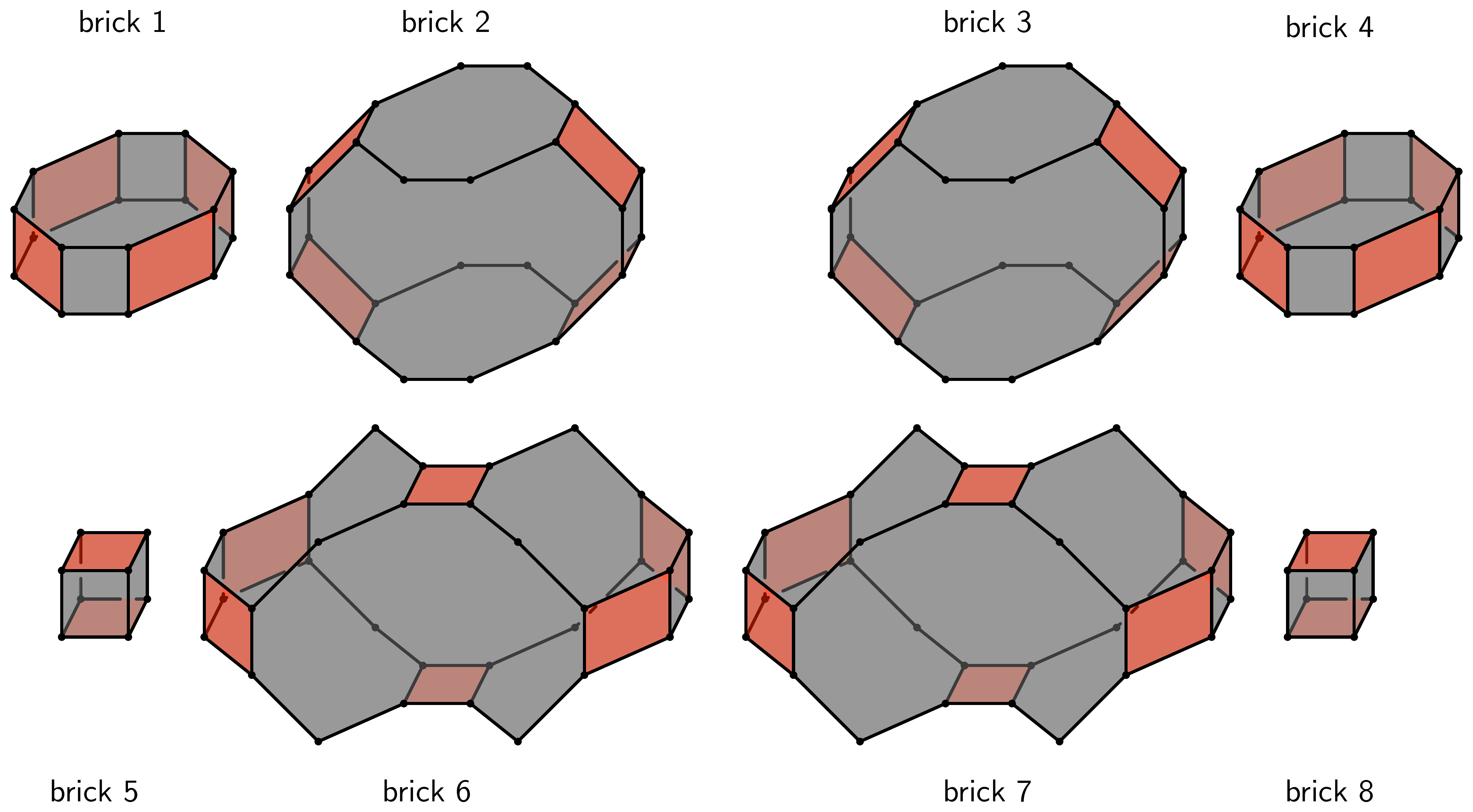}
}  
\caption{
Brane bricks for phase C of $Q^{1,1,1}/\mathbb{Z}_2$.
\label{fq111z2c-brick1}}
 \end{center}
 \end{figure} 

The chiral and Fermi field pieces of the phase boundary matrix are
\beal{q111z2c-hx}
\scriptsize
H_X =
\left(
\begin{array}{c|cccccccccccc}
\; &
\eta_{13} & \eta_{14} & \eta_{23} & \eta_{24} & \eta_{15} & \eta_{16} & \eta_{25} & \eta_{26} & \eta_{35} & \eta_{36} & \eta_{45} & \eta_{46} 
\\
\hline
X_{15}^{+} & 1 & 1 & 0 & 0 & 1 & 1 & 0 & 0 & 0 & 0 & 0 & 0 \\
X_{15}^{-} & 0 & 0 & 1 & 1 & 0 & 0 & 1 & 1 & 0 & 0 & 0 & 0 \\
X_{37}^{+} & 1 & 1 & 0 & 0 & 1 & 1 & 0 & 0 & 0 & 0 & 0 & 0 \\
X_{37}^{-} & 0 & 0 & 1 & 1 & 0 & 0 & 1 & 1 & 0 & 0 & 0 & 0 \\
X_{62}^{+} & 1 & 1 & 0 & 0 & 1 & 1 & 0 & 0 & 0 & 0 & 0 & 0 \\
X_{62}^{-} & 0 & 0 & 1 & 1 & 0 & 0 & 1 & 1 & 0 & 0 & 0 & 0 \\
X_{84}^{+} & 1 & 1 & 0 & 0 & 1 & 1 & 0 & 0 & 0 & 0 & 0 & 0 \\
X_{84}^{-} & 0 & 0 & 1 & 1 & 0 & 0 & 1 & 1 & 0 & 0 & 0 & 0 \\
X_{13}^{+} & -1 & 0 & -1 & 0 & 0 & 0 & 0 & 0 & 1 & 1 & 0 & 0 \\
X_{13}^{-} & 0 & -1 & 0 & -1 & 0 & 0 & 0 & 0 & 0 & 0 & 1 & 1 \\
X_{24}^{+} & -1 & 0 & -1 & 0 & 0 & 0 & 0 & 0 & 1 & 1 & 0 & 0 \\
X_{24}^{-} & 0 & -1 & 0 & -1 & 0 & 0 & 0 & 0 & 0 & 0 & 1 & 1 \\
X_{57}^{+} & -1 & 0 & -1 & 0 & 0 & 0 & 0 & 0 & 1 & 1 & 0 & 0 \\
X_{57}^{-} & 0 & -1 & 0 & -1 & 0 & 0 & 0 & 0 & 0 & 0 & 1 & 1 \\
X_{68}^{+} & -1 & 0 & -1 & 0 & 0 & 0 & 0 & 0 & 1 & 1 & 0 & 0 \\
X_{68}^{-} & 0 & -1 & 0 & -1 & 0 & 0 & 0 & 0 & 0 & 0 & 1 & 1 \\
X_{21}^{+} & 0 & 0 & 0 & 0 & -1 & 0 & -1 & 0 & -1 & 0 & -1 & 0 \\
X_{21}^{-} & 0 & 0 & 0 & 0 & 0 & -1 & 0 & -1 & 0 & -1 & 0 & -1 \\
X_{43}^{+} & 0 & 0 & 0 & 0 & -1 & 0 & -1 & 0 & -1 & 0 & -1 & 0 \\
X_{43}^{-} & 0 & 0 & 0 & 0 & 0 & -1 & 0 & -1 & 0 & -1 & 0 & -1 \\
X_{76}^{++} & -1 & 0 & -1 & 0 & -1 & 0 & -1 & 0 & 0 & 1 & -1 & 0 \\
X_{76}^{+-} & -1 & 0 & -1 & 0 & 0 & -1 & 0 & -1 & 1 & 0 & 0 & -1 \\
X_{76}^{-+} & 0 & -1 & 0 & -1 & -1 & 0 & -1 & 0 & -1 & 0 & 0 & 1 \\
X_{76}^{--} & 0 & -1 & 0 & -1 & 0 & -1 & 0 & -1 & 0 & -1 & 1 & 0 \\
\end{array}
\right) ~,~
\eea
\beal{q111z2c-hl}
\scriptsize
H_\Lambda = 
\left(
\begin{array}{c|cccccccccccc}
\; &
\eta_{13} & \eta_{14} & \eta_{23} & \eta_{24} & \eta_{15} & \eta_{16} & \eta_{25} & \eta_{26} & \eta_{35} & \eta_{36} & \eta_{45} & \eta_{46} 
\\
\Lambda_{56}^{+} & -1 & -1 & -1 & -1 & -1 & 0 & -1 & 0 & 0 & 1 & 0 & 1 \\
\Lambda_{56}^{-} & -1 & -1 & -1 & -1 & 0 & -1 & 0 & -1 & 1 & 0 & 1 & 0 \\
\Lambda_{78}^{+} & -1 & -1 & -1 & -1 & -1 & 0 & -1 & 0 & 0 & 1 & 0 & 1 \\
\Lambda_{78}^{-} & -1 & -1 & -1 & -1 & 0 & -1 & 0 & -1 & 1 & 0 & 1 & 0 \\
\Lambda_{64}^{++} & 0 & 1 & -1 & 0 & 1 & 1 & 0 & 0 & 1 & 1 & 0 & 0 \\
\Lambda_{64}^{+-} & 1 & 0 & 0 & -1 & 1 & 1 & 0 & 0 & 0 & 0 & 1 & 1 \\
\Lambda_{64}^{-+} & -1 & 0 & 0 & 1 & 0 & 0 & 1 & 1 & 1 & 1 & 0 & 0 \\
\Lambda_{64}^{--} & 0 & -1 & 1 & 0 & 0 & 0 & 1 & 1 & 0 & 0 & 1 & 1 \\
\Lambda_{17}^{++} & 0 & 1 & -1 & 0 & 1 & 1 & 0 & 0 & 1 & 1 & 0 & 0 \\
\Lambda_{17}^{+-} & 1 & 0 & 0 & -1 & 1 & 1 & 0 & 0 & 0 & 0 & 1 & 1 \\
\Lambda_{17}^{-+} & -1 & 0 & 0 & 1 & 0 & 0 & 1 & 1 & 1 & 1 & 0 & 0 \\
\Lambda_{17}^{--} & 0 & -1 & 1 & 0 & 0 & 0 & 1 & 1 & 0 & 0 & 1 & 1 \\
\Lambda_{23}^{++} & -1 & 0 & -1 & 0 & -1 & 0 & -1 & 0 & 0 & 1 & -1 & 0 \\
\Lambda_{23}^{+-} & -1 & 0 & -1 & 0 & 0 & -1 & 0 & -1 & 1 & 0 & 0 & -1 \\
\Lambda_{23}^{-+} & 0 & -1 & 0 & -1 & -1 & 0 & -1 & 0 & -1 & 0 & 0 & 1 \\
\Lambda_{23}^{--} & 0 & -1 & 0 & -1 & 0 & -1 & 0 & -1 & 0 & -1 & 1 & 0 \\
\end{array}
\right)~.~
\eea

\newpage

\subsection*{Phase D}

The periodic quiver for phase D is given in \fref{fq111z2d-quiver}. It has 8 gauge groups, 24 chiral fields and 16 Fermi fields.

\begin{figure}[H]
\begin{center}
\resizebox{0.6\hsize}{!}{
\includegraphics[trim=0cm 0cm 0cm 0cm,totalheight=10 cm]{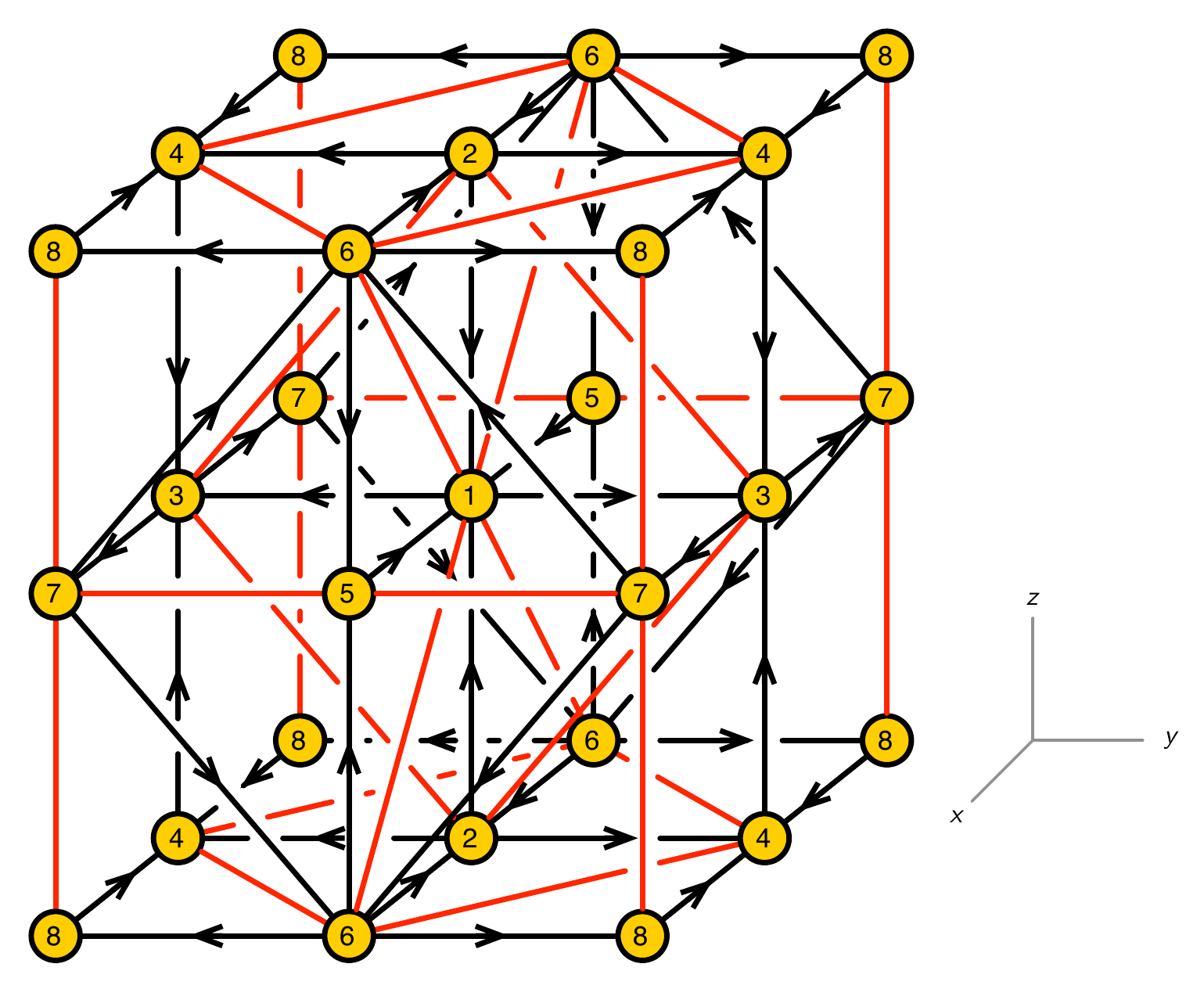}
}  
\caption{
Periodic quiver for phase D of $Q^{1,1,1}/\mathbb{Z}_2$.
\label{fq111z2d-quiver}}
 \end{center}
 \end{figure} 
 
The $J$- and $E$-terms for this theory are
{\small
\beq
\begin{array}{lcccc}
& &  J  & &  E \\  
\Lambda_{75}^{+} & :\ \ \ &  
X_{51}^{-} \cdot X_{13}^{-} \cdot X_{37}^{+} - X_{51}^{+} \cdot X_{13}^{-} \cdot X_{37}^{-} 
& \ \ \ \ &      
X_{76}^{++} \cdot X_{65}^{-} - X_{76}^{+-} \cdot X_{65}^{+}   
\\ 
\Lambda_{75}^{-} & :\ \ \ &  
X_{51}^{+} \cdot X_{13}^{+}\cdot X_{37}^{-} -
X_{51}^{-} \cdot X_{13}^{+} \cdot X_{37}^{+}    
& \ \ \ \ &      
X_{76}^{-+} \cdot X_{65}^{-} - X_{76}^{--} \cdot X_{65}^{+}  
\\ 
\Lambda_{78}^{+} & :\ \ \ &  
X_{84}^{+} \cdot X_{43}^{-} \cdot X_{37}^{-}   
-
X_{84}^{-} \cdot X_{43}^{-} \cdot X_{37}^{+}
& \ \ \ \ &
X_{76}^{++} \cdot X_{68}^{-} - X_{76}^{-+} \cdot X_{68}^{+}  
\\ 
\Lambda_{78}^{-} 
& :\ \ \ &  
X_{84}^{-} \cdot X_{43}^{+} \cdot X_{37}^{+} -
X_{84}^{+} \cdot X_{43}^{+} \cdot X_{37}^{-} 
& \ \ \ \ &     
X_{76}^{+-} \cdot X_{68}^{-} - X_{76}^{--} \cdot X_{68}^{+}    
\\  
\Lambda_{64}^{++} 
& :\ \ \ &  
X_{43}^{+} \cdot X_{37}^{-} \cdot X_{76}^{--} - X_{43}^{-} \cdot X_{37}^{-} \cdot X_{76}^{-+}    
& \ \ \ \ &     
X_{62}^{+} \cdot X_{24}^{+} - X_{68}^{+} \cdot X_{84}^{+}    
\\  
\Lambda_{64}^{--} 
& :\ \ \ &  
X_{43}^{+} \cdot X_{37}^{+} \cdot X_{76}^{+-}    
-
X_{43}^{-} \cdot X_{37}^{+} \cdot X_{76}^{++} & \ \ \ \ &     
X_{62}^{-} \cdot X_{24}^{-} - X_{68}^{-} \cdot X_{84}^{-}    
\\  
\Lambda_{64}^{+-} 
& :\ \ \ &  
X_{43}^{-} \cdot X_{37}^{-} \cdot X_{76}^{++} - X_{43}^{+} \cdot X_{37}^{-} \cdot X_{76}^{+-}    
& \ \ \ \ &     
X_{62}^{+} \cdot X_{24}^{-} - X_{68}^{-} \cdot X_{84}^{+}    
\\  
\Lambda_{64}^{-+} 
& :\ \ \ &  
X_{43}^{-} \cdot X_{37}^{+} \cdot X_{76}^{-+}    
-
X_{43}^{+} \cdot X_{37}^{+} \cdot X_{76}^{--} & \ \ \ \ &     
X_{62}^{-} \cdot X_{24}^{+} - X_{68}^{+} \cdot X_{84}^{-}    
\\ 
\Lambda_{23}^{++} 
& :\ \ \ &  
X_{37}^{+} \cdot X_{76}^{--} \cdot X_{62}^{-} - X_{37}^{-} \cdot X_{76}^{--} 
\cdot X_{62}^{+}
& \ \ \ \ &     
X_{24}^{+} \cdot X_{43}^{+} - X_{21}^{+} \cdot X_{13}^{+}   
\\ 
\Lambda_{23}^{--} 
& :\ \ \ &  
X_{37}^{+} \cdot X_{76}^{++} \cdot X_{62}^{-} - X_{37}^{-} \cdot X_{76}^{++} \cdot X_{62}^{+}    
& \ \ \ \ &     
X_{24}^{-} \cdot X_{43}^{-} - X_{21}^{-} \cdot X_{13}^{-} 
\\ 
\Lambda_{23}^{+-} 
& :\ \ \ &  
X_{37}^{-} \cdot X_{76}^{-+} \cdot X_{62}^{+}    
-
X_{37}^{+} \cdot X_{76}^{-+} \cdot X_{62}^{-} & \ \ \ \ &     
X_{24}^{+} \cdot X_{43}^{-} - X_{21}^{-} \cdot X_{13}^{+}   
\\ 
\Lambda_{23}^{-+} 
& :\ \ \ &  
X_{37}^{-} \cdot X_{76}^{+-} \cdot X_{62}^{+}   
-
X_{37}^{+} \cdot X_{76}^{+-} \cdot X_{62}^{-} & \ \ \ \ &     
X_{24}^{-} \cdot X_{43}^{+} - X_{21}^{+} \cdot X_{13}^{-}  
\\ 
\Lambda_{61}^{++} 
& :\ \ \ &  
X_{13}^{-} \cdot X_{37}^{-} \cdot X_{76}^{+-} - X_{13}^{+} \cdot X_{37}^{-} \cdot X_{76}^{--} 
& \ \ \ \ &     
X_{62}^{+} \cdot X_{21}^{+} - X_{65}^{+} \cdot X_{51}^{+}     
\\ 
\Lambda_{61}^{--} 
& :\ \ \ &  
X_{13}^{-} \cdot X_{37}^{+} \cdot X_{76}^{++}  -
X_{13}^{+} \cdot X_{37}^{+} \cdot X_{76}^{-+} 
& \ \ \ \ &     
X_{62}^{-} \cdot X_{21}^{-} - X_{65}^{-} \cdot X_{51}^{-}    
\\ 
\Lambda_{61}^{+-} 
& :\ \ \ & 
X_{13}^{+} \cdot X_{37}^{-} \cdot X_{76}^{-+}  -
X_{13}^{-} \cdot X_{37}^{-} \cdot X_{76}^{++} 
& \ \ \ \ &     
X_{62}^{+} \cdot X_{21}^{-} - X_{65}^{-} \cdot X_{51}^{+}     
\\ 
\Lambda_{61}^{-+} 
& :\ \ \ &  
X_{13}^{+} \cdot X_{37}^{+} \cdot X_{76}^{--} - X_{13}^{-} \cdot X_{37}^{+} \cdot X_{76}^{+-}   
& \ \ \ \ &     
X_{62}^{-} \cdot X_{21}^{+} - X_{65}^{+} \cdot X_{51}^{-} 
\\ 
\end{array}
\label{q111z2d-je}
\eeq
}

The brick matchings are given by
\beal{q111z2d-px}
\scriptsize
P =
\left(
\begin{array}{c|cccccc|ccccccccccccccc}
\; & p_1 & p_2 & p_3 & p_4 & p_5 & p_6 & s_1 & s_2 & s_3 & s_4 & s_5 & s_6 & s_7 & s_8 & s_9 & s_{10} & s_{11} & s_{12} & s_{13} & s_{14} & s_{15}\\
\hline
X_{37}^{+} & 1 & 0 & 0 & 0 & 0 & 0 & 1 & 0 & 0 & 0 & 0 & 0 & 0 & 0 & 0 & 0 & 0 & 0 & 0 & 0 & 0 \\
X_{37}^{-} & 0 & 1 & 0 & 0 & 0 & 0 & 1 & 0 & 0 & 0 & 0 & 0 & 0 & 0 & 0 & 0 & 0 & 0 & 0 & 0 & 0 \\
X_{51}^{+} & 1 & 0 & 0 & 0 & 0 & 0 & 0 & 0 & 0 & 0 & 0 & 1 & 0 & 1 & 1 & 0 & 0 & 1 & 1 & 0 & 0 \\
X_{51}^{-} & 0 & 1 & 0 & 0 & 0 & 0 & 0 & 0 & 0 & 0 & 0 & 1 & 0 & 1 & 1 & 0 & 0 & 1 & 1 & 0 & 0 \\
X_{62}^{+} & 1 & 0 & 0 & 0 & 0 & 0 & 0 & 0 & 0 & 0 & 0 & 0 & 0 & 1 & 1 & 1 & 1 & 0 & 0 & 0 & 0 \\
X_{62}^{-} & 0 & 1 & 0 & 0 & 0 & 0 & 0 & 0 & 0 & 0 & 0 & 0 & 0 & 1 & 1 & 1 & 1 & 0 & 0 & 0 & 0 \\
X_{84}^{+} & 1 & 0 & 0 & 0 & 0 & 0 & 0 & 0 & 0 & 0 & 1 & 0 & 0 & 0 & 1 & 0 & 1 & 0 & 1 & 0 & 1 \\
X_{84}^{-} & 0 & 1 & 0 & 0 & 0 & 0 & 0 & 0 & 0 & 0 & 1 & 0 & 0 & 0 & 1 & 0 & 1 & 0 & 1 & 0 & 1 \\
X_{13}^{+} & 0 & 0 & 1 & 0 & 0 & 0 & 0 & 1 & 0 & 1 & 1 & 0 & 0 & 0 & 0 & 0 & 0 & 0 & 0 & 0 & 0 \\
X_{13}^{-} & 0 & 0 & 0 & 1 & 0 & 0 & 0 & 1 & 0 & 1 & 1 & 0 & 0 & 0 & 0 & 0 & 0 & 0 & 0 & 0 & 0 \\
X_{24}^{+} & 0 & 0 & 1 & 0 & 0 & 0 & 0 & 0 & 0 & 1 & 1 & 0 & 0 & 0 & 0 & 0 & 0 & 1 & 1 & 1 & 1 \\
X_{24}^{-} & 0 & 0 & 0 & 1 & 0 & 0 & 0 & 0 & 0 & 1 & 1 & 0 & 0 & 0 & 0 & 0 & 0 & 1 & 1 & 1 & 1 \\
X_{68}^{+} & 0 & 0 & 1 & 0 & 0 & 0 & 0 & 0 & 0 & 1 & 0 & 0 & 0 & 1 & 0 & 1 & 0 & 1 & 0 & 1 & 0 \\
X_{68}^{-} & 0 & 0 & 0 & 1 & 0 & 0 & 0 & 0 & 0 & 1 & 0 & 0 & 0 & 1 & 0 & 1 & 0 & 1 & 0 & 1 & 0 \\
X_{21}^{+} & 0 & 0 & 0 & 0 & 1 & 0 & 0 & 0 & 0 & 0 & 0 & 1 & 1 & 0 & 0 & 0 & 0 & 1 & 1 & 1 & 1 \\
X_{21}^{-} & 0 & 0 & 0 & 0 & 0 & 1 & 0 & 0 & 0 & 0 & 0 & 1 & 1 & 0 & 0 & 0 & 0 & 1 & 1 & 1 & 1 \\
X_{43}^{+} & 0 & 0 & 0 & 0 & 1 & 0 & 0 & 1 & 0 & 0 & 0 & 1 & 1 & 0 & 0 & 0 & 0 & 0 & 0 & 0 & 0 \\
X_{43}^{-} & 0 & 0 & 0 & 0 & 0 & 1 & 0 & 1 & 0 & 0 & 0 & 1 & 1 & 0 & 0 & 0 & 0 & 0 & 0 & 0 & 0 \\
X_{65}^{+} & 0 & 0 & 0 & 0 & 1 & 0 & 0 & 0 & 0 & 0 & 0 & 0 & 1 & 0 & 0 & 1 & 1 & 0 & 0 & 1 & 1 \\
X_{65}^{-} & 0 & 0 & 0 & 0 & 0 & 1 & 0 & 0 & 0 & 0 & 0 & 0 & 1 & 0 & 0 & 1 & 1 & 0 & 0 & 1 & 1 \\
X_{76}^{++} & 0 & 0 & 1 & 0 & 1 & 0 & 0 & 0 & 1 & 0 & 0 & 0 & 0 & 0 & 0 & 0 & 0 & 0 & 0 & 0 & 0 \\
X_{76}^{+-} & 0 & 0 & 1 & 0 & 0 & 1 & 0 & 0 & 1 & 0 & 0 & 0 & 0 & 0 & 0 & 0 & 0 & 0 & 0 & 0 & 0 \\
X_{76}^{-+} & 0 & 0 & 0 & 1 & 1 & 0 & 0 & 0 & 1 & 0 & 0 & 0 & 0 & 0 & 0 & 0 & 0 & 0 & 0 & 0 & 0 \\
X_{76}^{--} & 0 & 0 & 0 & 1 & 0 & 1 & 0 & 0 & 1 & 0 & 0 & 0 & 0 & 0 & 0 & 0 & 0 & 0 & 0 & 0 & 0 \\
\end{array}
\right)~,~
\nn\\
\eea
\beal{q111z2d-pl}
\scriptsize
P_\Lambda =
\left(
\begin{array}{c|cccccc|ccccccccccccccc}
\; & p_1 & p_2 & p_3 & p_4 & p_5 & p_6 & s_1 & s_2 & s_3 & s_4 & s_5 & s_6 & s_7 & s_8 & s_9 & s_{10} & s_{11} & s_{12} & s_{13} & s_{14} & s_{15}\\
\hline
\Lambda_{75}^{+} & 0 & 0 & 1 & 0 & 1 & 1 & 0 & 0 & 1 & 0 & 0 & 0 & 1 & 0 & 0 & 1 & 1 & 0 & 0 & 1 & 1 \\
\Lambda_{75}^{-} & 0 & 0 & 0 & 1 & 1 & 1 & 0 & 0 & 1 & 0 & 0 & 0 & 1 & 0 & 0 & 1 & 1 & 0 & 0 & 1 & 1 \\
\Lambda_{78}^{+} & 0 & 0 & 1 & 1 & 1 & 0 & 0 & 0 & 1 & 1 & 0 & 0 & 0 & 1 & 0 & 1 & 0 & 1 & 0 & 1 & 0 \\
\Lambda_{78}^{-} & 0† & 0 & 1 & 1 & 0 & 1 & 0 & 0 & 1 & 1 & 0 & 0 & 0 & 1 & 0 & 1 & 0 & 1 & 0 & 1 & 0 \\
\Lambda_{64}^{++} & 1 & 0 & 1 & 0 & 0 & 0 & 0 & 0 & 0 & 1 & 1 & 0 & 0 & 1 & 1 & 1 & 1 & 1 & 1 & 1 & 1 \\
\Lambda_{64}^{+-} & 1 & 0 & 0 & 1 & 0 & 0 & 0 & 0 & 0 & 1 & 1 & 0 & 0 & 1 & 1 & 1 & 1 & 1 & 1 & 1 & 1 \\
\Lambda_{64}^{-+} & 0 & 1 & 1 & 0 & 0 & 0 & 0 & 0 & 0 & 1 & 1 & 0 & 0 & 1 & 1 & 1 & 1 & 1 & 1 & 1 & 1 \\
\Lambda_{64}^{--} & 0 & 1 & 0 & 1 & 0 & 0 & 0 & 0 & 0 & 1 & 1 & 0 & 0 & 1 & 1 & 1 & 1 & 1 & 1 & 1 & 1 \\
\Lambda_{23}^{++} & 0 & 0 & 1 & 0 & 1 & 0 & 0 & 1 & 0 & 1 & 1 & 1 & 1 & 0 & 0 & 0 & 0 & 1 & 1 & 1 & 1 \\
\Lambda_{23}^{+-} & 0 & 0 & 1 & 0 & 0 & 1 & 0 & 1 & 0 & 1 & 1 & 1 & 1 & 0 & 0 & 0 & 0 & 1 & 1 & 1 & 1 \\
\Lambda_{23}^{-+} & 0 & 0 & 0 & 1 & 1 & 0 & 0 & 1 & 0 & 1 & 1 & 1 & 1 & 0 & 0 & 0 & 0 & 1 & 1 & 1 & 1 \\
\Lambda_{23}^{--} & 0 & 0 & 0 & 1 & 0 & 1 & 0 & 1 & 0 & 1 & 1 & 1 & 1 & 0 & 0 & 0 & 0 & 1 & 1 & 1 & 1 \\
\Lambda_{61}^{++} & 1 & 0 & 0 & 0 & 1 & 0 & 0 & 0 & 0 & 0 & 0 & 1 & 1 & 1 & 1 & 1 & 1 & 1 & 1 & 1 & 1 \\
\Lambda_{61}^{+-} & 1 & 0 & 0 & 0 & 0 & 1 & 0 & 0 & 0 & 0 & 0 & 1 & 1 & 1 & 1 & 1 & 1 & 1 & 1 & 1 & 1 \\
\Lambda_{61}^{-+} & 0 & 1 & 0 & 0 & 1 & 0 & 0 & 0 & 0 & 0 & 0 & 1 & 1 & 1 & 1 & 1 & 1 & 1 & 1 & 1 & 1 \\
\Lambda_{61}^{--} & 0 & 1 & 0 & 0 & 0 & 1 & 0 & 0 & 0 & 0 & 0 & 1 & 1 & 1 & 1 & 1 & 1 & 1 & 1 & 1 & 1 \\
\end{array}
\right)~.~
\eea

Figures \ref{fq111z2d-brick2} and \ref{fq111z2d-brick1} show the brane brick model and individual bricks for this theory.

\begin{figure}[H]
\begin{center}
\resizebox{0.5\hsize}{!}{
\includegraphics[trim=0cm 0cm 0cm 0cm,totalheight=10 cm]{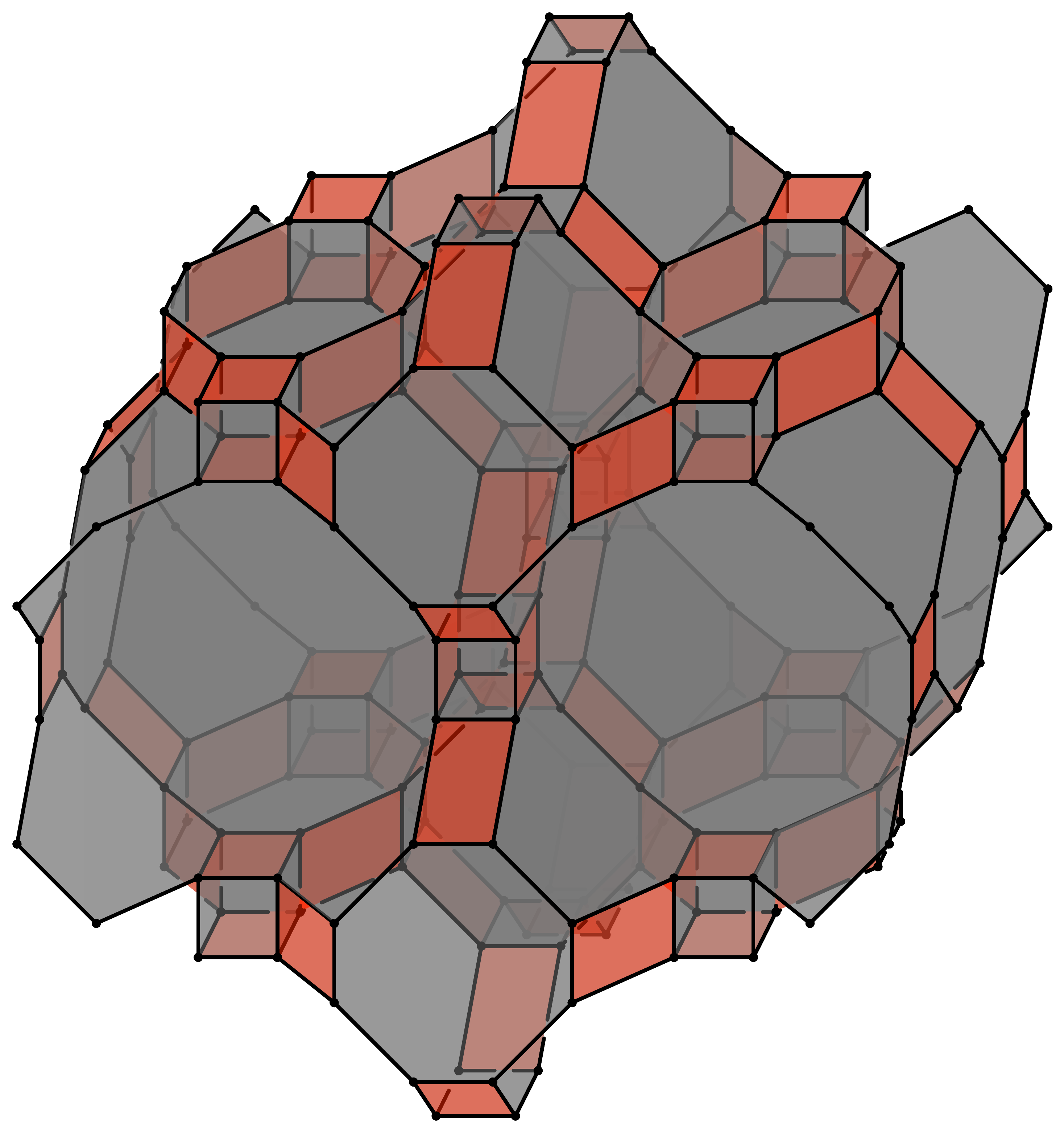}
}  
\caption{
Brane brick model for phase D of $Q^{1,1,1}/\mathbb{Z}_2$.
\label{fq111z2d-brick2}}
 \end{center}
 \end{figure} 

\begin{figure}[H]
\begin{center}
\resizebox{0.7\hsize}{!}{
\includegraphics[trim=0cm 0cm 0cm 0cm,totalheight=10 cm]{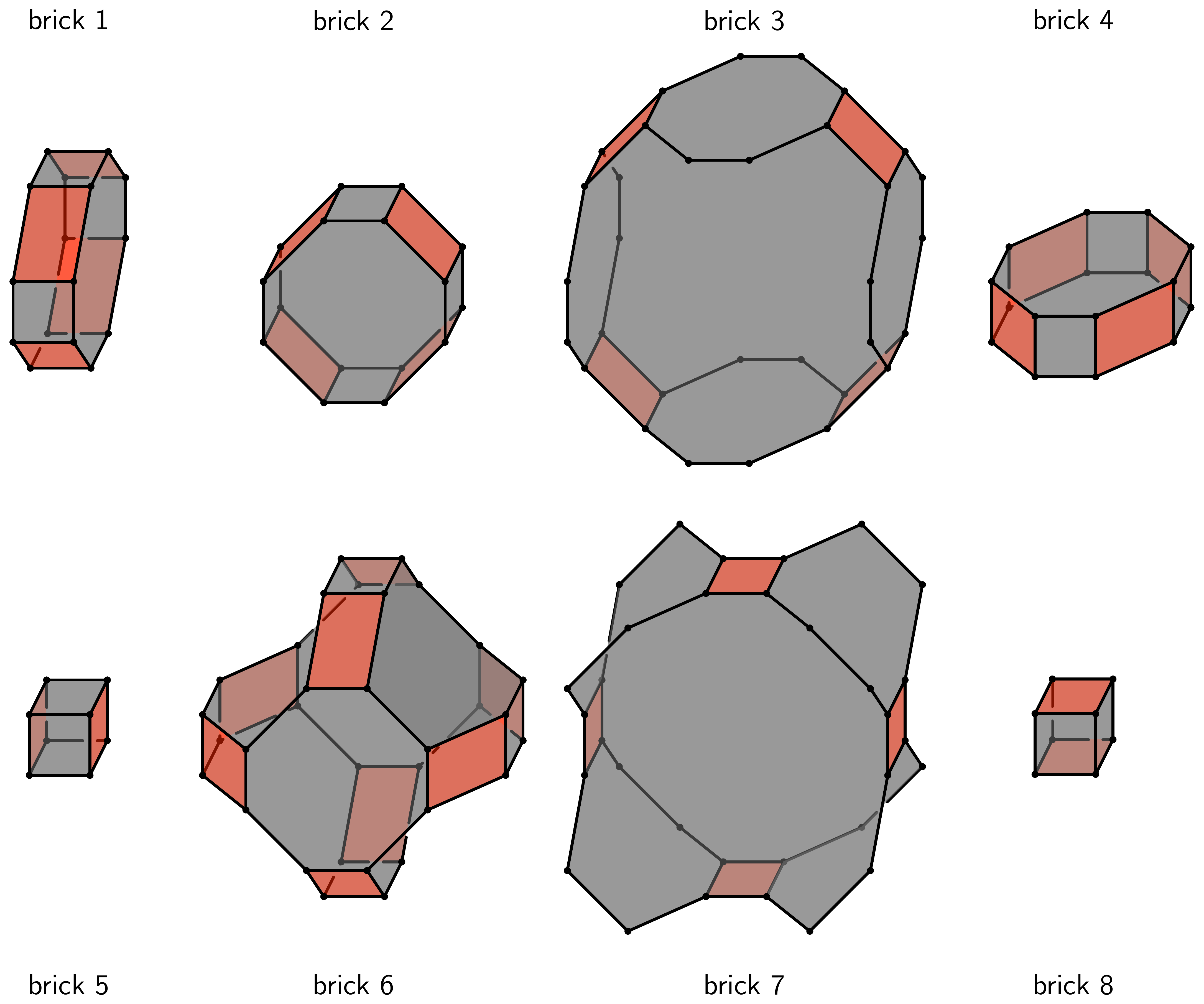}
}  
\caption{
Brane bricks for phase D of $Q^{1,1,1}/\mathbb{Z}_2$.
\label{fq111z2d-brick1}}
 \end{center}
 \end{figure} 

The chiral and Fermi field pieces of the phase boundary matrix are
\beal{q111z2d-hx}
\scriptsize
H_X=
\left(
\begin{array}{c|cccccccccccc}
\; &
\eta_{13} & \eta_{14} & \eta_{23} & \eta_{24} & \eta_{15} & \eta_{16} & \eta_{25} & \eta_{26} & \eta_{35} & \eta_{36} & \eta_{45} & \eta_{46} 
\\
\hline
X_{37}^{+} & 1 & 1 & 0 & 0 & 1 & 1 & 0 & 0 & 0 & 0 & 0 & 0 \\
X_{37}^{-} & 0 & 0 & 1 & 1 & 0 & 0 & 1 & 1 & 0 & 0 & 0 & 0 \\
X_{51}^{+} & 1 & 1 & 0 & 0 & 1 & 1 & 0 & 0 & 0 & 0 & 0 & 0 \\
X_{51}^{-} & 0 & 0 & 1 & 1 & 0 & 0 & 1 & 1 & 0 & 0 & 0 & 0 \\
X_{62}^{+} & 1 & 1 & 0 & 0 & 1 & 1 & 0 & 0 & 0 & 0 & 0 & 0 \\
X_{62}^{-} & 0 & 0 & 1 & 1 & 0 & 0 & 1 & 1 & 0 & 0 & 0 & 0 \\
X_{84}^{+} & 1 & 1 & 0 & 0 & 1 & 1 & 0 & 0 & 0 & 0 & 0 & 0 \\
X_{84}^{-} & 0 & 0 & 1 & 1 & 0 & 0 & 1 & 1 & 0 & 0 & 0 & 0 \\
X_{13}^{+} & -1 & 0 & -1 & 0 & 0 & 0 & 0 & 0 & 1 & 1 & 0 & 0 \\
X_{13}^{-} & 0 & -1 & 0 & -1 & 0 & 0 & 0 & 0 & 0 & 0 & 1 & 1 \\
X_{24}^{+} & -1 & 0 & -1 & 0 & 0 & 0 & 0 & 0 & 1 & 1 & 0 & 0 \\
X_{24}^{-} & 0 & -1 & 0 & -1 & 0 & 0 & 0 & 0 & 0 & 0 & 1 & 1 \\
X_{68}^{+} & -1 & 0 & -1 & 0 & 0 & 0 & 0 & 0 & 1 & 1 & 0 & 0 \\
X_{68}^{-} & 0 & -1 & 0 & -1 & 0 & 0 & 0 & 0 & 0 & 0 & 1 & 1 \\
X_{21}^{+} & 0 & 0 & 0 & 0 & -1 & 0 & -1 & 0 & -1 & 0 & -1 & 0 \\
X_{21}^{-} & 0 & 0 & 0 & 0 & 0 & -1 & 0 & -1 & 0 & -1 & 0 & -1 \\
X_{43}^{+} & 0 & 0 & 0 & 0 & -1 & 0 & -1 & 0 & -1 & 0 & -1 & 0 \\
X_{43}^{-} & 0 & 0 & 0 & 0 & 0 & -1 & 0 & -1 & 0 & -1 & 0 & -1 \\
X_{65}^{+} & 0 & 0 & 0 & 0 & -1 & 0 & -1 & 0 & -1 & 0 & -1 & 0 \\
X_{65}^{-} & 0 & 0 & 0 & 0 & 0 & -1 & 0 & -1 & 0 & -1 & 0 & -1 \\
X_{76}^{++} & -1 & 0 & -1 & 0 & -1 & 0 & -1 & 0 & 0 & 1 & -1 & 0 \\
X_{76}^{+-} & -1 & 0 & -1 & 0 & 0 & -1 & 0 & -1 & 1 & 0 & 0 & -1 \\
X_{76}^{-+} & 0 & -1 & 0 & -1 & -1 & 0 & -1 & 0 & -1 & 0 & 0 & 1 \\
X_{76}^{--} & 0 & -1 & 0 & -1 & 0 & -1 & 0 & -1 & 0 & -1 & 1 & 0 \\
\end{array}
\right) ~,~
\eea
\beal{q111z2d-hl}
\scriptsize
H_\Lambda = 
\left(
\begin{array}{c|cccccccccccc}
\; &
\eta_{13} & \eta_{14} & \eta_{23} & \eta_{24} & \eta_{15} & \eta_{16} & \eta_{25} & \eta_{26} & \eta_{35} & \eta_{36} & \eta_{45} & \eta_{46} 
\\
\hline
\Lambda_{75}^{+} & -1 & 0 & -1 & 0 & -1 & -1 & -1 & -1 & 0 & 0 & -1 & -1 \\
\Lambda_{75}^{-} & 0 & -1 & 0 & -1 & -1 & -1 & -1 & -1 & -1 & -1 & 0 & 0 \\
\Lambda_{78}^{+} & -1 & -1 & -1 & -1 & -1 & 0 & -1 & 0 & 0 & 1 & 0 & 1 \\
\Lambda_{78}^{-} & -1 & -1 & -1 & -1 & 0 & -1 & 0 & -1 & 1 & 0 & 1 & 0 \\
\Lambda_{64}^{++} & 0 & 1 & -1 & 0 & 1 & 1 & 0 & 0 & 1 & 1 & 0 & 0 \\
\Lambda_{64}^{+-} & 1 & 0 & 0 & -1 & 1 & 1 & 0 & 0 & 0 & 0 & 1 & 1 \\
\Lambda_{64}^{-+} & -1 & 0 & 0 & 1 & 0 & 0 & 1 & 1 & 1 & 1 & 0 & 0 \\
\Lambda_{64}^{--} & 0 & -1 & 1 & 0 & 0 & 0 & 1 & 1 & 0 & 0 & 1 & 1 \\
\Lambda_{23}^{++} & -1 & 0 & -1 & 0 & -1 & 0 & -1 & 0 & 0 & 1 & -1 & 0 \\
\Lambda_{23}^{+-} & -1 & 0 & -1 & 0 & 0 & -1 & 0 & -1 & 1 & 0 & 0 & -1 \\
\Lambda_{23}^{-+} & 0 & -1 & 0 & -1 & -1 & 0 & -1 & 0 & -1 & 0 & 0 & 1 \\
\Lambda_{23}^{--} & 0 & -1 & 0 & -1 & 0 & -1 & 0 & -1 & 0 & -1 & 1 & 0 \\
\Lambda_{61}^{++} & 1 & 1 & 0 & 0 & 0 & 1 & -1 & 0 & -1 & 0 & -1 & 0 \\
\Lambda_{61}^{+-} & 1 & 1 & 0 & 0 & 1 & 0 & 0 & -1 & 0 & -1 & 0 & -1 \\
\Lambda_{61}^{-+} & 0 & 0 & 1 & 1 & -1 & 0 & 0 & 1 & -1 & 0 & -1 & 0 \\
\Lambda_{61}^{--} & 0 & 0 & 1 & 1 & 0 & -1 & 1 & 0 & 0 & -1 & 0 & -1 \\
\end{array}
\right) ~.~
\eea

\newpage

\bibliographystyle{JHEP}
\bibliography{mybib}


\end{document}